\RequirePackage{etoolbox}
\csdef{input@path}%
{%
 {sty/},
 {img/},
}

\documentclass{elsevierbook}

\usepackage{bm}
\usepackage{amsmath}
\usepackage{hyperref}

\usepackage{natbib}

\begin{document}

\setcounter{chapter}{5}

\Frontmatter
 
\Mainmatter

  \begin{frontmatter}

\chapter{MHD waves in the partially ionized plasma: from single to multi-fluid approach}\label{PI-waves}

{\bf Elena Khomenko$^{a,b}$ and David Mart{\'\i}nez-G{\'o}mez$^{a,b}$}

\noindent {\it (a) Instituto de Astrofísica de Canarias, La Laguna, Tenerife, Spain; \\
(b) Dpto de Astrofísica, Universidad de La Laguna, La Laguna, Tenerife, Spain}

 \minitoc

\begin{abstract}
 This Chapter outlines the basic properties of waves in solar partially ionized plasmas. It provides a summary of the main sets of equations, from the single-fluid formalism, to the multi-fluid one, giving examples for purely hydrogen, and for hydrogen-helium plasmas. It then discusses the solutions for waves under the single-fluid frame: the influence of the ambipolar diffusion, diamagnetic effect, and the Hall effect on the propagation, dissipation, and mode conversion of the magnetohydrodynamic waves. The Chapter continues by outlining the wave solutions in the multi-fluid formalism: the influence of the elastic inter-particle collisions into the propagation, damping and dissipation of different magnetohydrodynamic modes. Both parts discuss linear and non-linear wave solutions, and the effects of the gravitational stratification of the solar atmosphere.
\end{abstract}

\begin{keywords}
\kwd{Waves}
\kwd{Energy transfer}
\kwd{Sun}
\kwd{Atmosphere}
\end{keywords}

\end{frontmatter}

\section{Single-fluid and multi-fluid formalism} \label{sec1}

\subsection{Single-fluid multi-species plasma}
\label{sec:1f}    
   
Partially ionized (PI) plasmas are a common type of plasmas in the cosmic and in the laboratory environments. PI plasmas are usually rather cold and/or rarefied media with temperatures that do not allow hydrogen to ionize. In the Sun, partially ionized plasma fills the whole volume of the photosphere and the chromosphere, and may even be present at some locations in the (much hotter) transition regions and solar corona. Propagation of waves and shocks in such plasmas gives rise to a series of new effects, such as damping, dissipation, dispersion, or appearance of the new modes with mixed properties. The solar plasma is composed by particles of different atomic elements in different ionization stages. The mathematical description of such partially ionized plasma depends on the degree of the collision coupling of different neutral and charged components. Where collision coupling is strong enough, a single fluid approach can be used. In that case, the conservation equations take the following form.


\begin{equation}\label{eq:continuity-1fluid}
\frac{\partial \rho}{\partial t} + \nabla\cdot \left(\rho\bm{V}\right) =  0,
\end{equation}
\begin{equation}\label{eq:momentum-1fluid}
\frac{\partial (\rho\bm{V})}{\partial t} + \nabla\cdot (\rho\bm{V} \bm{V} +P \mathbb{I})  = \bm{J}\times\bm{B} + \rho\bm{g},
\end{equation}
\begin{equation}\label{eq:energy-1fluid}
\frac{\partial }{\partial t}\left(e + \frac{1}{2}\rho V^2 \right) + \nabla\cdot  \left( \bm{V}\, ( e + \frac{1}{2}\rho V^2) +P\bm{V} +  \bm{q} + \bm{F}_R \right)   = \bm{J} \cdot \bm{E}   + \rho\bm{V} \cdot \bm{g}.
\end{equation}

\noindent The macroscopic velocity, $\bm{V}$, pressure, $P$, internal energy, $e$, and heat flux, $\bm{q}$, are defined through the summation over all species composing the plasma, see \cite{Khomenko2014PhPl...21i2901K}: 

\begin{gather} \label{eq:1f-definitions}
\bm{V}  = \frac{\sum_{\alpha=1}^{2N+1}(\rho_{\alpha}\bm{V}_{\alpha})}{\rho},  \\
P = \sum_{\alpha=1}^{2N+1}\left(P_{\alpha} + \frac{1}{3}\rho_{\alpha}w_{\alpha}^2\right), \nonumber \\
e=\frac{3}{2}P + \sum_{\alpha=1}^{2N}\chi_\alpha, \nonumber \\
\bm{q} =  \sum_{\alpha=1}^{2N+1}\left(\bm{q}_{\alpha} + 
\frac{5}{2}P_{\alpha}\bm{w}_{\alpha} +\frac{1}{2}\rho_{\alpha}w_{\alpha}^2\bm{w}_{\alpha} \right), \nonumber
\end{gather}
where the quantities with sub-index $\alpha$ refer to individual plasma components ($N$ ions, $N$ neutrals of different chemical elements, and electrons), $\rho$ is the total plasma density, $\bm{w}_{\alpha}$ is the drift velocity of each specie taken with respect to $\bm{V}$, 
$\bm{w}_{\alpha} = \bm{V}_{\alpha} - \bm{V}$, and 
$\chi_\alpha$ is the total potential energy of ionization level of a species $\alpha$ (the ionization energy does not apply to electrons for obvious reasons). 

For completeness, equations above include radiative energy flux, $\bm{F}_R$. In the solar atmosphere, energy exchange by radiation plays a major role. A complex treatment for non-local equilibrium radiative transfer is especially relevant for the chromosphere, where the multi-fluid effects are also more pronounced. As for the heat conduction (also viscosity, not included in the equations above), under strong plasma magnetization, it acts differently in the directions parallel and perpendicular to the magnetic field. With weakening the collisions in the chromosphere, the classical approximations 
\citep[as those from standard plasma physics books by, e.g.,][]{Spitzer1962pfig.book.....S, 1965RvPP....1..205B, Bittencourt, Balescu} may become invalid, and more detailed treatment becomes necessary, see \cite{2022ApJS..260...26H}. In the following we will not discuss the effects of conductivity, viscosity, and radiation because they are analogous to those for waves in fully ionized plasmas. 



The system \ref{eq:continuity-1fluid}--\ref{eq:energy-1fluid} has the same form as an MHD system. The energy conservation equation is written for the sum of the internal and kinetic energies and its right hand side contains the Joule heating term, $\bm{J} \cdot \bm{E}$, where $\bm{J}$ is electric current,
\begin{equation}\label{eq:jota}
    \bm{J}=(\nabla\times\bm{B})/\mu_0 \,,
\end{equation}
and $\bm{E}$ is electric field. The form of the latter depends on the presence of extra species in a plasma. Therefore, both the energy conservation equation, and the induction equation will have additional terms due to neutrals, as described in the next section. 

\subsubsection{Generalized Ohm's law for single-fluid description}
\label{sect:GOL}

The expression for the electric field required for the single-fluid system \ref{eq:continuity-1fluid}--\ref{eq:energy-1fluid} is obtained from the generalized Ohm's law (GOL). Derivation of the GOL can be found in many plasma physics books, for example in \cite{1965RvPP....1..205B,  Krall+Trivelpiece1973, Bittencourt}. For the specific case of the Sun, derivation of the multi-species GOL is provided in, e.g., \cite{Khomenko2014PhPl...21i2901K} and \cite{2018SSRv..214...58B}. In the single-fluid description, the electric field is computed in the system of reference of the center of mass velocity of the whole plasma, including its neutral components, $\bm{V}$, and the GOL takes the following form,

\begin{equation} \label{eq:ohm-1fluid}
[\bm{E} + \bm{V}\times{\bm{B}}]= \eta\mu_0\bm{J} + \eta_H\mu_0\frac{[\bm{J} \times \bm{B}]}{|B|} - \eta_H\mu_0\frac{\nabla P_e}{|B|}   - \eta_A\mu_0\frac{[(\bm{J} \times \bm{B}) \times \bm{B}]}{|B|^2} + \frac{\xi_n}{\alpha_n}[\bm{G}\times \bm{B}],
\end{equation}

This expression contains at the right hand side the Ohmic, Hall, Biermann battery, ambipolar, and diamagnetic terms. In the expression for the battery term, $P_e$ is the electron pressure. The $\bm{G}$ vector is a combination of partial pressure gradients and is given by,
\begin{equation}
    \bm{G}=\xi_n\nabla (P_e+P_i) + (1-\xi_n)\nabla P_n,
\end{equation}
where $P_i$ and $P_n$ is the pressure of ions and neutrals, respectively. The coefficients of all the terms in Eq. \ref{eq:ohm-1fluid} are provided as follows, in the units of $[m^2 s^{-1}]$,
\begin{equation}\label{eq:etas}
\eta = \frac{\alpha_e}{\mu_0 e^2 n_e^2}; \,\,\,
\eta_A =\frac{\xi_n^2|B|^2}{\mu_0\alpha_n}; \,\,\,
\eta_H=\frac{|B|}{\mu_0 e n_e}, 
\end{equation}

\noindent with $\xi_n=\rho_n/\rho$ being the neutral fraction, $\rho_n$ neutral mass density, $n_e$ electron number density, $e$ electron charge. Coefficients $\alpha_n$ and $\alpha_e$ are neutral and electron collision parameters, that take into account collisions between all species in the plasma,

\begin{equation} \label{eq:alphan}
\alpha_n=\sum_{\beta=1}^N{\rho_e\nu_{en_\beta}} + \sum_{\alpha=1}^N\sum_{\beta=1}^N\rho_{i_\alpha}\nu_{i_\alpha n_\beta} \,\,;
\alpha_e=\sum_{\alpha=1}^N{\rho_e\nu_{ei_\alpha}} + \sum_{\beta=1}^N\rho_e\nu_{e n_\beta} \,.
\end{equation}
where sub-index $n_\beta$ indicate neutrals of the type $\beta$, $i_\alpha$ indicate ions of the type $\alpha$. The expressions for the collision frequencies between ions and neutrals ($\nu_{i_\alpha n_\beta}$) and  electrons with neutrals ($\nu_{en_\beta}$), can be taken from  \cite{Spitzer1962pfig.book.....S}, and the expressions for collisions between electrons and ions ($\nu_{ei_\alpha}$) can be taken from \cite{1965RvPP....1..205B}: 
\begin{equation} \label{eq:coll-freq-1fluid}
    \nu_{i_\alpha n_\beta}=n_{n_\beta}\sqrt{\frac{8k_BT}{\pi m_{i_\alpha n_\beta}}}\sigma_{in}; \,
    \nu_{en_\beta}=n_{n_\beta}\sqrt{\frac{8k_BT}{\pi m_{en_\beta}}}\sigma_{en}; \,
    \nu_{ei_\alpha}=\frac{n_ee^4\ln{\Lambda}}{3\epsilon_0^2m_e^2}\left(\frac{m_e}{2\pi k_BT}\right)^{3/2},
\end{equation}
where $m_{i_\alpha n_\beta}=m_{i_\alpha}m_{n_\beta}/(m_{i_\alpha}+m_{n_\beta})$, $m_{en_\beta}=m_em_{n_\beta}/(m_e+m_{n_\beta})$ are the reduced masses, $\sigma_{in}$ and $\sigma_{en}$ are collision cross sections \citep[see, e.g.,][]{Huba2013}, $\Lambda$ is the Coulomb logarithm.

The generalized induction equation is obtained by using GOL, Eq.~\ref{eq:ohm-1fluid}, together with the Faraday's law, 
\begin{equation} \label{eq:induction-1fluid}
\frac{\partial \bm{B}}{\partial t} = \nabla\times \left[{\bm V} \times {\bm B} -\eta\mu_0\bm{J} - \eta_A\mu_0\bm{J}_\perp + \frac{\nabla P_e}{e n_e} -\eta_H\mu_0\frac{[\bm{J} \times \bm{B}]}{|B|} - \frac{\xi_n}{\alpha_n}[\bm{G}\times \bm{B}] \right] ,
\end{equation}
where we have defined the current perpendicular to magnetic field as, $\bm{J}_\perp=-(\bm{J}\times\bm{B})\times\bm{B}/|B|^2$.  The relative importance of each of the non-ideal terms in the generalized induction equation, Eq.~\ref{eq:induction-1fluid}, can be evaluated by comparing the values of these coefficients in a solar atmospheric model.


  \begin{figure} [!t]
        \centering
        \includegraphics[width=0.95\hsize]{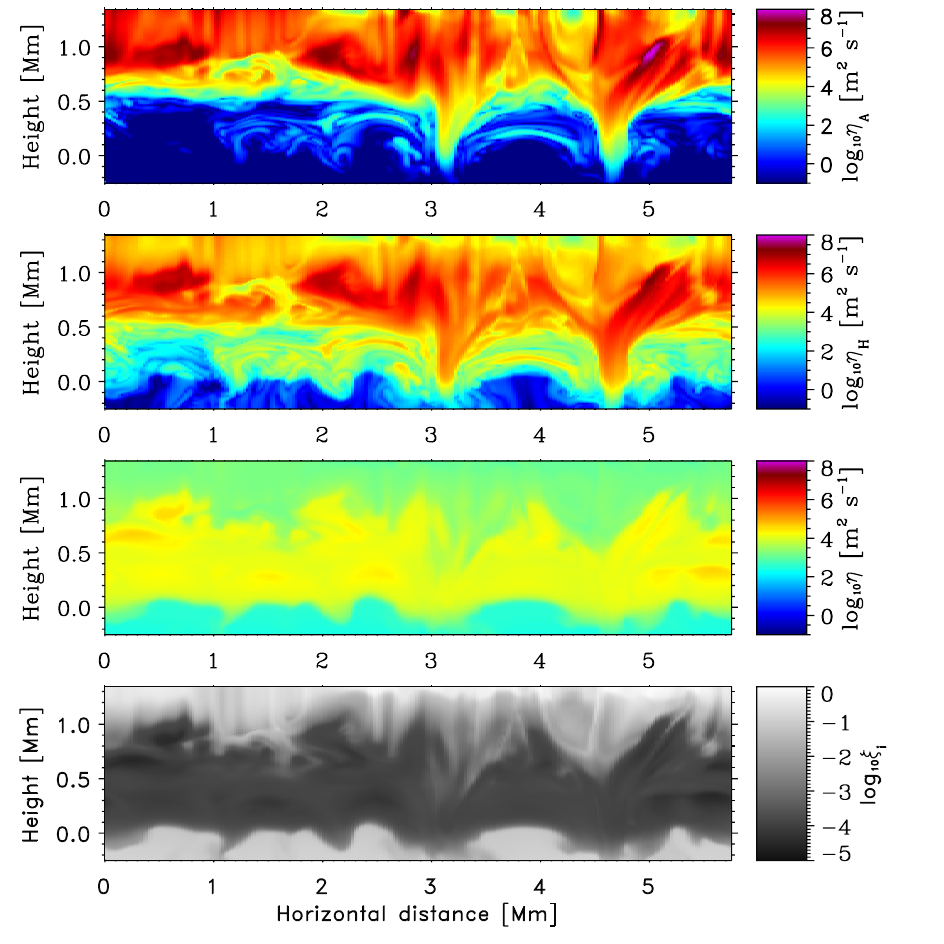}
        \caption{Three top panels: coefficients of the generalized induction equation defined in Eq.~\ref{eq:etas}, ambipolar, $\eta_A$, Hall, $\eta_H$, and Ohmic $\eta$. Bottom panel gives the ion fraction, $\xi_i=\rho_i/\rho$. All the quantities are shown along a horizontal direction and height in a 2D cut of a 3D magneto-convection model, computed with the Mancha3D code \citep{2018A&A...618A..87K}.}
        \label{fig:etas_snapshot}
    \end{figure}

Figure \ref{fig:etas_snapshot} shows the coefficients $\eta_A$, $\eta_H$, and $\eta$  as a function of horizontal distance and height in a numerically computed solar magneto-convection model with average magnetic field of 50 G \citep{2018A&A...618A..87K}. The height range covers from slightly below the photosphere to the middle chromosphere. The ion fraction (bottom panel) at these layers drops as low as $\xi_i=10^{-4}-10^{-5}$ (very weakly ionized), while recovering values around 1 (fully ionized) below the photosphere and in the upper chromosphere. As a consequence, both ambipolar, $\eta_A$, and Hall, $\eta_H$ coefficients take large values, exceeding by several orders of magnitude the Ohmic coefficient, $\eta$, from the middle photosphere up. The Hall coefficient dominates over the ambipolar one in the photosphere, at heights between 0 and 0.5 Mm, while the ambipolar coefficient dominates over all three at heights above 1 Mm. One can also observe strong horizontal variations in the values of the coefficients caused by the variations in the atmospheric parameters, temperature, density, and magnetic field. At locations with strong magnetic flux tubes, as those seen around horizontal coordinates 3 and 4.5 Mm, both $\eta_A$, and $\eta_H$ have high values through the whole photospheric layers. Therefore, it can be expected that both ambipolar and Hall effects would significantly affect wave behavior in magnetized photosphere and chromosphere.

Once the generalized induction equation is obtained, it can be used to derive the total energy conservation equation, including non-ideal terms,

\begin{gather} \label{eq:etot-1fluid}
\frac{\partial e_{\rm tot}}{\partial t}  + \nabla\cdot\Big[{\bm V}\Big(e_{\rm tot}+P+\frac{|{\bm B}|^2}{2 \mu_0}\Big)-\frac{\bm{B}({\bm V\cdot B})}{\mu_0} + \bm{q} + \bm{F}_R\Big] \\ \nonumber
+  \nabla\cdot\Big[(\eta_A + \eta)[\bm{J_\perp} \times \bm{B}] - \frac{\nabla P_e \times \bm{B}} {e n_e } + \eta_p\bm{G}_{\perp}\Big]=  
\rho\bm{V}\cdot\bm{g},
\end{gather}
with $e_{\rm tot}$ defined as,
\begin{equation}
e_{\rm tot}=\frac{1}{2}\rho \bm{V}^2+e+\frac{\bm{B}^2}{2\mu_0},
\end{equation}
and
\begin{equation}
    \bm{G}_{\perp} = -\frac{\bm{G}\times\bm{B}\times\bm{B}}{|B|^2}; \,\, \eta_p=\frac{\xi_n|B|^2}{\mu_0\alpha_n}.
\end{equation}

It can be seen that the Hall effect does not affect the energy of the system, while both Ohm and ambipolar effects provide a dissipation term, proportional to the sum of both coefficients, $\eta$ and $\eta_A$. Since $\eta_A$ exceeds $\eta$ by several orders of magnitude in the chromosphere, it can be expected that the ambipolar effect will be the main source of dissipation for chromospheric waves. The battery effect also has its corresponding counterpart in the energy equation, though in practice, its influence into the energy is rather small for the typical parameters of solar partially ionized plasma. The last term in the square brackets, the $\bm{G}_{\perp}$-term, is frequently (and unjustifiably) omitted.

Alternatively, energy conservation can be cast for the internal energy only,
\begin{equation} \label{eq:eint-1fluid}
\frac{\partial e}{\partial t}+ \nabla\cdot \left(\bm{V}e +\bm{q}+ \bm{F}_R \right) + P\nabla\cdot\bm{V} =  \eta\mu_0 J^2 + \eta_A\mu_0{J_\perp}^2 
- \bm{J}\cdot\frac{\nabla P_e}{e n_e} + \bm{J}\cdot\frac{\xi_n}{\alpha_n}[\bm{G}\times\bm{B}].
\end{equation}
The right hand side of this equation contains Joule heating terms, related to the Ohmic and ambipolar diffusion, as well as the battery and $G$-term. All these non-ideal terms directly affect the internal energy of the system. The Joule heating terms are always positive and lead to the internal energy increase. The ambipolar term is expected to be the leading one in the solar chromosphere. The Ohmic term is proportional to the square of total currents, $J^2$, while the ambipolar term only affects the currents perpendicular to the magnetic field, $J_\perp$. This makes the fundamental difference between both effects.

Equations \ref{eq:continuity-1fluid}, \ref{eq:momentum-1fluid}, \ref{eq:induction-1fluid} and either of Eq.~\ref{eq:etot-1fluid} or \ref{eq:eint-1fluid} provide a closed set of equations of single-fluid MHD that can be used for the wave analysis. The advantage of the single-fluid MHD is that one does specifically address the question of plasma composition except when computing the ambipolar, Hall and battery coefficients. Therefore, large scale simulations can be easily performed.


\subsection{Two-fluid multi-species plasma} \label{sec:eqs_2fluid}

Once the collision coupling weakens with height in the solar atmosphere, all individual species behave in a slightly different way, moving with different velocities. Strictly speaking, this requires considering particles of different chemical elements and ionization stages as separate fluids, see e.g.,  \cite{Khomenko2014PhPl...21i2901K}. However, such a description may result unpractical since many equations need to be solved. A good approximation is to assume that the difference in behavior between neutrals and charges is larger than between the neutrals/charges of different kinds themselves, since the latter feel the presence of the magnetic field and the former do not. This assumption allows to decrease the number of equations for different species to just two, for an average neutral particle and an average charged particle. This brings to the following system of equations,

\begin{equation} \label{eq:continuity-2fluid-n}
        \frac{\partial \rho_{\rm{n}}}{\partial t} + \nabla \cdot \left(\rho_{\rm{n}} \bm{V}_{\rm{n}}\right) = S_{n},
\end{equation}
\begin{equation} \label{eq:continuity-2fluid-c}
        \frac{\partial \rho_{\rm{c}}}{\partial t} + \nabla \cdot \left(\rho_{\rm{c}} \bm{V}_{\rm{c}}\right) = -S_{n},
\end{equation}
\begin{equation} \label{eq:momentum-2fluid-n}
        \frac{\partial \left(\rho_{\textrm{n}} \bm{V}_{\textrm{n}}\right)}{\partial t} + \nabla \cdot \left( \rho_{\rm{n}} \bm{V}_{\textrm{n}} \bm{V}_{\textrm{n}} + P_{\textrm{n}}\mathbb{I} \right) = \rho_{\rm{n}} \bm{g} + \bm{R}_{\textrm{n}},
\end{equation}
\begin{equation} \label{eq:momentum-2fluid-c}
        \frac{\partial \left(\rho_{\rm{c}} \bm{V}_{\rm{c}}\right)}{\partial t} + \nabla \cdot \left( \rho_{\rm{c}} \bm{V}_{\rm{c}} \bm{V}_{\rm{c}} + P_{\textrm{c}}\mathbb{I} \right) = \bm{J} \times \bm{B} + \rho_{\rm{c}} \bm{g}- \bm{R}_{\rm{n}},
\end{equation}
\begin{equation} \label{eq:energy-2fluid-n}
        \frac{\partial}{\partial t}\left(e_{\rm{n}} + \frac{1}{2}\rho_{\rm{n}} \bm{V}_{\rm{n}}^{2} \right) + \nabla \cdot \left( \bm{V}_{\rm{n}} (e_{\rm{n}} + \frac{1}{2} \rho_{\rm{n}} \bm{V}_{\rm{n}}^{2} ) + P_{\rm{n}} \cdot \bm{V}_{\rm{n}} + \bm{q}_{\rm n} +  \bm{F}_R^n\right) = -\rho_{\rm{n}} \bm{V}_{\rm{n}} \cdot \bm{g} + M_{\rm{n}},
\end{equation}
\begin{equation} \label{eq:energy-2fluid-c}
        \frac{\partial}{\partial t}\left(e_{\rm{c}} + \frac{1}{2}\rho_{\rm{c}} \bm{V}_{\rm{c}}^{2} \right) + \nabla \cdot \left( \bm{V}_{\rm{c}} (e_{\rm{c}} + \frac{1}{2} \rho_{\rm{c}} \bm{V}_{\rm{c}}^{2} ) + P_{\rm{c}} \cdot \bm{V}_{\rm{c}} + \bm{q}_{\rm c} +  \bm{F}_R^c\right) = -\rho_{\rm{c}} \bm{V}_{\rm{c}} \cdot \bm{g} + \bm{J} \cdot \bm{E} - M_{\rm{n}}.
\end{equation}
In these equations, the sub index ``n'' is for neutrals and ``c'' is for charges, the rest of the notations is the same as in the single-fluid case. The definitions of the partial pressures ($P_n$ and $P_c$), the heat flux vectors ($\bm{q}_{n}$ and  $\bm{q}_{c}$), the internal energies ($e_n$ and $e_c$), and the radiative energy fluxes ($\bm{F}_R^n$ and $\bm{F}_R^c$) all involve summation over the species present in the plasma, similar to Eq. \ref{eq:1f-definitions}, but separately for the charges (ions and electrons) and neutrals \citep{Khomenko2014PhPl...21i2901K}. 

\begin{figure} [!t]
    \centering
    \includegraphics[width=0.95\hsize]{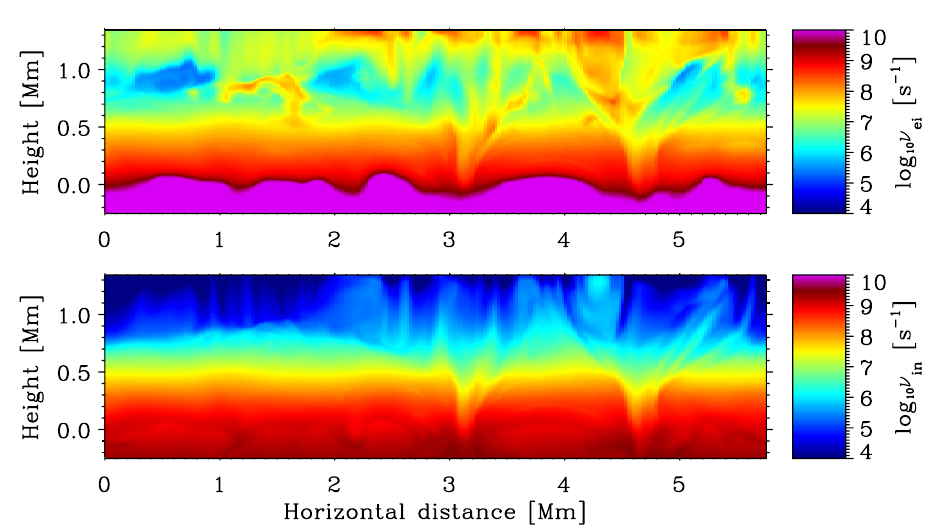}
    \caption{Collision frequencies between electrons and ions (top) and ions and neutrals (bottom) defined in Eq.~\ref{eq:coll-freq-1fluid} shown along a horizontal direction and height in a 2D cut of a 3D magneto-convection model, computed with the Mancha3D code \citep{2018A&A...618A..87K}.}
        \label{fig:collisions_snapshot}
    \end{figure}

The two-fluid description defined above requires a stronger coupling between charged particles than between charged and neutral particles \citep{Zaqarashvili2011A&A...534A..93Z}. An example calculation of collision frequencies between the charged and neutral particles is given in Fig.~\ref{fig:collisions_snapshot} for a 2D cut of a numerically computed model of solar magneto-convection, same as in Fig.~\ref{fig:etas_snapshot}. It can be observed that collisions between charged particles (electrons and ions, top panel) are orders of magnitude more frequent in the chromosphere, above $\sim$0.5 Mm, compared to the collisions between ions and neutrals (bottom panel). Therefore, neutrals can be considered much weakly coupled to the rest of the plasma. This justifies grouping the fluids into two,  charged and neutral, components. Under some circumstances, the coupling between different neutral components becomes weak enough to justify splitting the neutrals into more fluids. This approach has been applied in the case of hydrogen-helium plasma in \cite{Zaqarashvili2011A&A...534A..93Z}, and will be discussed below.

The two-fluid system of equations presents several additional terms at the right hand side, $S_n$, $\bm{R}_n$ and $M_n$. Those terms describe elastic and inelastic collisions between the fluid and couple each pair of equations. While the equations Eq. \ref{eq:continuity-2fluid-n}--\ref{eq:energy-2fluid-c} are generic, the collision terms are not. At this point one needs to specify the type of colliding particles, plasma composition, and the type of collisions. 
    
\subsection{Purely hydrogen plasma}
\label{sec:2f_collisions}
    
One of the widely used cases is the one for purely hydrogen plasmas \citep{2012ApJ...760..109L, 2014SSRv..184..107L, PopescuBraileanu2019A&A...627A..25P}. 

For a purely hydrogen plasma, the internal energies and partial pressures of the neutral and charged fluid are linked as follows,
\begin{equation} \label{eq:ienes_to_pressures}
    e_n=P_n/(\gamma -1) \,; \quad e_c=P_c/(\gamma-1)+n_e\phi_{\rm ion},
\end{equation}
where $\phi_{\rm ion}$ is hydrogen ionization energy.

The collision terms for this case can be specified in a relatively simple form:

\begin{equation} \label{eq:2f_sn}
        S_{\rm{n}} = \rho_{\rm{c}} \Gamma^{\rm{rec}} - \rho_{\rm{n}} \Gamma^{\rm{ion}},
\end{equation}
\begin{equation} \label{eq:2f_Rcoll}
        \bm{R}_{\rm{n}} = \rho_{\rm{c}} \bm{V}_{\rm{c}}\Gamma^{\rm{rec}} - \rho_{\rm{n}} \bm{V}_{\rm{n}} \Gamma^{\rm{ion}} + K_{\rm{col}} \rho_{\rm{c}} \rho_{\rm{n}} \left(\bm{V}_{\rm{c}} - \bm{V}_{\rm{n}} \right),
\end{equation}
\begin{gather} 
        M_{\rm{n}} = \left(\frac{1}{2}\Gamma^{\rm{rec}} \rho_{\rm{c}} \bm{V}_{\rm{c}}^{2} - \frac{1}{2}\rho_{\rm{n}} \bm{V}_{\rm{n}}^{2} \Gamma^{\rm{ion}} \right) + \frac{1}{\gamma - 1}\frac{k_{\rm{B}}}{m_{\rm{n}}}\left(\rho_{\rm{c}}T_{\rm{c}}\Gamma^{\rm{rec}} - \rho_{\rm{n}} T_{\rm{n}} \Gamma^{\rm{ion}} \right) \nonumber \\ 
        +\frac{1}{2} \left(\bm{V}_{\rm{c}}^{2} - \bm{V}_{\rm{n}}^{2}\right) K_{\rm{col}}\rho_{\rm{c}} \rho_{\rm{n}} + \frac{1}{\gamma - 1} \frac{k_{\rm{B}}}{m_{\rm{n}}}\left(T_{\rm{c}} - T_{\rm{n}}\right)K_{\rm{col}} \rho_{\rm{c}} \rho_{\rm{n}}.
        \label{eq:m-2fluid}
\end{gather} 

The collision term in the continuity equations, $S_n$, takes into account inelastic collisions between the particles, i.e., those where the particle identity is modified. For the purely hydrogen plasma those are ionization and recombination processes, with the corresponding rates, $\Gamma^{\rm{ion}}$ and $\Gamma^{\rm{rec}}$. The simplified treatment given by Eqs.~\ref{eq:continuity-2fluid-n} and \ref{eq:continuity-2fluid-c} only considers the hydrogen atom transferring between two, neutral and ionized states and does not explicitly include the excitation states. Both ionization and recombination can be collisional or radiative. For the complete treatment of hydrogen ionization balance, one can refer to the work by \cite{2007A&A...473..625L}. Otherwise, a simplified treatment is frequently used, as in \cite{2011PhDT.......208M}, \cite{2012ApJ...760..109L}, \cite{PopescuBraileanu2019A&A...627A..25P} or \cite{2022arXiv220511091M}. In those later works the excitation is by electron impact and recombination is spontaneous, these approximations have limited validity for the range of parameters found for partially ionized chromospheric and photospheric plasmas, and are mostly valid for coronal conditions. 

The collision term in the momentum equation, $\bm{R}_n$, Eq.~\ref{eq:2f_Rcoll}, includes the momentum exchange during ionization/recombination processes (terms proportional to $\Gamma$), and elastic collisions (term proportional to $K_{\rm{col}}$). The latter can be linked to the $\alpha_n$ collision parameter given by Eqs. \ref{eq:alphan}, but, more generically, it can also include the charge exchange contribution \citep{2012PhPl...19g2508M, PopescuBraileanu2019A&A...627A..25P},
\begin{equation}
K_{\rm{col}}=\frac{\alpha_n}{\rho_c\rho_n}+K^{\rm{cx}}_{\rm{col}}.
\end{equation}
Adding up the charge exchange collisions together with the rest of elastic collisions in a single collision parameter $K_{\rm col}$ is possible only for hydrogen plasma. The mathematical treatment of the charge exchange reaction is more complex once other chemical species are present.

The collision term in the energy equation, $M_n$, Eq.~\ref{eq:m-2fluid}, contains four groups of contributions: kinetic energy exchange in ionization/recombination, thermal energy exchange in ionization/recombination, kinetic energy exchange in elastic collisions and thermal exchange in elastic collisions. 

Since Eqs.~\ref{eq:continuity-2fluid-n}--\ref{eq:energy-2fluid-c} are written in conservation form, and the energy equation is written in terms of the sum of internal and kinetic energies, the collision terms are completely symmetric and sum to zero if the equations for neutrals and charges are added up. The role of collisions for heating the plasma can be better understood if one considers the conservation equations of internal energy only, 
\begin{equation} \label{eq:2f_iene_n}
         \frac{\partial e_{\rm{n}}}{\partial t} + \nabla \cdot \left(\bm{V}_{\rm{n}} e_{\rm{n}} + \bm{q}_n +  \bm{F}_R^n \right) + P_{\rm{n}} \nabla \cdot \bm{V}_{\rm{n}}  = Q_{\rm{n}},
\end{equation}
\begin{equation} \label{eq:2f_iene_c}
         \frac{\partial e_{\rm{c}}}{\partial t} + \nabla \cdot \left(\bm{V}_{\rm{c}} e_{\rm{c}} + \bm{q}_c +  \bm{F}_R^c \right) + P_{\rm{c}} \nabla \cdot \bm{V}_{\rm{c}} = \bm{J} \cdot \bm{E} + Q_{\rm{c}}.
\end{equation}
 
The collision terms, $Q$, in the internal energy equation have a different the form compared to those refined by Eqs.~\ref{eq:m-2fluid}. 
\begin{gather} 
        Q_{\rm{n}} = 
        \frac{1}{2}\Gamma^{\rm{rec}} \rho_{\rm{c}} \left(\bm{V}_{\rm{c}} - \bm{V}_{\rm{n}}\right)^{2} + \frac{1}{\gamma - 1}\frac{k_{\rm{B}}}{m_{\rm{n}}}\left(\rho_{\rm{c}} T_{\rm{c}} \Gamma^{\rm{rec}} - \rho_{\rm{n}} T_{\rm{n}} \Gamma^{\rm{ion}} \right) \nonumber \\
        + \frac{1}{2}\left(\bm{V}_{\rm{c}} - \bm{V}_{\rm{n}}\right)^{2} K_{\rm{col}} \rho_{\rm{c}} \rho_{\rm{n}} + \frac{1}{\gamma - 1} \frac{k_{\rm{B}}}{m_{\rm{n}}}\left(T_{\rm{c}} -T_{\rm{n}}\right) K_{\rm{col}} \rho_{\rm{c}} \rho_{\rm{n}},
        \label{eq:2f_mnprime}
\end{gather}
    
\begin{gather} 
        Q_{\rm{c}} =  \frac{1}{2}\Gamma^{\rm{ion}} \rho_{\rm{n}} \left(\bm{V}_{\rm{c}} - \bm{V}_{\rm{n}} \right)^{2} - \frac{1}{\gamma - 1} \frac{k_{\rm{B}}}{m_{\rm{n}}} \left(\rho_{\rm{c}} T_{\rm{c}} \Gamma^{\rm{rec}} - \rho_{\rm{n}} T_{\rm{n}} \Gamma^{\rm{ion}}\right) \nonumber \\
        + \frac{1}{2}\left(\bm{V}_{\rm{c}} - \bm{V}_{\rm{n}}\right)^{2} K_{\rm{col}} \rho_{\rm{c}} \rho_{\rm{n}} - \frac{1}{\gamma - 1} \frac{k_{\rm{B}}}{m_{\rm{n}}}\left(T_{\rm{c}} - T_{\rm{n}} \right) K_{\rm{col}} \rho_{\rm{c}} \rho_{\rm{n}}.
        \label{eq:2f_mcprime}
\end{gather}

While some of the contributions sum to zero (those corresponding to thermal energy exchange, second and forth group of terms), those corresponding to the kinetic energy exchange are always positive in both neutral and charges energy equations. 
According to these equations, if there is a velocity difference between charges and neutrals, it will produce frictional heating of the medium, contributing to the internal energy increase. 


Closing the two-fluid system of equations requires the generalized Ohm's law for the electron field, and generalized induction equation for the evolution of the magnetic field. The GOL for the two-fluid case is expressed in the frame of reference of the charged fluid ($\bm{V}_c\approx \bm{V}_i$), and has the following form,

\begin{equation} \label{eq:2f_efield}        
\bm{E} +\bm{V}_{\rm{c}} \times \bm{B} = \eta\mu_0 \bm{J}  + \eta_{\rm{H}}\mu_0 \frac{\bm{J} \times \bm{B}}{|B|} - \eta_{\rm{H}}\mu_0\frac{\nabla P_{\rm{e}}}{|B|}  - \eta_{\rm{D}} \left(\bm{V}_{\rm{c}} - \bm{V}_{\rm{n}} \right),
\end{equation}
with the coefficients
\begin{equation}
\eta=\frac{\rho_e(\nu_{ei}+\nu_{en})}{\mu_0 e^2 n_e^2}\,; \eta_H=\frac{|B|}{\mu_0 e n_e}\,; \eta_D=\frac{\rho_e(\nu_{en} - \nu_{in})}{en_e}.
\end{equation}
Notice that $\eta_D$ has different units (m/(t q)) from $\eta$ and $\eta_H$ (l$^2$/t).
This GOL has similar contributions as in the single-fluid case, Eq.~\ref{eq:ohm-1fluid}, i.e. Ohmic, Hall, and battery terms. However, due to the change of the system of reference, the ambipolar term is not present. Instead, the last term contributing to the electric field is the one proportional to the charges-neutral velocity difference. The coefficient multiplying this term, $\eta_D$ is proportional to the electron mass and it is generally very small in the solar context \citep{2021A&A...650A.123M}. This last term should not be confused with the ambipolar term.

\subsection{Hydrogen-helium plasmas} \label{sec:eqs_helium}

The consideration of hydrogen-only plasma is a good first approximation for the study of the solar atmosphere dynamics, since hydrogen is by far the most abundant element in the Sun. Nevertheless, a more accurate description can be achieved by including the second most abundant element, i.e., helium, whose abundance is about $10 \%$ of that of hydrogen (in terms of particles number). Since a helium particle is around 4 times more massive than a hydrogen particle, even such a low abundance of helium can have an important influence on the propagation of waves, and on the overall dynamics of the solar plasma in the multi-fluid description.
    
For the range of temperatures of the solar atmosphere, helium can be found in its three possible states of ionization: neutral, singly ionized, and doubly ionized. Together with the two ionization states of hydrogen, and electrons, the plasma is then composed by six different kinds of particles. This circumstance opens several options for a multi-fluid approach, depending on the degree of coupling assumed for each pair of species. Below we discuss two alternative descriptions of hydrogen-helium plasmas most frequently used for waves studies in the Sun: the three-fluid and the five-fluid models.

\subsubsection{Three-fluid model} \label{sec:eqs_3f}

The three-fluid model assumes that the temperature of the plasma is not high enough to allow for the presence of doubly-ionized helium (He\textsc{III}). It additionally considers that protons (p), singly-ionized helium (He\textsc{II}) and electrons (e) are strongly coupled, so they are treated as a single fluid. The other two fluids of the model are the neutral hydrogen (H) and neutral helium (He), respectively. Thus, the continuity, momentum, and energy conservation equations are given by \citep{Zaqarashvili2011A&A...534A..93Z},

\begin{equation}\label{eq:helium_continuity}
        \frac{\partial \rho_{\rm{s}}}{\partial t} + \nabla \cdot \left(\rho_{\rm{s}} \bm{V}_{\rm{s}} \right) = 0, \,\,\, s \in \{ \rm{c}, \rm{H}, \rm{He}\},
\end{equation}
\begin{gather} \label{eq:helium_momentum}
 \frac{\partial \left(\rho_{\rm{c}} \bm{V}_{\rm{c}}\right)}{\partial t} + \nabla \cdot \left( \rho_{\rm{c}} \bm{V}_{\rm{c}} \bm{V}_{\rm{c}} + P_{\textrm{c}}\mathbb{I} \right) = \bm{J} \times \bm{B} + \rho_{\rm{c}} \bm{g}+ \bm{R}_{\rm{c}}, \\ \nonumber
\frac{\partial \left(\rho_{\textrm{H}} \bm{V}_{\textrm{H}}\right)}{\partial t} + \nabla \cdot \left( \rho_{\rm{H}} \bm{V}_{\textrm{H}} \bm{V}_{\textrm{H}} + P_{\textrm{H}}\mathbb{I} \right) = \rho_{\rm{H}} \bm{g} + \bm{R}_{\textrm{H}},  \\ \nonumber
\frac{\partial \left(\rho_{\textrm{He}} \bm{V}_{\textrm{He}}\right)}{\partial t} + \nabla \cdot \left( \rho_{\rm{He}} \bm{V}_{\textrm{He}} \bm{V}_{\textrm{He}} + P_{\textrm{He}}\mathbb{I} \right) = \rho_{\rm{He}} \bm{g} + \bm{R}_{\textrm{He}},
\end{gather}
\begin{gather} \label{eq:helium_energy}
\frac{\partial e_{\rm{c}}}{\partial t} + \nabla \cdot \left(\bm{V}_{\rm{c}} e_{\rm{c}}\right) + P_{\rm{c}} \nabla \cdot \bm{V}_{\rm{c}} = \bm{J}\cdot\bm{E} + Q_{\rm{c}}, \\ \nonumber
\frac{\partial e_{\rm{H}}}{\partial t} + \nabla \cdot \left(\bm{V}_{\rm{H}} e_{\rm{H}}\right) + P_{\rm{H}} \nabla \cdot \bm{V}_{\rm{H}} = Q_{\rm{H}}, \\ \nonumber
\frac{\partial e_{\rm{He}}}{\partial t} + \nabla \cdot \left(\bm{V}_{\rm{He}} e_{\rm{He}}\right) + P_{\rm{He}} \nabla \cdot \bm{V}_{\rm{He}} = Q_{\rm{He}}.
\end{gather}

\noindent In these equations, the density of charges is given by $\rho_{\rm{c}} = \rho_{\rm{p}} + \rho_{\rm{He\sc{II}}} + \rho_{\rm{e}} \approx \rho_{\rm{p}} + \rho_{\rm{He\sc{II}}}$, and the center of mass velocity of the charged component is given by
\begin{equation} \label{eq:helium_vc}
        \bm{V}_{\rm{c}} = \frac{\rho_{\rm{p}}\bm{V}_{\rm{p}} + \rho_{\rm{He\sc{II}}} \bm{V}_{\rm{He\sc{II}}} + \rho_{\rm{e}} \bm{V}_{\rm{e}}}{\rho_{\rm{p}} + \rho_{\rm{He\sc{II}}} + \rho_{\rm{e}}} \approx \frac{\rho_{\rm{p}}\bm{V}_{\rm{p}} + \rho_{\rm{He\sc{II}}} \bm{V}_{\rm{He\sc{II}}}}{\rho_{\rm{c}}}.
\end{equation}
The internal energies $e_{\rm{s}}$ are defined without taking into account the ionization energy, i.e. simply by $e_{\rm{s}}=P_{\rm{s}}/(\gamma -1)$. Notice that, for simplicity, Eqs. \ref{eq:helium_energy} do not include thermal and radiative energy fluxes, compared to Eqs. \ref{eq:energy-2fluid-n}--\ref{eq:energy-2fluid-c}. 

The collision terms in the momentum Eq. \ref{eq:helium_momentum} are given by,
    \begin{gather} \label{eq:helium_rc}
        \bm{R}_{\rm{c}} = -\left(\alpha_{\rm{pH}} + \alpha_{\rm{pHe}} + \alpha_{\rm{He\sc{II}H}} + \alpha_{\rm{He\sc{II}He}}\right) \bm{V}_{\rm{c}} + \xi_{\rm{p}} \left(\alpha_{\rm{He\sc{II}He}} + \alpha_{\rm{He\sc{II}H}} \right) \bm{w} \nonumber \\
        -\xi_{\rm{He\sc{II}}} \left(\alpha_{\rm{pH}} + \alpha_{\rm{pHe}} \right) \bm{w}+\left(\alpha_{\rm{pH}} + \alpha_{\rm{He\sc{II}H}} \right) \bm{V}_{\rm{H}} +\left(\alpha_{\rm{pHe}} + \alpha_{\rm{He\sc{II}He}} \right) \bm{V}_{\rm{He}},
\end{gather}
\begin{equation} \label{eq:helium_rh}
        \bm{R}_{\rm{H}} = -\left(\alpha_{\rm{p}H} + \alpha_{\rm{He\sc{II}H}} + \alpha_{\rm{HeH}} \right)\bm{V}_{\rm{H}} + \alpha_{\rm{HeH}} \bm{V}_{\rm{He}} + \left( \alpha_{\rm{pH}} + \alpha_{\rm{He\sc{II}H}}\right) \bm{V}_{\rm{c}},
\end{equation}
\begin{equation} \label{eq:helium_rhe}
        \bm{R}_{\rm{He}} = -\left(\alpha_{\rm{pHe}} + \alpha_{\rm{He\sc{II}He}} + \alpha_{\rm{HeH}} \right) \bm{V}_{\rm{He}} + \alpha_{\rm{HeH}} \bm{V}_{\rm{H}} + \left(\alpha_{\rm{pHe}} + \alpha_{\rm{He\sc{II}He}}\right) \bm{V}_{\rm{c}}.
\end{equation}

Here, the mass fractions of protons and singly ionized Helium are defined as,  $\xi_{\rm{p}} = \rho_{\rm{p}}/\rho_{\rm{c}}$, $\xi_{\rm{He\sc{II}}} = \rho_{\rm{He\sc{II}}}/\rho_{\rm{c}}$, with respect to the total mass density of charges. The drift velocity of plasma is given by $\bm{w}=\bm{V}_{\rm{p}}-\bm{V}_{\rm{He\textsc{II}}}$. In practice, the terms proportional to $\bm{w}$ are dropped out arguing their smallness, since there is no consistent way for computing them in the frame of this model \citep{Zaqarashvili2011A&A...534A..93Z}.

The quantities $\alpha_{\rm{st}} \equiv \rho_{\rm{s}} \nu_{\rm{st}} = \rho_{\rm{t}} \nu_{\rm{ts}}$ represent the friction coefficients of collisions between species `s' and `t'. These parameters are similar to those defined in  Eq. \ref{eq:alphan}, but neglecting collisions with electrons due to their much lower mass. 

The collision frequencies between different species, $\nu_{\rm{st}}$, are modified compared to Eqs. \ref{eq:coll-freq-1fluid}, taking into account the possibility of different temperatures of the species. For collisions between one neutral and one ionized species,
\begin{equation}
\nu_{\rm{st}}=n_t \frac{m_{\rm t}}{m_{\rm t}+m_{\rm s}}\left(\frac{8k_BT_{\rm s}}{\pi m_{\rm s}} +\frac{8k_BT_{\rm t}}{\pi m_{\rm t}} \right)^{1/2}\sigma_{\rm st}.
\end{equation}
The choice of the cross-sections, $\sigma_{\rm{st}}$, plays an important role when it comes to quantitative conclusions regarding wave damping in the solar chromosphere \citep{2015A&A...573A..79S}. For collisions between two ions,
\begin{equation}
\nu_{\rm{st}}=\frac{n_{\rm t}Z^2_{\rm s}Z^2_{\rm t}e^4\ln\Lambda}{3\epsilon_0^2 m_{\rm s}m_{\rm st}}\left(\frac{2\pi k_BT_{\rm s}}{m_{\rm s}}+\frac{2\pi k_BT_{\rm t}}{m_{\rm t}} \right)^{-3/2},
\end{equation}
where $Z_{\rm{s,t}}$ is the charge number of the species ``s" or ``t'', and the Coulomb's logarithm, $\ln\Lambda$, is generalized for the case of plasma with several temperatures \citep[][]{Spitzer1962pfig.book.....S,Vranjes2013AA...554A..22V}.
The charge exchange collisions are neglected in this model.

The terms $Q_{\rm{c}}$, $Q_{\rm{H}}$ and $Q_{\rm{He\sc{II}}}$ represent the energy exchange due to elastic collisions between each pair of species. 

\begin{equation}
Q_{\rm{s}}=\sum_{t\ne s}Q_{st}; \,\,\, s,t \in \{ \rm{c}, \rm{H}, \rm{He}\},
\end{equation}
where
\begin{equation} \label{eq:5f_qst}
Q_{\rm{st}} =  \frac{1}{2}\left(\bm{V}_{\rm{t}} - \bm{V}_{\rm{s}}\right)^{2}K_{\rm{coll}}^{\rm{st}}\rho_{\rm{s}}\rho_{\rm{t}} + A_{\rm{st}}\frac{1}{\gamma -1} \frac{k_{\rm{B}}}{m_{\rm{s}} + m_{\rm{t}}} \left(T_{\rm{t}} - T_{\rm{s}}\right) K_{\rm{coll}}^{\rm{st}} \rho_{\rm{s}}\rho_{\rm{t}}.
\end{equation}
Here $A_{\rm{st}} = 4$ for collisions between electrons and neutral particles and $A_{\rm{st}} = 3$ for collisions between the remaining species \citep{Draine1986MNRAS.220..133D}. The relation between $K_{\rm{coll}}$ and $\alpha_{\rm{st}}$ is, $K_{\rm{coll}}^{\rm{st}} \rho_{\rm{s}}\rho_{\rm{t}}=\alpha_{\rm{st}}$. Note that Eq.~\ref{eq:5f_qst} is valid in the limit of small drift velocities between the species \citep[see the corrections made by][]{Schunk1977RvGSP..15..429S}. Ionization/recombination has been neglected in this model, so the corresponding terms in the continuity Eqs.~\ref{eq:helium_continuity} are null, compared to Eqs.~\ref{eq:continuity-2fluid-n} and \ref{eq:continuity-2fluid-c}.  
    
Finally, the generalized Ohm's law for the three-fluid Hydrogen-Helium model is given by,
\begin{equation} \label{eq:helium_efield}
        \bm{E}+\bm{V}_{\rm{c}} \times \bm{B} = \frac{\alpha_{\rm{ep}} + \alpha_{\rm{eHe\sc{II}}}}{e^{2}n_{\rm{e}}^{2}} \bm{J} - \frac{\nabla  P_{\rm{e}}}{e n_{\rm{e}}} + \frac{\bm{J} \times \bm{B}}{en_{\rm{e}}},
\end{equation}
where one recognizes Ohmic, battery and Hall terms. The smaller terms proportional to the drift velocities are removed.

\subsubsection{Five-fluid model} \label{sec:eqs_5f}

The five-fluid model treats each ionisation state of hydrogen and helium as a different fluids, which interact with each other by means of elastic collisions. No strong coupling is assumed between each pair of ionization states \citep{MartinezGomez2017ApJ...837...80M}. Therefore, each state $s \in \{\rm{p}, \rm{H}, \rm{He}, \rm{He\sc{II}}, \rm{He\sc{III}}\}$ has a corresponding full set of mass, momemtum and pressure evolution equations, which, following the works by \citet{Schunk1977RvGSP..15..429S} and \citet{Draine1986MNRAS.220..133D}, are given by:
\begin{equation} \label{eq:5f_continuity}
        \frac{\partial n_{\rm{s}}}{\partial t} + \nabla \cdot \left(n_{\rm{s}} \bm{V}_{\rm{s}} \right) = 0,
\end{equation}
\begin{equation} \label{eq:5f_momentum}
        \frac{\partial \left(\rho_{\rm{s}} \bm{V}_{\rm{s}}\right)}{\partial t} + \nabla \cdot \left(\rho_{\rm{s}} \bm{V}_{\rm{s}} \bm{V}_{\rm{s}} + P_{\rm{s}} \mathbb{I} \right) = q_{\rm{s}} n_{\rm{s}} \left(\bm{E} + \bm{V}_{\rm{s}} \times \bm{B} \right) + \rho_{\rm{s}} \bm{g} + \sum_{\rm{t} \ne \rm{s}} \bm{R}_{\rm{st}},
\end{equation}     
\begin{equation} \label{eq:5f_pressures}
\frac{\partial e_{\rm{s}}}{\partial t} + \nabla \cdot \left(\bm{V}_{\rm{s}} e_{\rm{s}}\right) + P_{\rm{s}} \nabla \cdot \bm{V}_{\rm{s}} = \sum_{\rm{t}\ne \rm{s}} Q_{\rm{st}},
\end{equation}
where $m_{\rm{s}}$ is the mass, $n_{\rm{s}}$ the number density, and $q_{\rm{s}}=Z_{\rm{s}}e$ is the electric charge (with $Z_{\rm{s}}$ being the signed charge number).
    
In this approach, electrons are not strongly coupled to any of the other charged species but their dynamics is not described by a full set of temporal evolution equations. The number density of electrons is computed from the assumption of quasi-neutrality of the plasma, $\sum_{\rm{s}} Z_{\rm{s}} n_{\rm{s}} \approx 0$, so that $n_{\rm{e}} \approx \sum_{\rm{i}}^{M} Z_{\rm{i}} n_{\rm{i}}$, where $M$ is the number of ionized species. Then, due to their low mass in comparison with that of the rest of species, the inertia of electrons is neglected and their momentum equation is used to compute the following expression for the electric field:
\begin{equation} \label{eq:5f_electric}
        \bm{E} + \bm{V}_{\rm{e}} \times \bm{B} = - \frac{\nabla P_{\rm{e}}}{e n_{\rm{e}}} +  \frac{m_{\rm{e}}}{e}\bm{g} + \frac{1}{en_{\rm{e}}}\sum_{\rm{j} \ne \rm{e}} \bm{R}_{\rm{ej}}.
\end{equation}
    
Furthermore, the current density is given by
\begin{equation} \label{eq:5f_current}
        \bm{J} = \sum_{\rm{s}} q_{\rm{s}}n_{\rm{s}}\bm{V}_{\rm{s}} = e n_{\rm{e}} \left(\bm{V}_i - \bm{V}_{\rm{e}} \right),
\end{equation}
where
\begin{equation} \label{eq:5f_ion_velocity}
        \bm{V}_i \equiv \frac{\sum_{\rm{j}}^{M} Z_{\rm{j}} n_{\rm{j}} \bm{V}_{\rm{j}}}{n_{\rm{e}}}
\end{equation}
is the velocity of ions. 
Then, the velocity of electrons can be expressed as $\bm{V}_{\rm{e}} = \bm{V}_i - \bm{J} / \left(e n_{\rm{e}} \right)$, and the electric field can be rewritten in the system of reference of the whole plasma as,
\begin{equation} \label{eq:5f_electric2}
        \bm{E} + \bm{V}_i \times \bm{B}=  \frac{\bm{J} \times \bm{B}}{e n_{\rm{e}}} - \frac{\nabla P_{\rm{e}}}{e n_{\rm{e}}} + \frac{m_{\rm{e}}}{e}\bm{g} + \frac{1}{e n_{\rm{e}}} \sum_{\rm{j} \ne \rm{e}} \bm{R}_{\rm{ej}}.
    \end{equation}
    
\noindent Hence, the only remaining evolution equation for electrons is the energy one,  Eq. \ref{eq:5f_pressures}.
    
The terms $\rm{R}_{\rm{st}}$ and $Q_{\rm{st}}$ in Eqs. \ref{eq:5f_momentum} and \ref{eq:5f_pressures} represent the momentum and the energy transfer due to elastic collisions between two species `s' and `t'. The $\rm{R}_{\rm{st}}$ term is given by
\begin{equation} \label{eq:5f_rst}
        \bm{R}_{\rm{st}} = \rho_{\rm{s}}\rho_{\rm{t}}K_{\rm{coll}}^{\rm{st}} \left(\bm{V}_{\rm{t}} - \bm{V}_{\rm{s}} \right) 
\end{equation}
and the $Q_{\rm{st}}$ is the one given by Eq.~\ref{eq:5f_qst}. The ionization/recombination is again neglected.

\section{Waves in strongly collision-coupled plasmas}
 
In the photosphere and low chromosphere of the Sun, the collision frequency between plasma species is much higher than the typical frequencies of the observed waves, see Fig. \ref{fig:collisions_snapshot}. This justifies the use of a single-fluid approximation. According to Sect. \ref{sect:GOL}, the presence of neutrals leads to the modified GOL, where the ambipolar term is expected to be the lead, followed by the modified Hall term. The ambipolar diffusion and Hall effect have been studied extensively in the context of solar atmospheric waves, both using the analytical theory, idealized, and realistic numerical simulations  \citep[see the review by][]{2018SSRv..214...58B}. The section below gradually builds the theory of wave propagation in partially ionized plasma under a single-fluid approximation, from linear wave theory to shocks and the effects of gravitational stratification. 

\subsection{Basic equations of linear theory}
\label{sec:1f-linearization}

In order to study linear waves, one has to consider Eqs. \ref{eq:continuity-1fluid}--\ref{eq:energy-1fluid} (or Eq. \ref{eq:eint-1fluid}), together with the induction Eq. \ref{eq:induction-1fluid}, and split the variables into the background component (variables with subindex ``0'') and a small perturbation (variables with subindex ``1''),
\begin{gather}
    \bm{V}=\bm{V_1(\bm{r}},t); \,\, \bm{B}=\bm{B}_0(\bm{r})+\bm{B}_1(\bm{r},t); \bm{J}=\bm{J_1(\bm{r}},t)=\nabla\times\bm{B_1}/\mu_0;\\ \nonumber
    P=P_0(\bm{r})+P_1(\bm{r},t); \,\, \rho=\rho_0(\bm{r})+\rho_1(\bm{r},t).
\end{gather}


\noindent Here we assumed that background magnetic field is current-free, i.e. $\bm{J_0}=\nabla\times\bm{B_0}/\mu_0=0$. This condition has repercussion for the energy conservation and the induction equation. We assume no velocity is present in the equilibrium. 

We will consider an ideal gas with no ionization/recombination, so that $e=P/(\gamma-1)$. In the following we will also neglect the battery term due to its smallness. The linearized set of equations becomes,

\begin{equation}\label{eq:continuity-linear-init}
    \frac{\partial \rho_1}{\partial t}+\rho_0\nabla\cdot \bm{V}=0,
\end{equation}
\begin{equation}
\rho_0\frac{\partial \bm{V}}{\partial t} + \nabla P_1 = [\bm{J_1}\times\bm{B_0}] + \rho_1\bm{g},
\end{equation}
\begin{equation}
    \frac{\partial P_1}{\partial t}+\gamma P_0\nabla\cdot \bm{V} + (\bm{V}\cdot\nabla)P_0= - (\gamma-1)\Big[\nabla\cdot(\bm{q}+\bm{F_R}) - \bm{J_1}\cdot\frac{\xi_n}{\alpha_n}[\bm{G_0}\times{\bm{B_0}}]\Big],
\end{equation}
\begin{gather}\label{eq:induction-linear-init}
        \frac{\partial \bm{B}_{1}}{\partial t} = \nabla\times \left\{
        [\bm{V} \times \bm{B}_0] +  \eta\mu_0\bm{J}_1 - \eta_H\mu_0\frac{\left[\bm{J_1} \times \bm{B}_{0} \right]}{ |B_0|}  + \right.\\ \nonumber
        \left. \eta_A\mu_0\frac{[ [\bm{J_1} \times \bm{B}_{0} ] \times \bm{B}_0]}{|B_0|^2}
         - \frac{\xi_n}{\alpha_n}[\bm{G_1}\times \bm{B}_0] \right\}.
\end{gather}
The equation of internal energy evolution has been rewritten in terms of pressure. Expressions for radiation and heat conduction will be specified below, when considering particular applications. The Ohmic and ambipolar heating terms in the energy equation are of the second order and therefore they have been neglected in the linear approximation.

These equations are generic and do not include specific assumptions about the equilibrium, except that it is current-free. If the ionization fraction can be assumed smoothly varying in space, the partial pressure $\bm{G}_{0,1}$ term can be simplified to
\begin{equation}
\frac{\xi_n}{\alpha_n}\bm{G}_{0,1}=\Xi \nabla P_{0,1}; \,\, \Xi=\frac{\xi_n^2\xi_i}{(1+\xi_i)\alpha_n}
\end{equation}
with $\xi_i=1-\xi_n$ being the ionization fraction. This way, unknown partial pressures of ions and neutrals, are eliminated in favor of the total plasma pressure.

Several particular analytical solutions  of the linearized system of equations have been developed, e.g., for the cases of a homogeneous unbounded plasma, and, in some particular cases, solutions for gravitationally stratified isothermal atmosphere. Below we consider some of the most common solutions. 

\subsection{Waves in homogeneous plasmas}
\label{sec:1f-homogeneous}

In MHD description, a homogeneous unbounded plasma supports three types of waves, fast and slow magnetoacoustic waves and the Alfv\'en wave. The single-fluid, quasi MHD description does not change this picture, and three distinct wave modes are still present. However, these modes acquire different properties, thanks to the additional damping and dispersion mechanisms present in the partially ionized plasmas. 

In the absence of gravity, one can choose the reference system such that the homogeneous magnetic field is directed along the $x$ axis, $\bm{B}_0=B_0\hat{x}$. All equilibrium parameters are constant in space, i.e. $P_0, \rho_0, T_0={\rm const.}$ Following \citet{2008A&A...492..223F}, we assume the presence of the heat conduction with $\bm{q}=-\kappa\cdot\nabla T$, and heat-loss function with $\nabla\cdot\bm{F_R}=\rho L(\rho, T)$. For the moment, we will neglect the Hall effect. Then, the linearized equations take the following form,

\begin{equation}\label{eq:cont-linear}
    \frac{\partial \rho_1}{\partial t}+\rho_0\nabla\cdot \bm{V}=0,
\end{equation}
\begin{equation}\label{eq:momentum-linear}
\rho_0\frac{\partial \bm{V}}{\partial t} + \nabla P_1 = \frac{1}{\mu_0}[[\nabla\times\bm{B_1}]\times\bm{B_0}],
\end{equation}
\begin{equation}\label{eq:energy-linear}
    \frac{\partial P_1}{\partial t}+\gamma P_0\nabla\cdot \bm{V} =  (\gamma-1)\Big[\nabla\cdot(\kappa\cdot\nabla T_1)-\rho L(\rho, T) \Big]
\end{equation}
\begin{gather}\label{eq:induction-linear}
        \frac{\partial \bm{B}_{1}}{\partial t} = \nabla\times \left\{
        [\bm{V} \times \bm{B}_0] +  \eta[\nabla\times\bm{B_1}] + \right.\\ \nonumber
        \left. \eta_A\frac{[ [[\nabla\times\bm{B_1}] \times \bm{B}_{0} ] \times \bm{B}_0]}{|B_0|^2}
         - \Xi[\nabla P_1\times \bm{B}_0] \right\}.
\end{gather}

In a homogeneous plasma the coefficients $\eta,\eta_A, \Xi, \kappa $ are constant. Therefore, one can look for the solution for perturbations in the form of plane waves. Without loss of generality, one can assume the wave vector $k$ lying in the $x-z$ plane, while all the vector quantities having 3 spatial dimensions. This way, all the perturbed quantities are proportional to, $f\sim\exp(-i\omega t + i k_x x + ik_z z)$, and it is trivial to obtain a linear system of the scalar equations for perturbations from Eq. \ref{eq:cont-linear}--\ref{eq:induction-linear}.

Partially ionized plasmas in the Sun are often studied in the context or solar prominences. In the work by  \citet{2004A&A...415..739C}, the heat loss function, $L(\rho, T)$ has been specified for prominence conditions as a difference between the optically thin radiative losses, and some generic heating function, as,
\begin{equation}
    L(\rho,T)=\chi^*\rho T^\alpha - h\rho^a T^b.
\end{equation}
The coefficients $\chi^*$ and $\alpha$ for the thin radiative losses depend on temperature \citep{1974SoPh...35..123H, 1978ApJ...220..643R, 1979ApJ...232..304M}, while the exponents $a$ and $b$ of the heating function depend on the heating mechanisms. Without the loss of generality, the linearized version of the heat-loss term becomes \citep{2004A&A...415..739C},
\begin{equation}
    \rho L(\rho, T)^{\rm lin}=\rho_1 L + \left( \frac{\partial L}{\partial \rho}\right)_T\rho_1\rho_0 + \left( \frac{\partial L}{\partial T}\right)_\rho T_1\rho_0 = \rho_1 L + L_T\rho_1\rho_0 + L_\rho T_1\rho_0.
\end{equation}

The heat conduction in partially ionized plasmas is anisotropic, therefore the heat conduction tensor can be decomposed into the parallel and perpendicular components with respect to the magnetic field, 
\begin{equation}\label{eq:kappa-generic}
    \kappa = \kappa_{||}\bm{\hat{b}}\bm{\hat{b}} + \kappa_{\perp}(\mathbb{I}-\bm{\hat{b}}\bm{\hat{b}}).
\end{equation}
Assuming that the dominant component of the heat conduction is the electron heat conduction along magnetic field lines, and that the neutral heat conduction is isotropic, \citet{2008A&A...492..223F} reduced Eq.\ref{eq:kappa-generic} to
\begin{equation}
    \kappa=\kappa_{||,e}\bm{\hat{b}}\bm{\hat{b}} +\kappa_n\mathbb{I}.
\end{equation}

Under these approximations, and for the chosen geometry, the dispersion relations for the Alfv\'en waves and for magneto-acoustic waves decouple from each other \citep{2008A&A...492..223F}. The one for the Alfv\'en waves has the following form,
\begin{equation}\label{eq:disp_alfven_1f}
    \omega^2-k_x^2\Gamma(\theta)^2=0,
\end{equation}
and the one for magneto-acoustic waves is,
\begin{equation}\label{eq:disp_ma_1f}
    a_5\omega^5+a_4\omega^4+a_3\omega^3+a_2\omega^2+a_1\omega+a_0=0.
\end{equation}

The coefficients of these dispersion relations are given by the following expressions,
\begin{gather}
    \Gamma(\theta)= c_A^2-i\omega(\eta_C + \eta \tan^2\theta);\\
    a_0= -\frac{ik^2k_x^2c_A^2}{\rho_0}(AT_0-H\rho_0);\\
    %
    a_1= -k^2\left[c_S^2c_A^2k_x^2-\frac{(AT_0-H\rho_0)\Psi}{\rho_0} \right];\\
    a_2=ik^2\left[\frac{(AT_0-H\rho_0)}{\rho_0} +\frac{AT_0c_A^2}{P_0} +c_S^2\Psi\right]; \\
    a_3=k^2\left[(c_S^2+c_A^2) + \frac{AT_0\eta_C}{P_0}  \right]; \\
    %
    a_4=-i\left(\frac{AT_0}{P_0} + k^2\eta_C \right); \\
    %
    a_5= -1,
\end{gather}
where we have introduced Cowling conductivity, $\eta_C=\eta+\eta_A$ \citep{1945RSPSA.183..453C}. The coefficients $A$, $H$ and $\Psi$ are defined as,
\begin{gather}
    A= (\gamma-1)(\kappa_{||,e}k_x^2 + \kappa_n k^2 + \rho_0 L_T);\\
    H=(\gamma-1)(L+\rho_0 L_{\rho});\\
    \Psi=k^2\eta_C - k_z^2c_A^2\rho_0\Xi.
\end{gather}%
In the equation above, $k^2=k_x^2+k_z^2$ is the total wave vector, $c_S=\sqrt{\gamma P_0/\rho_0}$ is the
sound speed and $c_A=B_0/\sqrt{\mu_0\rho_0}$ is the Alfv\'en speed taking the total plasma density. 

Below we discuss the solutions for the magneto-acoustic and Alfv\'en waves separately, starting by the former.

\subsubsection{Magneto-acoustic waves}
\label{sec:ma-1f-homogeneous}

Equation \ref{eq:disp_ma_1f} can be simplified by introducing a modified, complex, sound speed,
\begin{equation}
%
    \Lambda^2=\frac{\frac{T_0}{\rho_0}A-H-i\omega c_S^2}{\frac{T_0}{\rho_0}A-H-i\omega}.
\end{equation}
With this definition, the dispersion relation for magneto-acoustic waves shortens to,
\begin{equation} \label{eq:disp_ma_1f_proto}
%
\left(\omega^2 - k^2\Lambda^2 \right) \left(i\omega k^2\eta_C+\omega^2\right) - k^2c_A^2(\omega^2 - \Lambda^2k_x^2) +i\Lambda^2c_A^2k^2k_z^2\Xi\rho_0\omega =0.
\end{equation}

\noindent In the absence of heat conduction, heating and radiation losses, the modified sound speed $\Lambda$ is equal to the common sound speed $c_S$, and the dispersion relation, Eq.\ref{eq:disp_ma_1f_proto}, reduces to the one reported in \citet{2007A&A...461..731F}.
%
Furthermore, it can be verified that, by taking the neutral fraction to zero, and neglecting Ohmic diffusion, this dispersion relation simplifies to the known case for magneto-acoustic waves in homogeneous unbounded plasmas \citep{Priest}.
\begin{equation} \label{eq:disp_ma}
    \omega^4 -\omega^2k^2(c_S^2 + c_A^2)+ k^2k_x^2c_A^2c_S^2=0.
\end{equation}

The adiabatic dispersion relation for magneto-acoustic waves, i.e., Eq. \ref{eq:disp_ma_1f_proto} with $\Lambda=c_S$, is 4th order in $\omega$ and supports 4 wave modes. However, compared to the ideal case, Eq. \ref{eq:disp_ma}, the former equation is not bi-quadratic, and has imaginary terms. It means that waves traveling in the opposite directions will, generically, have different properties, and will be damped or amplified in time/space, depending on the conditions of the plasma.

In \citet{2007A&A...461..731F},  adiabatic Eq.~\ref{eq:disp_ma_1f_proto} was solved assuming temporal damping, that is, $\omega=\omega_R+i\omega_I$, with $\omega_I\ll\omega_R$ (weak damping approximation). Under this approximation, the real part of the wave frequency remains the same as for magneto-acoustic waves in a fully ionized plasma, i.e. one given by Eq. \ref{eq:disp_ma}. The approximate expression for the imaginary part of the frequency becomes,
\begin{align}
\omega_I\approx -\frac{k^2k_z^2c_A^2c_S^2\Xi\rho_0+\eta_Ck^2(\omega_R^2-k^2c_S^2)}{4\omega_R^2-2k^2(c_S^2+c_A^2)}.
\end{align}
In the limit of low plasma $\beta$, with $c_S\ll c_A$, and $\omega_R^{\rm{fast,slow}}\approx\{kc_A, k_xc_S\}$ for the fast and slow wave, respectively, the damping simplifies to
\begin{equation} \label{eq:omegai_fast_1f}
    2\omega_I^{\rm{fast}}\approx - \eta_C k^2 - \eta_A k_z^2\frac{c_S^2}{c_A^2} \approx -\eta_C k^2,
\end{equation}
for the fast wave, and
\begin{equation} \label{eq:omegai_slow_1f}
    2\omega_I^{\rm{slow}}\approx - k_z^2\frac{c_S^2}{c_A^2}\left(\eta + \eta_A\frac{1}{1+\xi_i}\right),
\end{equation}
for the slow wave. Notice that the diamagnetic effect (the term proportional to $\Xi$) only affects the slow wave in this approximation and that, without taking it into account, the damping for the slow wave would reduce to $2\omega_I^{\rm{slow}}\approx - k_z^2(c_S^2/c_A^2)\eta_C$. Similar expressions for the damping times were also derived in \citet{2009ApJ...699.1553S}.

It can be observed from the expressions for the $\omega_I^{\rm{fast}}$ and $\omega_I^{\rm{slow}}$ coefficients, that the damping is proportional to the Cowling diffusivity $\eta_C$, i.e. the sum of the Ohmic and ambipolar diffusivities, and that both cause similar effects. Since $c_S\ll c_A$, the damping is much stronger for the fast wave than for the slow wave.

 \begin{figure} [!t]
        \centering
        \includegraphics[width=0.95\hsize]{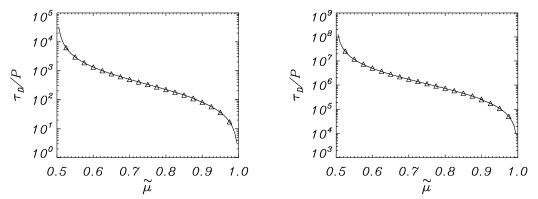}
        \caption{Damping time over wave period for the fast magneto-acoustic wave (left) and for the slow magneto-acoustic wave (right) as a function of ionization fraction, defined as $\tilde{\mu}=1/(1+\xi_i)$. The data are computed according to expressions Eq. \ref{eq:omegai_fast_1f} and \ref{eq:omegai_slow_1f} (triangles) and from the numerical solution of the adiabatic dispersion relation for magneto-acoustic waves (Eq.~\ref{eq:disp_ma_1f_proto}). The computations are done for conditions typical for solar prominences:  $B_0=10 G$, $T_0=8$ kK, $\rho_0=5\times 10^{-11}$ kg m$^{-3}$ ($c_A=126$ km s$^{-1}$, $c_S=10.5$  km s$^{-1}$), $k_x x_0=\pi/2$, $k_z x_0=0.1$, $x_0=3$ Mm. Figure from \citet{2007A&A...461..731F}.}
        \label{fig:Forteza2007}
    \end{figure}

\citet{2007A&A...461..731F} computed the damping times over the wave period for the typical solar prominence conditions in the adiabatic case. Their calculations are illustrated in Fig.~\ref{fig:Forteza2007} for the fast wave (left) and for the slow wave (right). The results are presented as a function of ionization fraction, where the value of $\tilde{\mu}$=0.5 means fully ionized plasma and $\tilde{\mu}=1$ means fully neutral gas. It can be observed that, indeed, the fast mode damping due to neutrals is several orders of magnitude larger than that of the slow mode. Nevertheless, even for the fast mode, significant damping, $\tau_D/P\approx 1$, is only achieved for very low ionization fractions. Most contribution into this damping comes from the ambipolar diffusion, since for prominence parameters, $\eta_A$ values significantly exceed the values of the Ohmic diffusion coefficient, $\eta$.

\begin{figure} [!t]
\centering
\includegraphics[width=0.95\hsize]{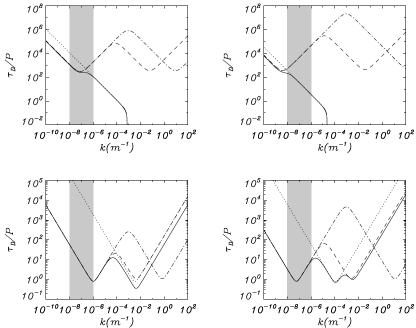}
        \caption{Damping time over wave period for the fast magneto-acoustic wave (top) and for the slow magneto-acoustic wave (bottom) as a function of wave number $k$ in a non-adiabatic partially ionized plasma, calculated by numerically solving the dispersion relation, Eq.\ref{eq:disp_ma_1f}, adopted from \citet{2008A&A...492..223F}. The following parameters were used for the calculation: ionization fraction $\tilde{\mu}=1/(1+\xi_i)=0.8$ (left panels) and $\tilde{\mu}=0.99$ (right panels), $B_0=10 G$, $\theta=45^{o}$, $T_0=8$ kK, $\rho_0=5\times 10^{-11}$ kg m$^{-3}$ and the parameters of the heating-cooling function from \citet{1974SoPh...35..123H}. Different lines correspond do different combinations of damping mechanisms. Solid line: ion-neutral collisions and thermal mechanisms; dotted line: only ion-neutral collisions; dashed line: only thermal mechanisms; dashed-dotted line: only heating-cooling and electron thermal conduction.}
        \label{fig:Forteza2008_fig4}
    \end{figure}


When non-adiabatic effects are present, the dispersion relation, Eq.\ref{eq:disp_ma_1f_proto} is of the 5th order. In addition to the fast and slow magneto-acoustic modes, there is a thermal mode, as described by \citet{2004A&A...415..739C}
Since the properties of the thermal mode in partially ionized plasma are similar to those of the fully ionized case, here we do not discuss them further.


Figure \ref{fig:Forteza2008_fig4} illustrates the damping over period times for the fast 
(top) and slow (bottom) non-adiabatic magneto-acoustic waves as a function of the wave number $k$, as computed by numerically solving Eq.\ref{eq:disp_ma_1f_proto} by \citet{2008A&A...492..223F}. The shaded area in this figure marks the range of the wave numbers typically detected in solar prominences. The influence of different damping mechanisms is shown by different lines styles. The damping of the fast wave is dominated by radiative cooling at small $k$, and by ion-neutral collisions at large $k$ for both neutral fractions considered (upper panels). Variation of the ionization fraction produces the change of the wavelength of the dominant damping mechanisms. For prominence conditions, ion-neutral collisions become the dominant damping mechanism for large neutral fraction, while radiation plays significant role at lower neutral fractions (shaded area at the upper panel).

For the slow wave and large neutral fraction $\tilde{\mu}=0.99$ (bottom right panel) the ratio of the damping time to period shows three minima (maximum attenuation) associated to the typical scales of the different damping mechanisms. That at smallest $k$ is due to radiative cooling effects, the one at intermediate $k$ is due to ion-neutral collisions, and the one at largest $k$ is due to neutral thermal conduction. For $\tilde{\mu}=0.8$, the latter two peaks merge. For the prominence conditions, radiation plays the most important role in the damping of the slow wave, regardless the neutral fraction.

There is another important effect that can be observed in Fig.~\ref{fig:Forteza2008_fig4}. Namely, that the fast waves disappear after a certain wavelength. This effect is related to the presence of the cut-off wavelength for the fast mode in partially ionized plasma. It can be illustrated by considering propagation parallel to the magnetic field, i.e., $k=k_x$. In this case the dispersion relation Eq. \ref{eq:disp_ma_1f_proto} reduces to,
\begin{equation}
%
    (\omega^2 - k_x^2\Lambda^2)(i\omega k_x^2\eta_C +\omega^2 - k_x^2c_A^2)=0. 
\end{equation}
The fast waves decouples from the slow and thermal waves, and its dispersion relation (second bracket in the last expression) can be solved to obtain the complex temporal frequency,
\begin{equation}
    \omega=-\frac{ik_x^2\eta_C}{2}\pm\frac{k_x}{2}\sqrt{4c_A^2 - k_x^2\eta_C^2}.
\end{equation}
Therefore, in order to have non-zero real part of the wave frequency, $\omega_R\ne 0$, one has to fulfill the condition,
\begin{equation}
    k_x<\frac{2c_A}{\eta_C}=k_c,
\end{equation}
where $k_c$ is the cutoff wavelength. For wave numbers greater that the critical wavelength, the wave is completely damped and is not propagating in time domain. This effect will be discussed in more details when considering magneto-acoustic waves in a two-fluid approximation and it will be shown that the cutoff wavelength disappears.

\begin{figure} [!t]
\centering
\includegraphics[width=0.45\hsize]{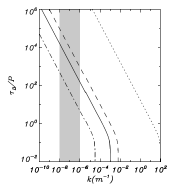}
\caption{Damping time over wave period for the Alfv\'en wave as a function of wave number $k$ in a non-adiabatic partially ionized plasma, calculated by numerically solving the dispersion relation, Eq.\ref{eq:disp_alfven_1f}, adopted from \citet{2008A&A...492..223F}. The atmospheric parameters used for the calsulation are the same as in Fig.\ref{fig:Forteza2008_fig4}. The ionization fraction  $\tilde{\mu}=0.5$ (dotted line),  $\tilde{\mu}=0.6$ (dashed line),  $\tilde{\mu}=0.8$ (solid line),  $\tilde{\mu}=0.99$  (dashed-dotted line).}
\label{fig:Forteza2008_fig8}
\end{figure}
    
\subsubsection{Alfv\'en waves}
\label{sec:alfven-1f-homogeneous}


According to the dispersion relation Eq.\ref{eq:disp_alfven_1f}, the Alfv\'en waves are not affected by the thermal effects and are only affected by the ion-neutral collisions. By solving the dispersion relation, the real and imaginary parts of the wave frequency can be obtained,
\begin{gather} \label{eq:disp_alfven_1f_proto}
%
\omega=\omega_R+i\omega_I = -\frac{ik_x^2(\eta_C+\eta \tan^2\theta)}{2}\pm \frac{k_x}{2}\sqrt{4c_A^2 - k_x^2(\eta_C+\eta\tan^2\theta)^2}.
\end{gather}

The behavior of the Alfv\'en wave is similar to that of the fast wave in the adiabatic case, see Fig.\ref{fig:Forteza2008_fig8}. In the non-adiabatic case, the damping of the fast wave is modified by the thermal effects at lower frequencies, compared to the Alfv\'en wave. By comparing the curves for different ionization fraction in Fig.\ref{fig:Forteza2008_fig8}, one can observe that the ratio of the damping time to period decreases when going to higher neutral fractions, similar to the fast wave. For the typical prominence conditions, the Alfv\'en waves are only weakly damped (shaded area in Fig.\ref{fig:Forteza2008_fig8}).

One can also observe the presence of a cutoff wave number after which the Alfv\'en wave does not propagate. This cutoff number can be computed from the dispersion relation, Eq. \ref{eq:disp_alfven_1f_proto} and the following result is obtained

\begin{equation}
    k_x<\frac{2c_A}{\cos\theta(\eta_C+\eta\tan^2\theta)}=k_c^A.
\end{equation}
This cutoff number depends on the ionization fraction through $\eta_C$, and on the propagation angle $\theta$ with respect to the magnetic field. Usually $k_c^A> k_c$, and both become equal for the parallel propagation, then the fast and Alfv\'en waves become degenerate and cannot be distinguished. By considering the fully ionized plasmas, $\eta_C=\eta$, the value of $k_c^A$ for the resistive plasmas can be recovered \citep{1961itmf.book.....F}.

Similarly as for the fast wave, the sharp cutoff wave number for the Alfv\'en waves disappears in the two-fluid description, as discussed below.

\subsubsection{Aflv\'en waves and the Hall effect}
\label{sec:alfven-hall-1fluid}

The Hall effect is a non-ideal effect that does not necessarily require the presence of neutrals. In a fully ionized plasma this effect is due to the different gyro-frequencies of ions, $\Omega_{\rm{ci}}$, and electrons, $\Omega_{\rm{ce}}$, i.e., mainly the difference in ion and electron mass. For a range of ion-electron collision frequencies, the electrons are magnetized ($\Omega_{\rm{ce}} > \nu_{\rm{ie}}$) while the ions are not ($\Omega_{\rm{ci}} < \nu_{\rm{ie}}$). The Hall effect in fully ionized plasmas becomes important at frequencies of the order of the ion-cyclotron frequency.

The Hall effect in partially ionized plasmas can be enhanced by collisions between ions and neutrals, which can work to demagnetize the ions further. In order to describe the action of the Hall effect, let us consider a linearized set of equations, neglecting the diamagnetic effect, radiation and thermal conduction, and considering a homogeneous isothermal plasma. We use linear continuity and momentum Eqs.\ref{eq:cont-linear} and \ref{eq:momentum-linear}, together with the induction equation containing the Ohmic, ambipolar and Hall terms \citep{2008MNRAS.385.2269P},


\begin{gather}\label{eq:induction-linear-hall}
        \frac{\partial \bm{B}_{1}}{\partial t} = \nabla\times \left\{
        [\bm{V} \times \bm{B}_0] +  \eta[\nabla\times\bm{B_1}] -\eta_H\frac{ [\nabla\times\bm{B_1}]\times\bm{B_0}}{|B_0|}+ \right.\\ \nonumber
        \left. \eta_A\frac{[ [[\nabla\times\bm{B_1}] \times \bm{B}_{0} ] \times \bm{B}_0]}{|B_0|^2}
          \right\}.
\end{gather}
The system is closed by an isothermal energy equation $P_0=c_S^2\rho_0$. 

By comparing in order of magnitude the Hall and convective terms, one can see that the Hall effect in partially ionized plasma becomes important at frequencies of the order of the Hall frequency \citep{2008A&A...492..223F, 2015ApJ...814..106C}, 
\begin{equation}
\omega_H=\xi_i\Omega_{\rm{ci}}.
\end{equation}
The Hall frequency can be several orders of magnitude lower than the ion-cyclotron frequency at locations with very cold plasma  and low ionization fractions near the solar temperature minimum. Following the discussion in Section \ref{sect:GOL}, the Hall effect is expected to be the dominant non-ideal effect in the middle-upper photosphere of the Sun. 

\citet{2008MNRAS.385.2269P} and \citet{2012A&A...544A.143Z} derived the dispersion relation for magneto-acoustic waves and Alfv\'en waves, including the Hall effect, in the form,
\begin{gather}
%
%
%
\left\{ \left[ \omega^2 - (c_A^2 - i\omega\eta_A) k^2\cos^2\theta + i\omega k^2\eta \right] + k^2\sin^2\theta \left[\frac{\omega}{\tilde{\omega}}c_A^2 + i \omega\eta_A\right]    \right\}\times \\ \nonumber
\left[ \omega^2 - (c_A^2 - i\omega\eta_A) k^2\cos^2\theta + i\omega\eta k^2 \right]  - 
\eta_H^2 k^4\omega^2\cos^2\theta=0,
\end{gather}
where $\tilde{\omega}^2=\omega^2-k c_S^2$. It can be verified that, in the absence of the Hall effect, one recovers the dispersion relations for the fast and Alfv\'en waves previously discussed in the sections above (\ref{sec:ma-1f-homogeneous}, \ref{sec:alfven-1f-homogeneous}), see \citet{2007A&A...461..731F, 2008A&A...492..223F}. 


For propagation parallel to magnetic field, $\theta=0$, the dispersion relation for the Alfv\'en waves from \citet{1999MNRAS.307..849W} is recovered,
\begin{equation}
    \omega^2 + i\omega\eta_C k^2-k^2c_A^2=\pm \eta_H\omega k^2.
\end{equation}
In the low frequency limit, for waves with frequencies much lower than the Alfv\'en frequency, $\omega \ll k c_A$, this dispersion relation has the following solution for the real and imaginary parts of the wave frequency \citep{1999MNRAS.307..849W, 2008A&A...492..223F, 2012A&A...544A.143Z}
\begin{gather}
    \omega_R=\pm c_A^2\frac{\eta_H}{\eta_C^2+\eta_H^2} = \pm \omega_H\frac{\eta_H^2}{\eta_C^2+\eta_H^2}, \\
    \omega_I=-c_A^2\frac{\eta_C}{\eta_C^2+\eta_H^2}=-\omega_H\frac{\eta_H\eta_C}{\eta_C^2+\eta_H^2}.
\end{gather}
It can be observed that, in the presence of the Hall current, the real part of the wave frequency is small, but always non-zero. It means that the sharp cutoff wavelength, present in partially ionized plasmas due to ion-neutral collisions, disappears in the presence of the Hall effect. An explanation for that is the following. When the Hall current is included, electrons can have a different dynamics to those of the ions. Ions may not be able to follow the magnetic field fluctuations due to the effect of ion-neutral collisions, but it is easier for electrons to remain coupled to the magnetic field, i.e., to stay magnetized. Therefore, ion-neutral collisions cannot completely suppress the fluid oscillations because of the distinct behavior of electrons when Hall's current and/or electron inertia are included \citep[see the discussion in][]{2008MNRAS.385.2269P}.




\subsection{Effects of gravitational stratification}

Solar atmosphere is a strongly gravitationally stratified medium. The effects of neutrals, considered before, apply for homogeneous plasmas and are valid only locally. There are only few attempts to consider fully analytical solutions in partially ionized gravitationally stratified medium. Even without taking into account partial ionization, the height dependence of the coefficients in the wave equation allows for analytical solution only in some limited cases. For example, the case of an isothermal gravitationally stratified atmosphere with arbitrary inclined constant magnetic field has an exact solution in terms of Meier function or hypergeometric functions \citep{1984A&A...132...45Z, 2001ApJ...548..473C}.

Another approach to attack the problem of the wave propagation in a gravitationally stratified atmosphere is by means of a local dispersion relation. This technique has been widely used to study acoustic-gravity waves, as well as adiabatic MHD waves. In this case, the atmosphere is approximated by layers with locally homogeneous properties. This approach gives approximately precise results as far as the wavelength of the perturbation is much smaller than density/pressure scale height (0-order WKB approximation). Nevertheless, gradients caused by gravitational stratification introduce new effects and the local dispersion relation approach may not always describe the variations of the waveforms with height. 

The presence of gradients and strong vertical stratification allows for the process of wave-mode transformation. The classical mode transformation happens when the fast magneto-acoustic mode travelling through the solar interior (a $p$-mode, which is essentially acoustic below the photosphere) emerges at the solar surface and encounters the layer, located somewhere in the photosphere or chromosphere, where acoustic and Alfv\'en speeds are similar.  At this layer, part of the energy of the fast magneto-acoustic mode is transformed into a slow magneto-acoustic mode (also essentially acoustic) propagating along the magnetic field, and a fast magneto-acoustic mode (now essentially magnetic) that will eventually refract and reflect back to the surface due to the gradients of the Alfv\'en speed  \citep{2006RSPTA.364..333C, 2006ApJ...653..739K}. At the upper turning point of the fast (magnetic) mode, an Alfv\'en mode can be generated through the secondary mode transformation \citep{2008SoPh..251..251C}. This transformation can be considered as a  geometry-induced mode transformation since it results from the stratification of the solar atmosphere and its efficiency depends on the orientation of the magnetic field. Mode transformation is expected to play an important role for waves propagating in solar active regions \citep{2015LRSP...12....6K}.
Partial ionization affects both, propagation of different wave modes through the stratified atmosphere, and the mode transformation process. We discuss these effects in the sections below.

\subsubsection{Local dispersion relation}

In this section, a local dispersion relation for waves in a gravitationally stratified atmosphere is derived, taking into account Ohm, ambipolar and Hall effects. 
It considers a particular case, which includes the majority of the modes and effects discussed above, but now adding a stratification. The diamagnetic effect, heat conduction and radiation are neglected.

A Cartesian reference system is chosen such that spatially constant equilibrium magnetic field vector lies in the $x-z$ plane, 
$$\bm{B}=\{B_{x0}, 0, B_{z0}\}.$$ 
In this system, a perturbation propagates in the vertical $z$ direction, and the perturbed vector quantities have the following components,
$$\bm{V}=\{V_x(z), V_y(z), V_z(z) \}, \,\,
\bm{B_1}=\{B_{x1}(z), B_{y1}(z),0\}.$$

\noindent Since the horizontal magnetic field components are functions of the vertical coordinate only, in order to fulfill the divergence-free condition, the vertical component must be constant with $z$, therefore, 
$B_{z1}=0$. 

In the spirit of the local dispersion relation we assume that all the background quantities are locally constant with height withing a narrow layer. One then can apply the equations layer by layer, by varying the values of the background quantities. 
By imposing all the first-order quantities varying as $f\sim\exp(-i\omega t + ik_z z)$, the system Eq. \ref{eq:continuity-linear-init}--\ref{eq:induction-linear-init} becomes \citep{2006ApJS..166..613K},

\begin{gather}
-i\omega \rho_1 + ik_zV_z=0, \\
-i\omega\rho_0 V_x - \frac{ik_zB_{x1}B_{z0}}{\mu_0}=0, \\
-i\omega\rho_0 V_y - \frac{ik_zB_{y1}B_{z0}}{\mu_0}=0, \\
-i\omega\rho_0 V_x +ik_zP_1+\rho_1 g + \frac{ik_zB_{x1}B_{x0}}{\mu_0}=0, \\
-i\omega B_{x1} - ik_zV_xB_{z0} + ik_zV_zB_{x0}+\eta_Ck_z^2 B_{x1} + \eta_H\cos\theta k_z^2 B_{y1}=0,\\
-i\omega B_{y1}- ik_zV_yB_{z0} + (\eta+\eta_A\cos^2\theta)k_z^2 B_{y1} -\eta_H\cos\theta k_z^2 B_{x1}=0,\\
P_1=c_S^2\rho_1,
\end{gather}
where $c_S^2=\gamma P_0/\rho_0$ is the background sound speed and $\theta$ is magnetic field inclination angle with respect to the vertical $z$ direction. After combining these equations a single dispersion relation can be obtained,
\begin{gather}\label{eq:dispersion-1fluid-stratified}
\left( \omega^2 - k_z^2c_{Az}^2  + i\eta_C k_z^2\omega \right)\left( \omega^2 - k_z^2c_{Az}^2  + i
(\eta+\eta_A\cos^2\theta)k_z^2\omega \right)\left(\omega^2 - k_z^2c_S^2 + igk_z\right) - \\ \nonumber
- \left( k_z^2\omega\eta_H\cos\theta \right)^2\left(\omega^2 - k_z^2c_S^2 + igk_z \right) 
\\ \nonumber
-k_z^2\omega^2 c_{Ax}^2\left( \omega^2 - k_z^2c_{Az}^2  + i (\eta+\eta_A\cos^2\theta)k_z^2\omega \right)=0.
\end{gather}

This equation is 6th order in $\omega$ and allows for three wave modes with properties resembling fast, slow and Alfv\'en modes.  In the case of the absence of the magnetic field, it reduces to,

\begin{equation}
\left( \omega^2 + i\eta k_z^2\omega \right)^2 \left(\omega^2 - k_z^2c_S^2 + igk_z\right)=0.
\end{equation}
The first bracket contains an entropy mode ($\omega=0$) and an evanescent perturbation in $B_{x1}$ and $B_{y1}$ with $\omega=-ik_z^2\eta$. The second bracket is a usual acoustic-gravity mode propagating vertically in a stratified atmosphere. The propagation of this mode is subject to the acoustic cut-off with the value defined by temperature ($\omega_c=g/2c_S$). Under approximations of the local dispersion relation, the propagation of this mode is unaffected by any of the non-ideal effects. Notice that, this conclusion goes in line with the discussion from Section \ref{sec:ma-1f-homogeneous}. According to Section \ref{sec:ma-1f-homogeneous}, slow magneto-acoustic waves propagating along the magnetic field are unaffected by non-ideal effects to the first order (this can be seen by setting $\Xi=0$ and $k=k_z$ in the adiabatic Eq. \ref{eq:disp_ma_1f_proto}).

Another particular case is obtained by setting $B_{x0}=0$. This way we consider waves propagating vertically along the vertical magnetic field. The dispersion relation Eq. \ref{eq:dispersion-1fluid-stratified} reduces to,
\begin{equation}
\left\{\left( \omega^2 - k_z^2c_{Az}^2 + i\eta_C k_z^2\omega \right)^2 - \left( k_z^2\omega\eta_H\right)^2\right\}\left(\omega^2 - k_z^2c_S^2  +igk_z\right)=0.
\end{equation}
Yet again we obtain longitudinally propagating  acoustic-gravity mode in the second bracket. The first bracket corresponds to the Alfv\'en wave (or fast wave, which is indistinguishable in this limit), affected by the Ohmic and ambipolar dissipation and Hall effect, with the dispersion relation,

\begin{equation}
\omega^2 - k_z^2c_{Az}^2 + k_z^2\omega(i\eta_C\pm\eta_H)=0.
\end{equation}
It can be seen that, under the approximations done to derive the local dispersion relation, the propagation of this wave is not affected by gravity, and we recover the case described in Section \ref{sec:alfven-hall-1fluid}.

Finally, by setting $B_{z0}=0$ we recover the case of waves propagating transverse to the magnetic field in the direction of stratification,
\begin{equation}
(\omega^2 + i\eta k_z^2\omega)\left[ (\omega^2 + i \eta_C k_z^2\omega)(\omega^2 - k_z^2c_S^2  +igk_z)- k_z^2\omega^2c_{Ax}^2\right]=0.
\end{equation}
Notice that the Hall term, due to its particular shape, does not affect waves in this case. Yet again, the fist bracket in the dispersion relation corresponds to a fully damped perturbation. The second bracket describes modes with mixed properties, and the analytical solution is impractical. It represents the generalization of the fast magneto-acoustic mode affected by resistive dissipation and gravity. In the absence of resistive effects, it reduces to the one of the fast mode in a gravitationally stratified atmosphere, with properties analogous to the acoustic-gravity modes, but with the propagation speed, and the gravitational cut-off, dependent on the magnetic field,
\begin{equation}
    \omega^2-k_z^2(c_S^2+c_{Ax}^2)+ik_z g=0.
\end{equation}

The dispersion equation for acoustic gravity or fast modes can be solved for $k_z$ recovering its real and imaginary parts, $k_z=k_{\rm{zR}}+ik_{\rm{zI}}$. The imaginary part gives the wave amplitude growth with height, $k_{\rm{zI}}=g/\sqrt{4(c_S^2+c_{Ax}^2)}.$ Wave amplitude growth in a stratified atmosphere is a usual property derived from the conservation of the kinetic energy. Since $\eta_C$ introduces damping, it can be expected that the effects of the gravitational stratification and damping will be competing with each other. Depending on the height, magnetic field strength, and wave frequency, the amplitude growth can be overcome by the damping.

\subsubsection{Propagation of Alfv\'en waves in a stratified atmosphere}
\label{sec:alfven-1f-stratified}


Local dispersion relation is usually not a good approximation for fast and Alfv\'en waves because their wavelength in the solar atmosphere strongly increases with height due to the Alfv\'en speed growth. Consider an example of torsional Alfv\'en waves propagating along a an expanding magnetic flux tube. Using cylindrical coordinate system, the following coupled linearized equations for $V_{\theta}$ and $B_{1\theta}$ components can be obtained \citep{2013A&A...549A.113Z}.
\begin{gather}
    \frac{\partial}{\partial t}(rV_{\theta})=\frac{\bm{B_0}\cdot\nabla}{\mu_0\rho_0}(rB_{1\theta}), \\
    \frac{\partial B_{1\theta}}{\partial t}=r (\bm{B_0}\cdot\nabla)\left[\frac{V_{\theta}}{r}+\frac{\eta_C}{B_0^2}\frac{\bm{B_0}\cdot\nabla}{r^2}(rB_{1\theta}) \right].
\end{gather}
Following the original work by \citet{2013A&A...549A.113Z}, the Cowling diffusion coefficient is defined taking into account collisions between electrons, neutral and ionized Hydrogen and Helium. By defining $U_{\theta}=V_{\theta}/r$ these two equations can be combined into a single wave equation,
\begin{equation}
    \frac{\partial^2 U_{\theta}}{\partial t^2}=\frac{B_{0s}}{\mu_0\rho_0 r^2}\frac{\partial}{\partial s}\left[r^2 B_{0s}\frac{\partial}{\partial s} \left(U_{\theta} + \frac{\mu_0\rho_0\eta_C}{B_{0s}^2}\frac{\partial U_{\theta}}{\partial t} \right) \right],
\end{equation}
with the derivatives taken along the magnetic field direction $s$, and being $B_{0s}$ the magnetic field along $s$. This equation can be simplified at locations close to the flux tube axis, so that $B_{0s}r^2\approx \rm{const}$:
\begin{equation}\label{eq:torsional-alfven-stratified}
    \frac{\partial^2 U_{\theta}}{\partial t^2}=c_A^2\frac{\partial^2}{\partial s^2}\left[ \left(1 + \frac{\eta_C}{ c_A^2}\frac{\partial }{\partial t} \right)U_{\theta} \right].
\end{equation}
Equation~\ref{eq:torsional-alfven-stratified} can be Fourier-analyzed in the temporal domain, but not in the spatial domain, since the coefficients depend on the stratification. Nevertheless, based on approximate calculations in the FAL93-F model \citep{1993ApJ...406..319F}, the ratio ${\eta_C}/{c_A^2}$ does not vary much with height in the chromosphere, and can be assumed approximately constant \citep{2013A&A...549A.113Z}. Additionally, if a thin flux tube approximation can be assumed valid, then the density fall off with height is compensated by the magnetic field dependence, making the Alfv\'en speed constant. Under these conditions, Eq. \ref{eq:torsional-alfven-stratified} reduces to a homogeneous Alfv\'en equation in a form similar to Eq. \ref{eq:disp_alfven_1f}, and the conclusions about the damping and cutoff wave number hold valid \citep{2008A&A...492..223F}, 
\begin{equation}
    \omega^2+i\eta_C k_s^2\omega-k_s^2c_A^2=0; \,\, \omega=\pm k_s c_A\sqrt{1-\frac{\eta_C^2 k_s^2}{4 c_A^2}}-i\frac{\eta_C k_s^2}{2}.
\end{equation}
Given the atmospheric parameters from FAL93-F, \citet{2013A&A...549A.113Z} evaluated the typical values of the damping times over the wave period, $\tau_d/T_0$, to be between $\approx 1-70$ for wave periods of $1-60$ sec. When collisions with Helium are taken into account, the damping is about 20-30\% stronger. 

\begin{figure} [!t]
\centering
\includegraphics[width=0.49\hsize]{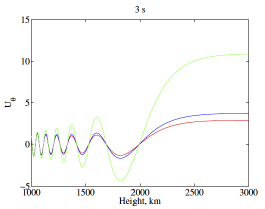}
\includegraphics[width=0.49\hsize]{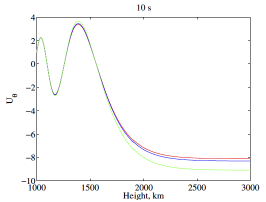}
\caption{Height dependence of the velocity of an upward propagating torsional Alfv\'en wave in the atmosphere with parameters from FAL93-F model atmosphere. Green lines: fully ionized plasma; blue lines: partially ionized hydrogen plasma; red lines: partially ionized hydrogen-helium plasma. Left panel: period of 3 s, right panel: period of 10 s. Figure from \citet{2013A&A...549A.113Z}}
\label{fig:Zaqarashvili2013_fig6}
\end{figure}

At the upper part of magnetic flux tubes the field expands and the thin flux tube approximation is not valid anymore. In this situation one has to consider exponentially changing Alfv\'en speed in Eq. \ref{eq:torsional-alfven-stratified}. In such a case, assuming $c_A=c_{A0}\exp(-s/2H)$, the Fourier-transformed Eq. \ref{eq:torsional-alfven-stratified} reduces to,
\begin{equation}
    \frac{\partial^2 U_{\theta}}{\partial s^2}+\exp{\left(-\frac{s}{H}\right)}k_{s0}^2U_{\theta}=0; \,\,\, k_{s0}^2=\frac{\omega^2}{c_{A0}^2 -i\omega\eta_{C0}}.
\end{equation}
The values with sub-index ``0'' are taken at a reference height. The solution of this equation is a standard Bessel or Hankel function.
%
Figure \ref{fig:Zaqarashvili2013_fig6} illustrates the solution for $U_{\theta}$ for short-period waves. It can be observed that wave amplitudes increase with height due to the effects of stratification. However, the wave forms in partially ionized plasma are visibly damped at higher heights, compared to the case of the fully ionized solution (green lines). The damping is more significant when the Helium atoms are taken into account. The shorter period waves experience a stronger dependence on the ion-neutral collisions, as expected from analytical solutions for homogeneous plasma. It can be also observed that, due to the effects of stratification, the effective wavelength of perturbations becomes very large, so that these waves turn out to be almost evanescent. This MHD effect is not changed by partial ionization, however, the amplitudes of these long-wavelength waves are visibly reduced due to ion-neutral collisions. Overall, the work by \citet{2013A&A...549A.113Z} allows to quantify that short-period (<5 s) torsional Alfv\'en waves damp quickly in conditions close to those in chromospheric network, owing to ion-neutral collision, while this damping is not so important for waves with longer periods.

\subsubsection{Propagation of magneto-acoustic waves in a stratified atmosphere}
\label{sec:ma-1f-stratified}

In Section \ref{sec:ma-1f-homogeneous}, it was shown how ambipolar diffusion affects fast and slow magneto-acoustic waves, concluding that fast waves can be significantly damped. Slow waves are also susceptible to ambipolar damping, but to a lower degree. These conclusions were obtained for conditions close to those in solar prominences. When it comes to the propagation of fast and slow magneto-acoustic waves in the gravitationally stratified solar chromosphere, analytical solutions become impractical and numerical approach allows to get better insights.   

\begin{figure} [!t]
\centering
\includegraphics[width=0.95\hsize]{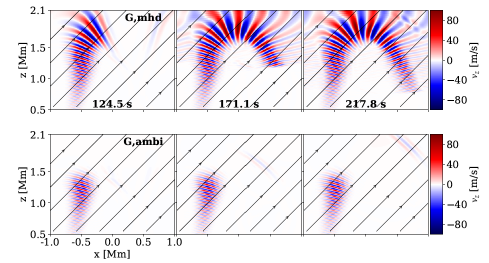}
\caption{Snapshots of the simulation of magneto-acoustic wave pulse excited at the top of the photosphere and propagating through the chromosphere. Panels from left to right show different moments of time. Upper panels: pure MHD case, bottom panels: ambipolar diffusion is included. The ambipolar diffusion coefficient changes almost linearly on log$_{10}$ scale from 10$^3$ m$^2$ s$^{-1}$ at 0.5 Mm to 10$^{11}$ m$^2$ s$^{-1}$ at 2.1 Mm. Black lines are magnetic field lines. Magnetic field has a strength of 17.4 G and is inclined by 45$^o$ to the vertical gravity direction. The wave period is of 5 sec. Plasma $\beta$ changes from $\beta>1$ to $\beta<1$ at about 1.25 Mm. Figure from \citet{2021A&A...653A.131P}.}
\label{fig:PopescuKeppens_fig8}
\end{figure}

\citet{2021A&A...653A.131P} considered a fully numerical solution for the propagation of magneto-acoustic waves launched from the upper photosphere upwards and propagating through the stratified chromosphere permeated by a constant inclined magnetic field, and with thermodynamic conditions given, approximately, by solar 1D model VALC \citep{Vernazza1981ApJS...45..635V}. An illustrative example from these numerical experiments is shown in Figure \ref{fig:PopescuKeppens_fig8}. The upper panel shows how a Gaussian pulse at the bottom boundary generates a fast magneto-acoustic wave package. On its way to the chromosphere this wave package suffers a refraction and reflection due to the gradients of the Alfv\'en speed, a typical behavior seen in many previous works \citep[e.g.,][]{2003ApJ...599..626B, 2006ApJ...653..739K}. The plasma $\beta=1$ layer is located around 1.25 Mm in their experiment, so mode transformation is taking place \citep{2006RSPTA.364..333C}. In this particular case, the fast wave propagating through $\beta=1$ layer does not generate significant slow wave due to the mode transformation, because of the relatively large field inclination and high wave frequency \citep{2006RSPTA.364..333C}. It is interesting to compare the ideal MHD case to the case when the ambipolar diffusion was included in the model (bottom panel). In the latter case, it is clear that the fast wave is "eaten out" by the ambipolar diffusion and is unable to complete the refraction trajectory. According to \citet{2021A&A...653A.131P}, in most cases considered in their work, the high frequency fast waves are significantly damped before they are reflected. This damping is stronger if the waves propagate across the magnetic field. The damping increases with wave frequency, and magnetic field strength, due to the increase of $\eta_A$. The results of these numerical experiments are in qualitative agreement with the analytical wave theory in homogeneous plasmas \citep{2007A&A...461..731F, 2008A&A...492..223F}, and they allow to quantify the effect of ambipolar diffusion on waves for the solar case. 

\subsubsection{Hall-induced mode transformation}

The geometrical fast-to-Alfv\'en mode transformation is intrinsically a 3D process. This transformation cannot happen when the wave vector lies in a plane defined by the magnetic field vector and the direction of the stratification (gravity or strong pressure gradient). So necessarily, the wave propagation should happen with an angle to the plane. In partially ionized plasmas, the Hall effect is able to assist the process of the mode transformation by naturally adding a 3rd dimension to the problem. It happens as a result of the Hall effect producing perturbations of the current in the direction perpendicular to the plane defined by the magnetic field and the gravity. 

To illustrate this effect, following \citet{2015ApJ...814..106C}, consider a single-fluid system of equations of mass, momentum and induction with only the Hall effect, and cold plasma ($P \ll P_{\rm mag}$, $c_S\approx0$) approximation (linearized Eqs. \ref{eq:cont-linear}, \ref{eq:momentum-linear} and induction equation, Eq. \ref{eq:induction-linear-hall} with only the Hall term). 
%
%
The velocity is substituted by the Lagrangian displacement, $\bm{V}=\partial \bm{\xi}/\partial t$. 
The plasma is gravitationally stratified along the $x$ direction, and an inclined magnetic field is contained in the $x-z$ plane, while the wave vector is directed outside of this plane. Using the Fourier transform of the variables for the coordinates where no stratification exist, $\mathbf{\xi}(x,y,z,t)=\mathbf{\xi}(x)\exp[i(k_yy+k_zz-\omega t)]$, the following coupled system of equations for the displacement perturbation are obtained:
\begin{gather} \nonumber
\left(\partial_\parallel^2+\partial_\perp^2+\frac{\omega^2}{c_A^2}\right)\xi_\perp = -i k_y \partial_\perp\xi_y +i \left[ i\,k_y\partial_\perp(\epsilon\,\xi_\perp)-(\partial_\parallel^2+\partial_\perp^2)(\epsilon\,\xi_y)  \right], \\ \nonumber
\left(\partial_\parallel^2+\frac{\omega^2}{c_A^2}-k_y^2\right)\xi_y = -i k_y \partial_\perp\xi_\perp +i \left[ (\partial_\parallel^2-k_y^2)(\epsilon\,\xi_\perp)-i\,k_y\partial_\perp(\epsilon\,\xi_y) \right].     
\end{gather}

This system describes the propagation of coupled fast and Alfv\'en waves. The equations use a change of the reference system linked to the wave propagation directions, $(\bm{e}_\perp,\bm{e}_y,\bm{e}_\parallel)$. 
%
%
%
The $(\parallel,\perp)$ plane is the same as $(x,z)$ plane, the $\parallel$ and $\perp$ directions are the one along the magnetic field and perpendicular to it. 

\begin{figure}
\center
\includegraphics[width=0.48\hsize]{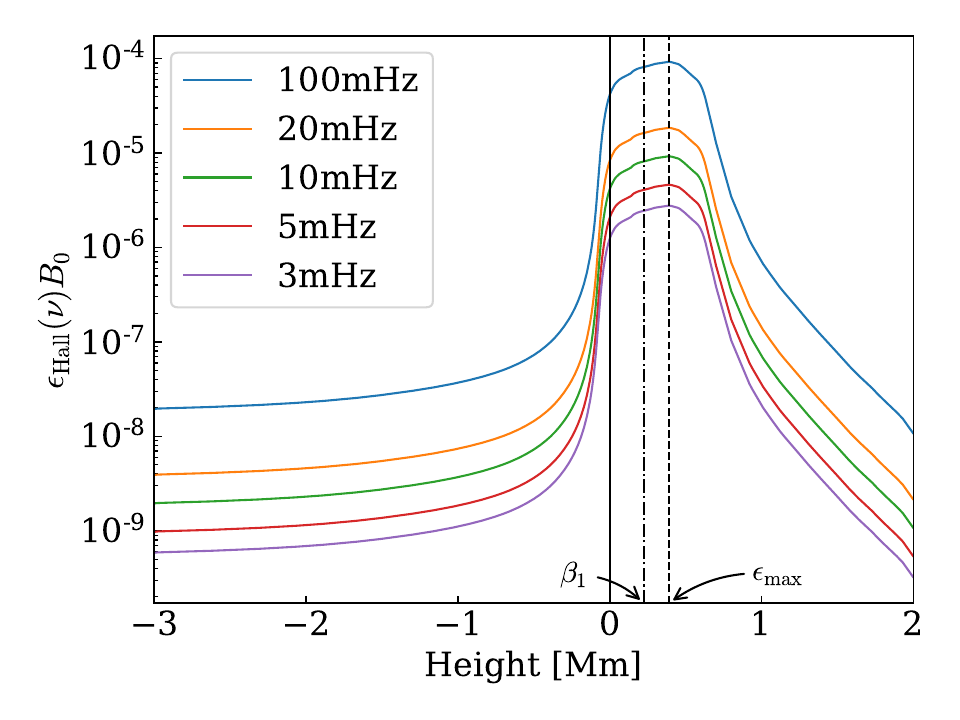}
\includegraphics[width=0.48\hsize]{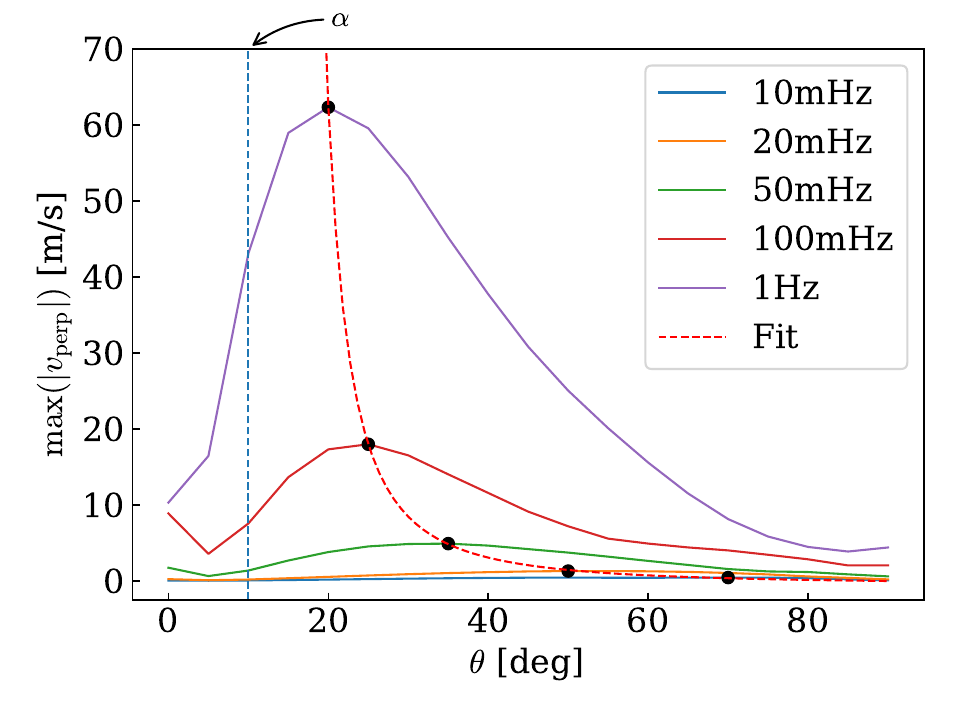}
\caption{\footnotesize Left panel: Hall parameter, $\epsilon$, as a function of height for different wave frequencies in the solar atmosphere. The vertical line corresponds to the photospheric level, $z=0$ km, negative heights are below solar surface. Right panel: Velocity $v_\mathrm{perp}=v_y$ of the Alfv\'en wave generated after the Hall-induced mode transformation as a function of  magnetic field inclination angle. The magnetic field in this experiment was inclined by 10 degrees with respect to the stratification direction. Figure from \citet{2019ApJ...870...94G}. }\label{fig:ehall}
\end{figure}

In these equations $\epsilon$ is the Hall parameter, defined as a ratio between the wave frequency and the Hall frequency,
\begin{equation}
    \epsilon=\omega/(\Omega_{\rm{ci}} \xi_i)=\omega/\omega_H.
\end{equation}
%
%
If the Hall effect is absent ($\epsilon=0$), the equations are only coupled through non-zero $k_y$, i.e. for wave propagation outside of the $x-z$ plane. However, if the Hall effect is present, the equations for the fast and Alfv\'en waves are coupled even if $k_y=0$. As discussed in Sect.~\ref{sec:alfven-hall-1fluid}, a small ionization fraction around the temperature minimum allows for the significant decrease in the frequency of waves for which the Hall effect becomes important. It can be expected that typical thermodynamic conditions in the Sun's atmosphere create the so-called Hall window, with the extension and amplitude as the one shown at the left panel of Figure \ref{fig:ehall}, taken from \citet{2019ApJ...870...94G}. These values of $\epsilon$ were  computed using the parameters from the Model-S of \citet{1996Sci...272.1286C} (below the photosphere) with the chromospheric model VAL-C by \citet{Vernazza1981ApJS...45..635V}. 
Inside the Hall window, Hall coupling produces a continuous oscillation between the fast-mode and Alfv\'en-mode states \citep{2015ApJ...814..106C}. 

\begin{figure}[t]
\begin{center}
\includegraphics[width=0.95\hsize]{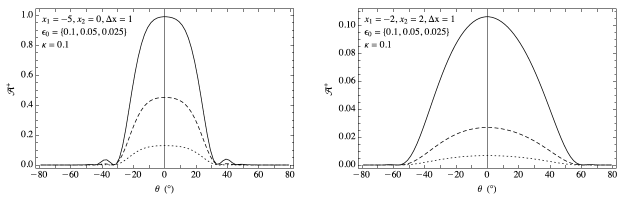}
\caption{Conversion coefficient from fast to Alfv\'en wave for the Hall-induced conversion model by \citet{2015ApJ...814..106C}, as a function of magnetic field inclination angle and $k_y=0$. Left and right panels are different by the location and thickness of the Hall window, indicated in the figure. Solid, dashed and dotted curves are for progressively smaller value of non-dimensional Hall coefficient $\epsilon$ parameter, indicated in the figure.}
\label{fig:CallyKhomenko2015_fig5}
\end{center}
\end{figure}

According to the theoretical mode conversion model in a cold plasma by \citet{2015ApJ...814..106C}, the Hall coupling preferentially occurs where the wave vector is nearly parallel to the guide field. This is illustrated in Figure \ref{fig:CallyKhomenko2015_fig5}. There, the Hall-induced mode conversion coefficient is shown as a function of the magnetic field inclination, for several parameters of the Hall window, and the values of $\epsilon$. The conversion efficiency has a broad maximum for the vertical fields. It is also sensitive to the location and extension of the Hall window, and its amplitude decreases with decreasing the value of $\epsilon$.

Numerical calculations by \citet{2019ApJ...870...94G}, performed for parameters applicable to the solar atmosphere are in a general agreement with the cold plasma model of the Hall-induced conversion. The right panel of Figure \ref{fig:ehall} shows the amplitudes of the Alfv\'en waves generated after Hall-induced transformation at chromospheric heights, as a function of inclination angle between the magnetic field and the wave propagation direction (in this experiment, the field was inclined by 10 degrees to the vertical). The amplitude of the Alfv\'en waves exponentially increases with frequency, and is a sensitive function of the inclination angle. The maximum amplitudes are reached for waves with frequencies of 1Hz, and make up to $\sim10$\% of the amplitude of the fast waves entering the Hall window. 

In a stratified solar atmosphere, the Hall-induced conversion is a two-step process \citep{2019ApJ...870...94G, 2019SoPh..294..147R}.  Firstly, a classic geometrical mode transformation at the $c_S=c_A$ layer has to happen, producing fast and slow magneto-acoustic waves from essentially acoustic solar $p$ modes. After that, the newly produced fast (magnetic) wave is able to convert to the Alfv\'en wave though the Hall effect. For this two-stage process to be effective,  particular conditions must be fulfilled, namely, the $c_S=c_A$ layer must be located below the Hall window layer (see the left panel of Fig. \ref{fig:ehall}), this can be fulfilled for intermediate field strengths of the order of hG.

\citet{2019SoPh..294..147R} demonstrated how the classic and the Hall-induced mode transformation work together in a stratified solar atmosphere to produce both up-going and down-going Alfv\'en waves. According to their results, a down-going slow wave, produced after the incident fast wave reflection/conversion, can couple to the down-going Alfv\'en wave through the Hall effect. This coupling will be strongest for the horizontal wave numbers oriented opposite to
the field inclination, and magnetic field strength of the order of 100 G in order to place the Hall window at the optimum location. Alternatively, an up-going slow wave injected from below can couple to the up-going Alfv\'en wave. Unlike the Hall-mediated fast-Alfvén coupling, the slow-Alfv\'en coupling occurs lower at the atmosphere and for much lower wave frequencies (those at with typical acoustic-gravity waves are evanescent).

\subsubsection{Influence of ambipolar diffusion on the mode transformation}


Ambipolar diffusion in the solar atmosphere is expected to be the largest in the middle-upper chromosphere. Right in the same regions, plasma $\beta$ is also expected to be around unity, and therefore geometrical mode transformation is expected to take place. Here we follow a theoretical study by \citet{2018ApJ...856...20C} of how ambipolar diffusion affects the process of the mode transformation from fast to Alfv\'en waves, using a cold plasma approximation.

\begin{figure}
\center
\includegraphics[width=0.99\hsize]{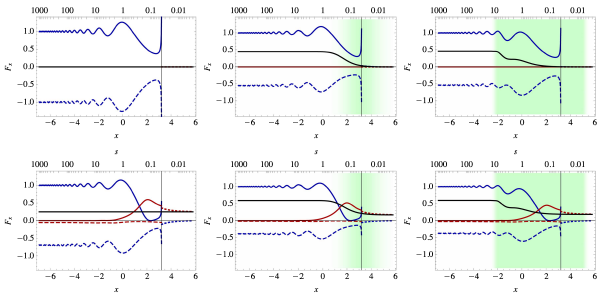}
\caption{\footnotesize  Vertical energy fluxes computed for three different prescriptions of ambipolar diffusion (columns): $\eta_A=0$ (left), Gaussian ambipolar diffusion region with $\epsilon_A=\epsilon_{A0}\exp{(-(x-x_0)^2/2\sigma^2)}$, $\epsilon_{A0}=\sigma=1$ (middle); $\epsilon_A=0.05$ between $x=-2$ and $x=5$ (right). Top row: $\theta=30^{o}$, $\phi=0^{o}$ (no Alfv\'en conversion). Bottom row: $\theta=30^{o}$, $\phi=30^{o}$ (conversion to mostly up-going Alfv\'en waves). Blue/red curves are for fast/Alfv\'en wave flux, solid/dashed lines are for upward/downward flux. Black curve is the total flux. Figure from \citet{2018ApJ...856...20C}. }\label{fig:CallyKhomenko2018_fig3}
\end{figure}

Consider a uniform background magnetic field, inclined with respect to the gravity $x$-direction by an angle of $\theta$, and contained in the $x-z$ plane. The wave vector lies outsize of the plane formed by the gravity and the magnetic field, forming an azimuth angle $\phi$. It is again convenient to consider three characteristic directions, $(\bm{e}_\perp,\bm{e}_y,\bm{e}_\parallel)$. The non-ideal induction equation contains only the ambipolar term. In this case, the wave equation for the displacement $\bm{\xi}$ takes the following form,
\begin{equation} \label{eq:conversion-ambi}
    \partial_t^2\bm{\xi} = c_A^2(\partial_{||}^2 -\nabla_{\perp}\nabla\cdot)\left(\bm{\xi} + \frac{\eta_A}{c_A^2}\partial_t\bm{\xi}\right),
\end{equation}
where the derivative $\nabla_{\perp}$ is taken in the direction perpendicular to the magnetic field. It can be seen that there is no structural change in the Eq. \ref{eq:conversion-ambi}, compared to the standard wave equation without $\eta_A$, i.e. there are no additional derivatives due to the ambipolar effect, that would lead to a different physics \citep{2018ApJ...856...20C}. In this regard, the ambipolar diffusion differs greatly from the Hall effect considered by \citet{2015ApJ...814..106C}, which produced its own mode conversion mechanism. 

Assuming a vertically stratified atmosphere such that $\omega H^2/c_A^2=\exp(-x/H)=s$, and introducing a dimensionless ambipolar diffusion parameter $\epsilon_A(s)=\omega \eta_A/c_A^2$, which is a function of $s$ only, Eq. \ref{eq:conversion-ambi} can be rewritten as,

\begin{equation} \label{eq:conversion-ambi1}
    ((1-i\epsilon_A)^{-1}\partial^2_t - c_A^2\partial_{||}^2)\bm{X}=c_A^2\nabla_{\perp}\nabla\cdot\bm{X},
\end{equation}
where $\bm{X}=(1-i\epsilon_A)\bm{\xi}$. Dropping the compression term at the right hand side this equation reduces to the ambipolar-damped Alfv\'en wave equation discussed in Section \ref{sec:alfven-1f-homogeneous}.

Equation \ref{eq:conversion-ambi1} can be Fourier analyzed in the $y$ and $z$ directions perpendicular to the stratification direction, and then it splits into two equations for the $\perp$ and $y$ components of $\bm{X}$,
\begin{gather}
    \left(\partial_x^2+\frac{\omega^2}{(1-i\epsilon)c_A^2}-k_z^2\right)X_{\perp}=-ik_y(\sin\theta\partial_x-ik_z\cos\theta)X_y,\\ \nonumber
\left((\cos\theta\partial_x+ik_z\sin\theta)^2+\frac{\omega^2}{(1-i\epsilon)c_A^2}-k_y^2 \right)X_y=-ik_y(\sin\theta\partial_x-ik_z\cos\theta)X_{\perp}.
\end{gather}

\noindent The magnetic Poynting flux can be obtained from the following expression,
\begin{equation}
    \bm{F}_{\rm mag}=\frac{1}{\mu_0}Re[\bm{E_1^*}\times \bm{B_1}]=\frac{\omega B_0}{\mu_0}Im[(\nabla\cdot\bm{X})\bm{X^*}+(\bm{X^*}\cdot\partial_{||}\bm{X})\bm{e}_{||}],
\end{equation}
with $\bm{E_1}=\bm{B}\times\partial_t(\bm{\xi}+\eta_A\bm{V}/c_A^2)$ being the 1st order perturbation of the electric current,  and $*$ denoting a complex conjugation. The computation of magnetic flux allows to explore the influence of the ambipolar effect on the mode conversion. The main conclusions of these calculations are summarized in Figure \ref{fig:CallyKhomenko2018_fig3}. Overall, the conversion picture stays the same in the presence of ambipolar diffusion. In the case of no Alfv\'en conversion ($\theta=30^o$, $\phi=0^o$, top row), the net wave flux is zero for $\eta_A=0$ case since the fast wave is perfectly reflected. When $\eta_A$ is distinct from zero, it can be observed the presence of an asymmetry between the up-going and down-going waves, the ambipolar diffusion reduces the non-converting fast mode flux by about 50\%. In the second case ($\theta=30^o$, $\phi=30^o$, bottom row), when Alfv\'en conversion is present, about 25\% of the energy is going to the upward Alfv\'en wave when $\eta_A=0$. This amount is significantly reduced in the cases of $\eta_A\ne 0$ (middle and right panels). As it can be seen from the height dependence of the fluxes, the upward propagating Alfv\'en waves, produced after the transformation, are almost immune to further ambipolar dissipation. This would reduce the impact for heating by waves produced by this mechanism.

Figure \ref{fig:CallyKhomenko2018_fig3} from \citet{2018ApJ...856...20C} is computed for non-dimensional $\epsilon_A$ coefficients. For the values of $\eta_A$ or $\epsilon_A$ computed as in Section \ref{sect:GOL} for the parameters of the solar atmosphere, the effect of ambipolar diffusion on the mode conversion would be too weak to measurably affect waves with frequencies typically observed in the Sun (around 3-5 mHz). In order to produce a measurable effect, one has to invoke a concept of turbulent ambipolar diffusion. 

\subsection{Non-linear perturbations and plasma heating}


Waves are one of the best candidates to bring energy to heat the upper solar atmosphere. In particular, heating theories based on Alfv\'en waves are very promising, because Alfv\'en waves do not shock at lower layers, and due to their incompressibility they are not affected by damping through viscosity, conductivity or radiation (Section \ref{sec:alfven-1f-homogeneous}). Efficient dissipation of these waves can be achieved through the ambipolar diffusion mechanism. The linear wave analysis, as the one considered above, does not allow to compute the dissipation of the wave energy from the equations, since the Joule dissipation term in the single-fluid approach is a second order term. Therefore, the linear analysis excludes important aspect of the wave heating. Many works suggest that Joule dissipation of currents by ambipolar mechanism has a potential to provide sufficient energy to maintain the temperature of the solar chromosphere \citep[see, e.g.,][]{1996ApJ...463..784G,  2004A&A...416.1159G, 2008ApJ...683L..87J, 2010ApJ...724.1542K,  2012ApJ...747...87K, 2012ApJ...753..161M}. MHD waves, such as fast or Alfv\'en waves, constantly produce currents across magnetic field lines, and these currents are susceptible to non-linear resistive dissipation, enhanced through ion-neutral interaction \citep{1998A&A...338..729D, 2000ApJ...533..501G,  2011ApJ...735...45G, 2006ApJS..166..613K, 2010ApJ...708..268G,  2016ApJ...819L..11S, 2018A&A...618A..87K}. 

In order to understand the role of neutrals in the wave plasma heating, the internal energy equation, Eq. \ref{eq:eint-1fluid}, can be expressed separating different heating terms on the right hand side,
\begin{equation}
    \frac{\partial e}{\partial t}+ \nabla\cdot \left(\bm{V}e \right) =  Q_J + Q_c +Q_{\rm visc} - Q_{\rm R} - \nabla\cdot \mathbf{q}.
\end{equation}
where $Q_J=\eta J^2 + \eta_A{J_\perp}^2$ is resistive Joule heating term, $Q_c=-P\nabla\cdot\bm{V}$ is shock compressional heating term, $Q_{\rm visc}$ is viscous heating, $Q_{\rm R}$ is radiative cooling, and the last term is total thermal energy flux. If the initial configuration of the magnetic field is current-free, then $Q_{J0}=0$, otherwise, $Q_{J0}\ne 0$. The latter value can be used as a reference value to estimate the importance of the resistive wave heating. It is also instructive to compare resistive heating to the other contributions. 


\citet{2010ApJ...708..268G} have studied propagation of shock wave perturbation developed from an initially sinusoidal fast-mode wave driver in an atmosphere with solar-like stratification (model FAL). In their 1.5D experiment the magnetic field is horizontal, $B_x(z)$, and it is stratified in the vertical direction due to gravity. The background magnetic field is not current-free. The magnitude of the current, and the associated resistive heating produced in the stationary atmosphere is used as a reference value for comparison with the model where waves are driven and develop shocks. The results of the simulation show that, at shock fronts the current density, $J_1$, can exceed by 2-3 orders of magnitude the background current density, $J_0$. The resistive heating rate, $Q_{J1}$, increases by 4-6 orders of magnitude in shocks relative to the stationary value $Q_{J0}$. 
%
The ratio between  the ambipolar and total resistive heating rates in the experiment by \citet{2010ApJ...708..268G} shows that almost all the resistive heating above 500 km is due to ambipolar diffusion. The compressional heating in shocks can exceed by 2-3 order of magnitude the background heating $Q_{J0}$. The values of the compressional shock heating are significantly above the resistive one in their experiment at all heights. The height-integrated heating rates differ almost a factor of 300. Therefore, resistive heating by itself stays significantly below the shock compresisonal heating. Similar conclusions were also reached in \citet{2016ApJ...817...94A}, who studied 1.5D propagation of a spectrum of Alfv\'en waves non-linearly coupled to the slow waves, producing shocks. Nevertheless, this conclusion does not mean that the resistive effects are not important. The resistivity defines the width of the shock fronts over which the compression happens. This way it sets the total heating rates through the atmosphere. 

In the solar chromosphere, the ambipolar diffusion dominates by orders of magnitude the Ohmic diffusion, therefore it can be considered the dominant resistive mechanism. The non-linear action of the ambipolar diffusion on the propagation of the magnetic Poynting flux can be determined by writing the total energy conservation equation as,
\begin{equation} \label{eq:etot}
\frac{\partial e_{\rm tot}}{\partial t} + \nabla\cdot\bm{S} = 0.
\end{equation}
Therefore, according to the divergence theorem, the time variation of the total energy in the closed volume is zero if no energy flux enters/exists through the boundary. The total energy flux, including heat flux and radiative energy flux is given by
\begin{equation}
\bm{S}=\left( \frac{\rho V^2}{2} + P + e \right) \bm{V} + \bm{S}_{\rm EM} +  \bm{q}  + \bm{F}_R.
\end{equation}
This expression includes the electromagnetic Poynting flux
$$\bm{S}_{\rm EM}=\frac{\bm{E}\times\bm{B}}{\mu_0}.$$
Taking the expression for the electric field from the generalized Ohm's law, Eq. \ref{eq:ohm-1fluid}, (without the battery and diamagnetic terms),
\begin{equation}
\bm{S}_{\rm EM}=- \frac{[\bm{V} \times \bm{B}]\times\bm{B}}{\mu_0}+ \eta\bm{J} \times\bm{B} + \eta_A \bm{J_{\perp}}\times\bm{B} = \bm{S}_{\rm EM}^{\rm ideal} + \bm{S}_{\rm EM}^{\rm non-ideal}.
\end{equation}
can be split into the ideal and resistive contributions.
This expression shows that, since the total energy must be conserved, if a given amount of the ideal Poynting flux is generated at the boundary of a volume (for example by means of a wave perturbation), after propagating though a volume where ambipolar (and Ohmic, to a lower extent) diffusion are acting, the amount of $\bm{S}_{\rm EM}^{\rm ideal}$ will be decreased. The difference between $\bm{S}_{\rm EM}^{\rm ideal}$ at the entrance and exit of the domain remains in the volume and is converted into the thermal energy. 

\begin{figure}
\center
\includegraphics[width=0.44\hsize]{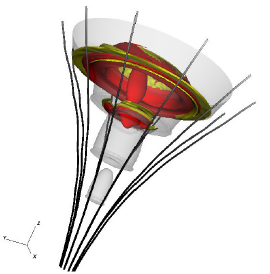}
\includegraphics[width=0.54\hsize]{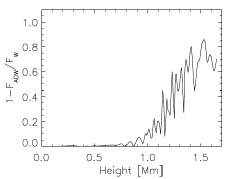}
\caption{\footnotesize Left: 3D rendering of the magnetic field lines in a flux tube covering from the photosphere to the chromosphere. Grey surfaces are the locations of plasma $\beta=1$. Yellow contours show the locations of the strong currents generated by propagating torsional Alfv\'en wave with a frequency of 25 mHz. Red contours show the locations of the temperature enhancement of 500 K relative to the ideal case without ambipolar diffusion. Right: Height dependence of the Poynting flux absorption, where 0 means no absorption, and 1 means total absorption.  Figure from \citet{2016ApJ...819L..11S}. }\label{fig:Shelyag2016}
\end{figure}

The absorption of the Poynting flux has been verified by means of simulations of torsional Alfv\'en waves propagating along magnetic flux tube from the photosphere to the chromosphere \citep{2016ApJ...819L..11S}, see Figure \ref{fig:Shelyag2016}. These simulations show how the locations with enhanced perpendicular current density spatially coincide with locations of the temperature enhancements (left panel). The absorption coefficient of the Poynting flux (right panel), defined as the ration between the Poynting flux in simulations with/without ambipolar diffusion, is a function of height, reaching maximum of about 80\% absorption in the chromosphere, where the ambipolar diffusion is the largest. Therefore, heating has been achieved by dissipation of torsional Alfv\'en waves. In a more complex situations, realistic numerical simulations also demonstrate the ability of the ambipolar diffusion to dissipate into heat incompressible magnetic waves (e.g., Alfv\'en waves), see \cite{2018A&A...618A..87K, 2020A&A...642A.220G, 2021RSPTA.37900176K}.

\section{Waves in two-fluid hydrogen plasmas}
    Previous sections have shown the importance of the presence of neutral species for the evolution of plasmas in the solar atmosphere. The interaction between the ionized and the neutral species plays a relevant role in processes such as the propagation and the mode transformation of MHD waves, or heating of solar plasma. These conclusions have been obtained through the application of single-fluid models, which assume that there is a strong coupling between all the components of the plasma. Therefore, they are strictly applicable to scenarios where the dynamics of the ionized and the neutral particles have very small differences. 
    
The strong-coupling assumption is perfectly valid for the photosphere, the lower regions of the chromosphere, or the cores of solar prominences, where the large densities lead to very high collisional frequencies. But it becomes less accurate, for instance, at the upper layers of the chromosphere or at the transition regions between solar prominences and the corona. There the plasma is more rarefied, and the charged-neutral collision frequencies might not be so much higher than the frequencies of the waves that propagate through those environments. As it was shown by the works of, e.g., \citet{Khomenko2016ApJ...823..132K}, \citet{Wiehr2019ApJ...873..125W,Wiehr2021ApJ...920...47W} or \citet{Zapior2022ApJ...934...16Z}, it is possible to measure drift velocities between the ionized and neutral species of the order of hundreds of meters per second at the edges or at rapidly evolving regions of solar prominences. These results point out the necessity of more general fluid models that allow for a larger ion-neutral decoupling than the single-fluid approach.
    
In the present section, we consider two-fluid hydrogen plasmas, with the charged and neutral components interacting by means of elastic collisions. Using the two-fluid model we reconsider some of the topics studied in the previous section and underline the differences with respect to the single-fluid model. We start by considering the linear regime in order to study the propagation of MHD waves in both homogeneous and stratified partially ionized plasmas. We then discuss the nonlinear regime in the context of plasma heating caused by collisions between different species, and the effects of the charged-neutral interaction on the formation and propagation of shocks in the solar chromosphere.
    
\subsection{Basic equations of linear theory}
   
Following the same procedure as in Section~~\ref{sec:1f-linearization}, let us assume that each variable $\bm{f}$ is the sum of an equilibrium value, $\bm{f}_{0}$, and a small-amplitude perturbation, $\bm{f}_{1}$. In addition, let us consider a static background, so there is no equilibrium flows, that is, $\bm{V}_{\alpha,0} = \bm{0}$. Then, applying these assumptions to Eqs.~\ref{eq:continuity-2fluid-n}--\ref{eq:energy-2fluid-c}, the following set of two-fluid linear equations is obtained:
    
    \begin{equation} \label{eq:2f_lin_rho_n}
        \frac{\partial \rho_{\rm{n},1}}{\partial t} +\rho_{\rm{n},0} \nabla \cdot \bm{V}_{\rm{n},1} + \bm{V}_{\rm{n},1} \cdot \nabla \rho_{\rm{n},0} = 0, 
    \end{equation}
    \begin{equation} \label{eq:2f_lin_rho_c}
        \frac{\partial \rho_{\rm{c},1}}{\partial t} +\rho_{\rm{c},0} \nabla \cdot \bm{V}_{\rm{c},1} + \bm{V}_{\rm{c},1} \cdot \nabla \rho_{\rm{c},0} = 0,
    \end{equation}
    \begin{equation} \label{eq:2f_lin_momn}
        \rho_{\rm{n},0} \frac{\partial \bm{V}_{\rm{n},1}}{\partial t} = -\nabla P_{\rm{n},1} + \rho_{\rm{n},1} \bm{g} + \alpha_{\rm{cn}} \left(\bm{V}_{\rm{c},1}  - \bm{V}_{\rm{n},1}\right),
    \end{equation}
    \begin{equation} \label{eq:2f_lin_momc}
        \rho_{\rm{c},0} \frac{\partial \bm{V}_{\rm{c},1}}{\partial t} = -\nabla P_{\rm{c},1} - \frac{\nabla\left(\bm{B}_{0} \cdot \bm{B}_{1}\right)}{\mu_{0}} + \frac{\left(\bm{B_{0}} \cdot \nabla\right) \bm{B}_{1}}{\mu_{0}} + \rho_{\rm{c},1} \bm{g} - \alpha_{\rm{cn}} \left(\bm{V}_{\rm{c},1} - \bm{V}_{\rm{n},1}\right),
    \end{equation}
    \begin{equation} \label{eq:2f_lin_presn}
        \frac{\partial P_{\rm{n},1}}{\partial t} + \gamma P_{\rm{n},0} \nabla \cdot \bm{V}_{\rm{n},1} + \bm{V}_{\rm{n},1} \cdot \nabla P_{\rm{n},0} = 0,
    \end{equation}
     \begin{equation} \label{eq:2f_lin_presc}
        \frac{\partial P_{\rm{c},1}}{\partial t} + \gamma P_{\rm{c},0} \nabla \cdot \bm{V}_{\rm{c},1} + \bm{V}_{\rm{c},1} \cdot \nabla P_{\rm{c},0} = 0,
    \end{equation}
    \begin{equation} \label{eq:2f_lin_induc}
        \frac{\partial \bm{B}_{1}}{\partial t} = \left(\bm{B}_{0} \cdot \nabla \right) \bm{V}_{\rm{c},1} - \bm{B}_{0} \nabla \cdot \bm{V}_{\rm{c},1} - \frac{1}{\mu_{0} e n_{\rm{e}}} \nabla \times \left[\left(\nabla \times \bm{B}_{1}\right) \times \bm{B}_{0} \right], 
    \end{equation}
where the evolution equations for the energies have been rewritten in terms of the pressure of each fluid (Eq.~\ref{eq:ienes_to_pressures}), and
    \begin{equation} \label{eq:2f_alphacn}
        \alpha_{\rm{cn}} = \rho_{\rm{c}} \nu_{\rm{cn}} = \rho_{\rm{n}} \nu_{\rm{nc}} = K_{\rm{col}} \rho_{\rm{c}} \rho_{\rm{n}},
    \end{equation}
is known as the friction coefficient. 
The friction coefficient generally depends on the complete densities of the two fluids, that is, the sum of the background and the perturbation values. However, for the linear approximation we assume that it only depends on the background densities, $\rho_{\rm{c},0}$ and $\rho_{\rm{n},0}$.
    
The equations above do not include any assumptions on the properties of the equilibrium state apart from being static. Therefore, they can be applied to homogeneous and inhomogeneous backgrounds, as in the cases of unbounded uniform plasmas or of gravitationally stratified atmospheres, respectively. Both scenarios are explored in the following sections. 
    
\subsection{Waves in homogeneous plasmas}

It has been shown in Section \ref{sec:1f-homogeneous} that MHD waves in homogeneous unbounded plasmas in a single-fluid description can be classified into two main categories: Alfv\'en waves and magnetoacoustic waves. 
A multi-fluid description increases the number of waves that are allowed in the system. In the two-fluid model analyzed in this section, the additional waves are related to acoustic modes of the neutral species. Due to a large number of available modes, their study can be cumbersome. It is helpful to make use of the general properties of the modes, which allows to simplify the mathematical description.
As it has been shown in \textbf{Chapter 5}, Alfv\'en waves are incompressible and propagate vorticity perturbations, while magnetoacoustic waves are compressible and they do not propagate vorticity. Therefore, by choosing an appropriate set of variables (vorticity for Alfv\'en waves and compressibility for magnetoacoustic waves), the equations that describe the dynamics of both kinds of waves decouple and they can be studied separately.

\subsubsection{Alfv\'en waves} \label{sec:2f_alfven}

This section follows \cite{Soler2013ApJ...767..171S} to derive the properties of Alfv\'en waves in the two-fluid homogeneous plasmas. The equilibrium state is given by a uniform and unbounded partially ionized plasma, unaffected by gravity, and embedded in a uniform and straight magnetic field oriented along the $z$-direction, $\bm{B} = B_{z} \hat{z}$. The Hall term is neglected in this section.

We perform a Fourier analysis by assuming that the spatial dependence of the perturbations is given by $\exp \left(i k_{x} x + i k_{y} y + i k_{z} \right)$, where $k_{x}$, $k_{y}$, and $k_{z}$ are the components of the wavenumber in the $x$-, $y$-, and $z$- directions, respectively. Then, we define the $z$-component of the vorticity of the neutral and charged fluids as
\begin{equation} \label{eq:2f_vort_n}
        \Gamma_{\rm{n}} = \left(\nabla \times \bm{V}_{\rm{n}}\right) \cdot \hat{z} = i k_{x} V_{\rm{n},y} - i k_{y} V_{\rm{n},x},
\end{equation}
\begin{equation} \label{eq:2f_vort_c}
        \Gamma_{\rm{c}} = \left(\nabla \times \bm{V}_{\rm{c}}\right) \cdot \hat{z} = i k_{x} V_{\rm{c},y} - i k_{y} V_{\rm{c},x}.
\end{equation}

By applying the rotational operator ($\nabla \times$) over Eqs.~\ref{eq:2f_lin_momn}, \ref{eq:2f_lin_momc}, and \ref{eq:2f_lin_induc}, and combining the resulting expressions, we obtain the following evolution equations for $\Gamma_{\rm{n}}$ and $\Gamma_{\rm{c}}$:
\begin{equation} \label{eq:2f_dtvort_n}
        \rho_{\rm{n}} \frac{\partial \Gamma_{\rm{n}}}{\partial t} + \alpha_{\rm{cn}} \Gamma_{\rm{n}} = \alpha_{\rm{cn}} \Gamma_{\rm{c}},
\end{equation}
\begin{equation} \label{eq:2f_dtvort_c}
        \rho_{\rm{c}} \frac{\partial^{2} \Gamma_{\rm{c}}}{\partial t^{2}} + \alpha_{\rm{cn}} \frac{\partial \Gamma_{\rm{c}}}{\partial t} + \rho_{\rm{c}}k_{z}^{2}c_{\rm{A}}^{2}\Gamma_{\rm{c}} = \alpha_{\rm{cn}} \frac{\partial \Gamma_{\rm{n}}}{\partial t},
\end{equation}
where $c_{\rm{A}} = B_{z}/\sqrt{\mu_{0} \rho_{\rm{c}}}$ is the Alfv\'en speed defined through the density of the charged component.
    
Then, assuming that the temporal dependence of the perturbations is proportional to $\exp \left(-i \omega t\right)$ and combining the previous equations, the dispersion relation for Alfv\'en waves is obtained:
    \begin{equation} \label{eq:2f_alfven_dr}
        \omega^{3} + i \frac{\alpha_{\rm{cn}}}{\rho \xi_{\rm{c}} \xi_{\rm{n}}} \omega^{2} - c_{\rm{A}}^{2} k_{z}^{2} \omega - i \frac{\alpha_{\rm{cn}}}{\rho \xi_{\rm{n}}} c_{\rm{A}}^{2}k_{z}^{2} = 0,
    \end{equation}
where $\rho = \rho_{\rm{c}} + \rho_{\rm{n}}$, $\xi_{\rm{c}} = \rho_{\rm{c}}/ \rho$, and $\xi_{\rm{n}} = \rho_{\rm{n}} / \rho$. The dispersion relation can also be written in terms of the ionization fraction, $\chi = \rho_{\rm{n}} / \rho_{\rm{c}}$, and the neutral-charged collision frequency, $\nu_{\rm{nc}} = \alpha_{\rm{cn}} / \rho_{\rm{n}}$:
    \begin{equation} \label{eq:2f_alfven_dr2}
        \omega^{3} + i \left(1+ \chi \right) \nu_{\rm{nc}} \omega^{2} - k_{z}^{2} c_{\rm{A}}^{2} \omega - i\nu_{\rm{nc}} k_{z}^{2} c_{\rm{A}}^{2} = 0.
    \end{equation}
    
To study the properties of standing Alfv\'en waves, we solve Eq. \ref{eq:2f_alfven_dr2} for a real wavenumber $k_{z}$ and allow for a complex temporal frequency, $\omega = \omega_{\rm{R}} + i \omega_{\rm{I}}$. Since Eq. \ref{eq:2f_alfven_dr2} is a cubic equation in $\omega$, it has three solutions. Exact analytic solutions are too complex to provide any useful information on the physics of these waves. However, some information can be extracted by performing the change of variable $\omega = i s$, which leads to the following expression

\begin{equation} \label{eq:2f_alfven_dr_s}
        s^{3} + \left(1 + \chi \right) \nu_{\rm{nc}}s^{2} + k_{z}^{2}c_{\rm{A}}^{2}s + \nu_{\rm{nc}}k_{z}^{2}c_{\rm{A}}^{2} = 0,
\end{equation}
and then computing its polynomial discriminant. Eq.~\ref{eq:2f_alfven_dr_s} is a cubic equation in which all the coefficients are real. Thus, its discriminant is given by:
\begin{equation} \label{eq:2f_alfven_disc}
        \Lambda = -k_{z}^{2}c_{\rm{A}}^{2} \left[4 \left(1+\chi\right)^{3} \nu_{\rm{nc}}^{4} - \left(\chi^{2} + 20\chi - 8 \right) \nu_{\rm{nc}}^{2}k_{z}^{2} c_{\rm{A}}^{2} + 4 k_{z}^{4}c_{\rm{A}}^4\right].
\end{equation}
    
Equation \ref{eq:2f_alfven_dr_s} has one real root and two complex conjugate roots when $\Lambda < 0$, a multiple real root when $\Lambda = 0$, and three distinct real roots when $\Lambda > 0$. The complex roots of Eq.~\ref{eq:2f_alfven_dr_s} correspond to damped oscillatory solutions of Eq.~\ref{eq:2f_alfven_dr2}, while the real solutions of Eq.~\ref{eq:2f_alfven_dr_s} correspond to evanescent solutions of Eq.~\ref{eq:2f_alfven_dr2}, that is, solutions with its real part of the frequency equal to zero.
    
In the absence of collisions ($\nu_{\rm{nc}} = 0$) the discriminant becomes $\Lambda = - 4k_{z}^{6}c_{\rm{A}}^{6} < 0$, which means that Eq.~\ref{eq:2f_alfven_dr_s} has one real root and two complex conjugate roots
\begin{equation} \label{eq:2f_alfven_sroots}
        s = \pm i k_{z} c_{\rm{A}}, \quad s = 0,
\end{equation}
which correspond to the frequencies
\begin{equation} \label{eq:2f_alfven_wroots}
        \omega = \pm k_{z} c_{\rm{A}}, \quad \omega = 0.
\end{equation}
The two non-zero solutions correspond to the solutions for Alfv\'en waves in a fully ionized plasma.
    
In a general case with $\nu_{\rm{nc}} \ne 0$, it is possible to find the wavenumbers that satisfy $\Lambda = 0$. At those wavenumbers, denoted by $k_{z}^{-}$ and $k_{z}^{+}$, the nature of the solutions changes. The values of $k_{z}^{-}$ and $k_{z}^{+}$ are given by, 
\begin{equation} \label{eq:2f_alfven_kpm}
        k_{z}^{\pm} = \frac{\nu_{\rm{nc}}}{c_{\rm{A}}} \left[\frac{\chi^{2} + 20\chi - 8}{8\left(1+\chi\right)^{3}} \pm \frac{\chi^{1/2}\left(\chi-8\right)^{3/2}}{8\left(1+\chi\right)^{3}}\right]^{-1/2}.
\end{equation}
    
Since $k_{z}$ was assumed real, it can be verified from Eq.~\ref{eq:2f_alfven_kpm} that there is a minimum value of the ionization fraction that allows for $\Lambda = 0$. This minimum value is $\chi =8$, which corresponds to the critical values of the wavenumbers $k_{z}^{+} = k_{z}^{-} = 3 \sqrt{3} \nu_{\rm{nc}}/c_{\rm{A}}$. For values larger than the minimum ionization fraction, the relation $k_{z}^{+} < k_{z}^{-}$ is fulfilled. Outside the interval $\left( k_{z}^{+}, k_{z}^{-} \right)$, the discriminant is negative, which means that two of the solutions of the dispersion relation correspond to damped Alfv\'en waves, while the remaining solution is evanescent. The interval $\left( k_{z}^{+}, k_{z}^{-} \right)$ is known as the cut-off region \citep{Kulsrud1969ApJ...156..445K,Soler2013ApJ...767..171S}, since for wavenumbers inside this interval $\Lambda > 0$, all three roots of Eq.~\ref{eq:2f_alfven_dr_s} are real, and the solutions of Eq.~\ref{eq:2f_alfven_dr2} are purely imaginary. Therefore, there are no propagating waves in the cut-off region.
    
When $k_{z} > k_{z}^{-}$, there is a weak coupling between the charged and neutral fluids and, thus, disturbances in the magnetic field affect only the charged fluid, as if the plasma were fully ionized. Conversely, when $k_{z} < k_{z}^{+}$ the collisional interaction is strong enough to couple both fluids, so they behave as a single fluid. In the intermediate situation, when $k_{z} \in \left(k_{z}^{+}, k_{z}^{-} \right)$, collisions between the charged and the neutral particles efficiently dissipate perturbations in the magnetic field before enough inertia is transferred to the neutral fluid. The result is that oscillations are suppressed inside that interval of wavenumbers.
    
\begin{figure} [t!]
    \centering
    \includegraphics[width=0.49\hsize]{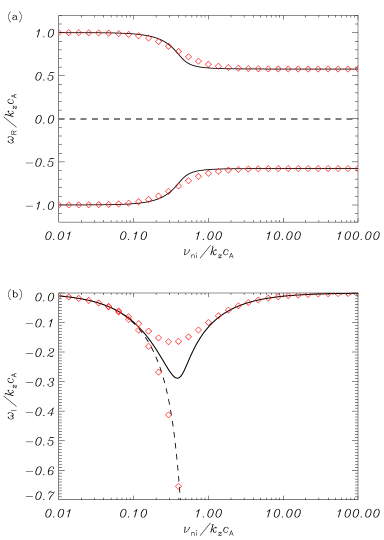}
    \includegraphics[width=0.49\hsize]{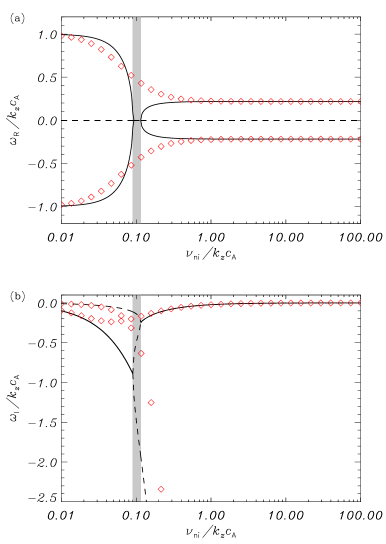}
    \caption{Normalized wave frequency, $\omega_{R} /k_{z}c_{\rm{A}}$ (top panels), and normalized damping rate, $\omega_{I}/k_{z}c_{\rm{A}}$ (bottom panels), as functions of the normalized collision frequency, $\nu_{\rm{nc}}/k_{z}c_{\rm{A}}$ for the Alfv\'en waves in partially ionized plasma. Solid and dashed lines represent numerical results of the oscillatory and evanescent modes, respectively. Symbols correspond to analytical approximations. Left panels show the case with $\chi = 2$ (intermediate plasma ionization), while right panels show the case with $\chi = 20$ (weak plasma ionization). Shaded areas on the right panels mark the position of the cut-off region. Figure from \citet{Soler2013ApJ...767..171S}.}
        \label{fig:2f_alfven}
    \end{figure}
    
Another path to obtain information on the properties of Alfv\'en waves in partially ionized plasmas without fully solving the dispersion relation is to consider  certain approximations. For that, one inserts the expression $\omega = \omega_{\rm{R}} + i \omega_{\rm{I}}$ into Eq. \ref{eq:2f_alfven_dr2} and assumes that the damping rate is small compared to the wave frequency, $|\omega_{\rm{I}}| \ll |\omega_{\rm{R}}|$. This leads to the following approximate expressions for $\omega_{\rm{R}}$ and $\omega_{\rm{I}}$ for two propagating modes,
\begin{equation} \label{eq:2f_alfven_approx_wr}
        \omega_{R} \approx \pm k_{z} c_{\rm{A}} \sqrt{\frac{k_{z}^{2}c_{\rm{A}}^{2} + \left(1 + \chi \right) \nu_{\rm{nc}}^{2}}{k_{z}^{2}c_{\rm{A}}^{2} + \left(1 + \chi \right)^{2} \nu_{\rm{nc}}^{2}}},
\end{equation}
\begin{equation} \label{eq:2f_alfven_approx_wi}
        \omega_{I} \approx -\frac{\chi \nu_{\rm{nc}}}{2 \left[k_{z}^{2}c_{\rm{A}}^{2}+ \left(1 + \chi \right)^{2} \nu_{\rm{nc}}^{2} \right]}k_{z}^{2}c_{\rm{A}}^{2}.
\end{equation}
The approximate solution for the remaining purely imaginary mode is $\omega = i \epsilon$, with
\begin{equation} \label{eq:2f_alfven_approx_eps}
        \epsilon \approx -\nu_{\rm{nc}} \frac{k_{z}^{2}c_{\rm{A}}^{2} + \left(1 + \chi \right)^{2} \nu_{\rm{nc}}^{2}}{k_{z}^{2}c_{\rm{A}}^{2} + \left(1 + \chi \right) \nu_{\rm{nc}}^{2}}.
\end{equation}
    
Now, approximate results for some interesting limits can be obtained. For instance, in the weak coupling regime, that is, when $\nu_{\rm{nc}} \ll k_{z} c_{\rm{A}}$, we get
\begin{equation} \label{eq:2f_alfven_approx_weak}
        \omega_{R} \approx \pm k_{z} c_{\rm{A}}, \quad \omega_{I} \approx -\frac{\chi \nu_{\rm{nc}}}{2}, \quad \epsilon \approx -\nu_{\rm{nc}},
\end{equation}
which shows that in this limit the damping of Alfv\'en waves is independent from the wavenumber of the perturbation. On the other hand, in the strong coupling limit ($\nu_{\rm{nc}} \gg k_{z} c_{\rm{A}}$), the approximate results are:
\begin{equation} \label{eq:2f_alfven_approx_strong}
        \omega_{R} \approx \pm \frac{k_{z}c_{\rm{A}}}{\sqrt{1 + \chi}}, \quad \omega_{I} \approx -\frac{\chi}{2 \left(1 + \chi \right)^{2}}\frac{k_{z}^{2}c_{\rm{A}}^{2}}{\nu_{\rm{nc}}}, \quad \epsilon \approx -\left(1 + \chi \right) \nu_{\rm{nc}}.
\end{equation}
In this regime, $\omega_{\rm{R}}$ is inversely proportional to the factor $\sqrt{1 + \chi}$, which implies that the presence of neutral species in the plasma reduces the frequency of the Alfv\'en waves in comparison to the fully ionized case. In addition, the damping is proportional to $k_{z}^{2}$ and, thus, the damping is more efficient for larger wavenumbers (or short wavelengths).
    
The assumption made to derive the previous approximations is not valid for waves with $k_{z} \in \left(k_{z}^{+}, k_{z}^{-} \right)$, because in the cut-off region $\omega_{\rm{R}} = 0$. The goodness of the approximate expressions can be checked in Fig.~\ref{fig:2f_alfven}, which shows a comparison between the exact solutions (black lines) and the approximations (red symbols). Figure \ref{fig:2f_alfven} also shows that the cut-off region is only present for large values of $\chi$ (large neutral fraction). Another important conclusion that can be extracted from the bottom panels of Fig. \ref{fig:2f_alfven} is that the damping of the propagating modes due to the charged-neutral collisions is more efficient around the value $\nu_{\rm{nc}} / \left(k_{z} c_{\rm{A}}\right) \approx 1$, that is, when the collision and the oscillation frequencies are of the same order of magnitude.
    
\subsubsection{Magneto-acoustic waves} \label{sec:2f_magnetoacoustic}

The same equilibrium conditions as in the previous section are assumed for magneto-acoustic waves. The spatial and temporal dependence of the perturbations are also the same. However, in this case one has to consider compressibility perturbations instead of the vorticity ones \citep{Soler2013ApJS..209...16S}. The compressibility of the neutral and the charged fluids, respectively, is defined as
\begin{equation} \label{eq:2f_div_vn}
        \Delta_{\rm{n}} = \nabla \cdot \bm{V}_{\rm{n}} = i k_{x} V_{\rm{n},x} + i k_{y} V_{\rm{n},y} + i k_{z} V_{\rm{n},z},
\end{equation}
\begin{equation} \label{eq:2f_div_vc}
        \Delta_{\rm{c}} = \nabla \cdot \bm{V}_{\rm{c}} = i k_{x} V_{\rm{c},x} + i k_{y} V_{\rm{c},y} + i k_{z} V_{\rm{c},z}.
\end{equation}
    
\noindent Then, one computes the divergence of Eqs. \ref{eq:2f_lin_momn}, \ref{eq:2f_lin_momc}, and \ref{eq:2f_lin_induc}, and combine the resulting expressions with Eqs. \ref{eq:2f_lin_rho_n}, \ref{eq:2f_lin_rho_c}, \ref{eq:2f_lin_presn}, and \ref{eq:2f_lin_presc} to obtain the two following coupled equations for $\nabla_{\rm{n}}$ and $\nabla_{\rm{c}}$:
\begin{equation} \label{eq:2f_dtdiv_vn}
        \left(\omega^{2} - k^{2} c_{\rm{S,n}}^{2} \right) \Delta_{\rm{n}} = -i \nu_{\rm{nc}} \left(\Delta_{\rm{n}} - \Delta_{\rm{c}} \right),
\end{equation}
\begin{gather} 
        \left(\omega^{4} - \omega^{2} k^{2} \left(c_{\rm{S,c}}^{2} + c_{\rm{A}}^{2} \right) + k^{2} k_{z}^{2} c_{\rm{A}}^{2} c_{\rm{S,c}}^{2}\right) \Delta_{\rm{c}} = -i \nu_{\rm{cn}} \omega^{3} \left(\Delta_{\rm{c}} - \Delta_{\rm{n}} \right) \nonumber \\
        + \frac{i \nu_{\rm{cn}}}{\omega + i \left(\nu_{\rm{nc}} + \nu_{\rm{cn}} \right)} k^{2} k_{z}^{2} c_{\rm{A}}^{2} \left(c_{\rm{S,c}}^{2} \Delta_{\rm{c}} - c_{\rm{S,n}}^{2} \Delta_{\rm{n}} \right)
        \label{eq:2f_dtdiv_vc},
\end{gather}
where $k^{2} = k_{x}^{2} + k_{y}^{2} + k_{z}^{2}$, and $c_{\rm{S,n}}$ and $c_{\rm{S,c}}$ are the sound speed of the neutral and the charged fluid, respectively, given by
    \begin{equation} \label{eq:2f_csc_csn}
        c_{\rm{S,c}} = \sqrt{\frac{\gamma P_{\rm{c},0}}{\rho_{\rm{c,0}}}} \quad \text{and} \quad c_{\rm{S,n}} = \sqrt{\frac{\gamma P_{\rm{n},0}}{\rho_{\rm{n},0}}}.
    \end{equation}
    
The combination of Eqs. \ref{eq:2f_dtdiv_vn} and \ref{eq:2f_dtdiv_vc} yields the dispersion relation for magneto-acoustic waves in a partially ionized plasma \citep{2018SSRv..214...58B},
\begin{gather}
        \Big[\left(\omega^{4} + i \nu_{\rm{cn}} \omega^{3} - k^{2} \left(c_{\rm{A}}^{2} + c_{\rm{S,c}}^{2}\right)\omega^{2}\right) \big(\omega + i\left(\nu_{\rm{cn}} + \nu_{\rm{nc}}\right)\big) \nonumber \\
        + k^{2}k_{z}^{2}c_{\rm{A}}^{2}c_{\rm{S,c}}^{2} \left(\omega + i \nu_{\rm{nc}}\right) \Big] \left(\omega^{2} - k^{2}c_{\rm{S,n}}^{2} + i \nu_{\rm{nc}} \omega \right) \nonumber \\
        + \nu_{\rm{cn}} \nu_{\rm{nc}} \omega \left[\omega^{3} \big(\omega + i\left(\nu_{\rm{cn}} + \nu_{\rm{nc}}\right) \big) - k^{2} k_{z}^{2} c_{\rm{A}}^{2} c_{\rm{S,n}}^{2} \right] = 0,
        \label{eq:2f_mawaves_dr}
\end{gather}
which is a seventh order equation of $\omega$, so it has seven different solutions. Due to its complexity, it must be solved numerically. Nevertheless, the study of limiting cases can provide useful information about the nature of the solutions.
    
By neglecting the collisional interaction, $\nu_{\rm{nc}} = \nu_{\rm{cn}} = 0$, the dispersion relation becomes
\begin{equation} \label{eq:2f_mawaves_drnu0}
        \omega \left(\omega^{4} - k^{2} \left(c_{\rm{A}}^{2} + c_{\rm{S,c}}^{2} \right) \omega^{2} + k^{2} k_{z}^{2}c_{\rm{A}}^{2}c_{\rm{S,c}}^{2}\right) \left(\omega^{2} - k^{2}c_{\rm{S,n}}^{2}\right) = 0.
\end{equation}
This expression shows that four of the seven modes are magneto-acoustic modes related to the charged fluid (forward and backward propagating slow and fast modes), two modes correspond to acoustic waves of the neutral fluid (forward and backward propagating acoustic modes). The remaining solution, with $\omega = 0$, is the entropy mode \citep{Goedbloed2004prma.book.....G}. 
    
In the opposite limit, when $\nu_{\rm{cn}} \to \infty$ and $\nu_{\rm{nc}} \to \infty$ (strong coupling between the fluids), the dispersion relation simplifies to,
\begin{equation} \label{eq:2f_mawaves_drcoupled}
        \omega^{3} \left(\omega^{4} - \omega^{2} k^{2} \frac{c_{\rm{A}}^{2} + c_{\rm{S,c}}^{2} + \chi c_{\rm{S,n}}^{2}}{1 + \chi} + k^{4} \frac{c_{\rm{A}}^{2} \left(c_{\rm{S,c}}^{2} + \chi c_{\rm{S,n}}^{2}\right)}{\left(1 + \chi \right)^{2}} \cos^{2} \theta \right) = 0,
\end{equation}
where $\theta$ is the angle between the wavevector, $\bm{k}$, and the equilibrium magnetic field, $\bm{B}$. There are three entropy modes with $\omega = 0$. The frequencies of the other four modes are given by
\begin{equation} \label{eq:2f_mawaves_coupled}
        \omega^{2} = k^{2} \frac{c_{\rm{A}}^{2} + c_{\rm{S,c}}^{2} + \chi c_{\rm{S,n}}^{2}}{2\left(1+\chi \right)} \pm k^{2} \frac{c_{\rm{A}}^{2} + c_{\rm{S,c}}^{2} + \chi c_{\rm{S,n}}^{2}}{2 \left(1 + \chi \right)}\left[1-\frac{4 c_{\rm{A}}^{2} \left(c_{\rm{S,c}}^{2} + \chi c_{\rm{S,n}}^{2} \right) \cos^{2} \theta}{\left(c_{\rm{A}}^{2} + c_{\rm{S,c}}^{2} + \chi c_{\rm{S,n}}^{2}\right)^{2}}\right]^{1/2},
\end{equation}
where the signs ``+/-'' correspond to the forward and backward modified fast/slow waves. Here, the term `modified' is used to underline that these modes have modified properties compared to the fully ionized case.
    
\begin{figure} [t!]
    \centering
    \includegraphics[width=\hsize]{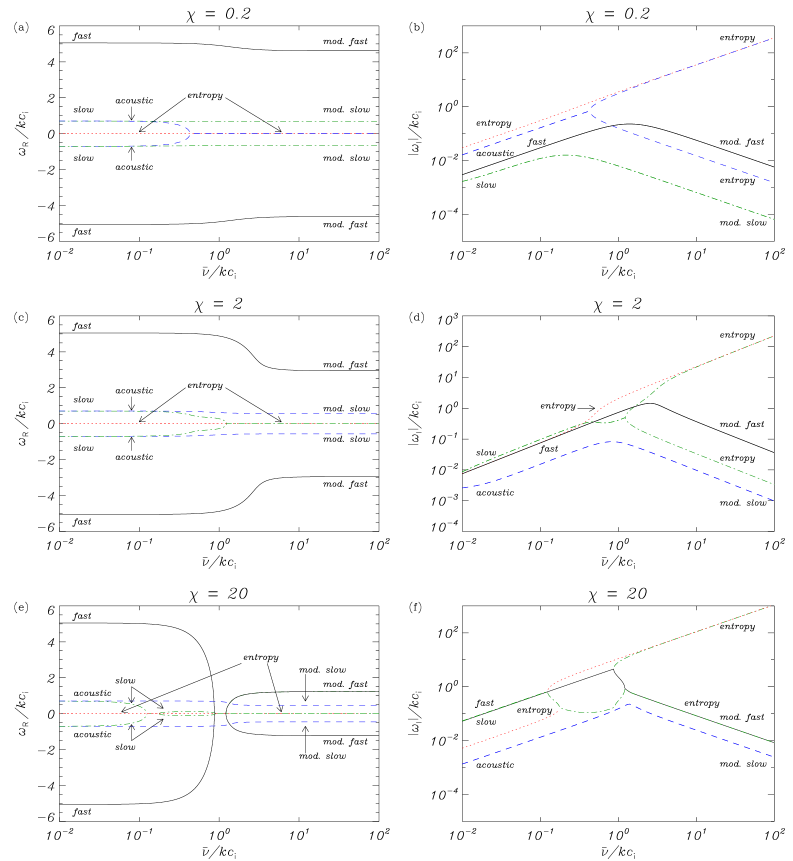}
    \caption{Real part (left) and absolute value of the imaginary part (right) of frequency of magneto-acoustic waves, as a function of the averaged collision frequency, $\tilde{\nu} \equiv 2 \alpha_{\rm{cn}}/\left(\rho_{\rm{c}} + \rho_{\rm{n}} \right)$, for oblique propagation at an angle $\theta = \pi/4$, with $\beta_{\rm{c}} \equiv c_{\rm{S,c}}^{2}/c_{\rm{A}}^{2} = 0.04$. From top to bottom, panels represent the cases with $\chi = 0.2$ (strong ionization), $\chi = 2$ (intermediate ionization), and $\chi = 20$ (weak ionization). All frequencies are normalized by $k c_{\rm{i}}$, with $c_{\rm{i}} \equiv c_{\rm{S,c}}$, respectively. Figure from \citet{Soler2013ApJS..209...16S}.}
     \label{fig:2f_mawaves_pi4}
\end{figure}
    
It is interesting to note that in the uncoupled case there are three different kinds of propagating modes (the slow and fast waves of the charged fluid plus the neutral acoustic mode), but in the strongly coupled case there are only two kinds of propagating modes (the modified versions of the slow and fast magnetoacoustic modes). The reason is that, in this limit, the neutral acoustic mode has transformed into an evanescent mode, with $\omega = 0$. Another remarkable feature of the strong coupling limit is that the wave frequencies are real, meaning the absence of collisional damping, as in the completely uncoupled case.
    
In an intermediate collisional coupling regime, the different modes have mixed properties, which strongly depend on the physical conditions of the two fluids (magnetic field strength, ionization degree) and the direction of propagation with respect to the background magnetic field. Therefore, for an arbitrary value of the collision frequency, $\nu_{\rm{nc}}$, there is no simple analytic solution of the dispersion relation and no general trend can be described for each mode. The solutions must be explored by numerically solving Eq. \ref{eq:2f_mawaves_dr}. 
    
To illustrate the numerical solutions that can be obtained from Eq. \ref{eq:2f_mawaves_dr} for a particular set of parameters, Fig.~\ref{fig:2f_mawaves_pi4} shows the results for magneto-acoustic waves obliquely propagating in a strongly magnetized plasma (that is, with a small value of plasma $\beta$). The top panels of Fig.~\ref{fig:2f_mawaves_pi4} show how, for the case of $\chi = 0.2$ (strong ionization), the slow and fast modes transform into the modified slow and fast modes as the averaged collision frequency, $\tilde{\nu} \equiv 2 \alpha_{\rm{cn}}/(\rho_{\rm{c}} + \rho_{\rm{n}})$, increases or how the acoustic mode related to the neutral fluid turns into an entropy mode. However, for an intermediate ionization degree of $\chi = 2$, the neutral acoustic mode transforms into the modified slow wave as $\tilde{\nu}$ increases, and the slow magneto-acoustic wave is the one that becomes an entropy mode. Finally, the bottom panels of Fig. \ref{fig:2f_mawaves_pi4} reveal the presence of cut-off regions of the slow and fast magneto-acoustic modes for the case of weak ionization, $\chi = 20$. Similar to Section \ref{sec:2f_alfven}, waves within the cut-off regions have $\omega_{\rm{R}} = 0$ and they do not propagate.

\subsection{Waves in stratified partially-ionized plasmas} \label{sec:2f_stratified}

The results from the previous section are applicable to scenarios where the equilibrium state of the plasma is homogeneous. Nevertheless, for the short-period waves (with wavelengths smaller than the pressure scale height), propagation through the stratified atmosphere can be approximately studied by locally solving Eqs. \ref{eq:2f_alfven_dr2} and \ref{eq:2f_mawaves_dr} for the parameters of the plasma corresponding to every height. For instance, this method has been used by \citet{Soler2013ApJS..209...16S} to study the propagation of magnetoacoustic waves in the solar chromosphere. They found that fast waves with wavelengths of $1-10$ km are  strongly damped due to charged-neutral collisions at heights of $1500-2000$ km above the photosphere, while the damping of slow waves is more important at heights of $1000-1500$ km.

Nevertheless, it has to be kept in mind that local application of the homogeneous dispersion relation is only an approximation. It misses two important effects related to the stratification of the background atmosphere, i.e. the existence of the gravitational cut-off frequency, and the increase of the wave amplitude with height caused by the exponential density fall-off \citep{1984oup..book.....M}. Fortunately, these two effects can still be addressed by the linear theory. Below we consider a particular case of how the charge-neutral interaction affects the propagation of fast magneto-acoustic waves in the solar chromosphere, taking also into account the influence of the gravitational stratification, following the work by \citet{PopescuBraileanu2019A&A...630A..79P}. 

The background atmosphere is gravitationally stratified in the $z$ direction, $\bm{g}=-g \hat{\bm{z}}$. The plasma temperature, and a purely horizontal magnetic field, $B_{0,x}$ both vary with height. Applying these physical conditions to Eqs. \ref{eq:2f_lin_rho_n}--\ref{eq:2f_lin_induc} we get that the linear evolution of fast magneto-acoustic waves is described by the following set of equations:
\begin{equation} \label{eq:2f_strat_rho_n}
        \frac{\partial \rho_{\rm{n},1}}{\partial t} + \rho_{\rm{n},0} \frac{\partial V_{\rm{n},z}}{\partial z} + V_{\rm{n},z} \frac{d \rho_{\rm{n},0}}{d z} = 0,
\end{equation}
\begin{equation} \label{eq:2f_strat_rho_c}
        \frac{\partial \rho_{\rm{c},1}}{\partial t} + \rho_{\rm{c},0} \frac{\partial V_{\rm{c},z}}{\partial z} + V_{\rm{c},z} \frac{d \rho_{\rm{c},0}}{d z} = 0,
\end{equation}
 \begin{equation} \label{eq:2f_strat_mom_n}
        \rho_{\rm{n},0} \frac{\partial V_{\rm{n},z}}{\partial t} = -\rho_{\rm{n},1} g - \frac{\partial P_{\rm{n},1}}{\partial z} + K_{\rm{col}} \rho_{\rm{n},0} \rho_{\rm{c},0} \left(V_{\rm{c},z} - V_{\rm{n},z} \right),
\end{equation}
\begin{gather} 
        \rho_{\rm{c},0} \frac{\partial V_{\rm{c},z}}{\partial t} = - \rho_{\rm{c},1}g - \frac{\partial P_{\rm{c},1}}{\partial z} - \frac{1}{\mu_{0}} \left(\frac{\partial B_{1,x}}{\partial z} B_{0,x} + \frac{d B_{0,x}}{d z} B_{1,x}\right) \nonumber \\
        + K_{\rm{col}} \rho_{\rm{n},0} \rho_{\rm{c},0} \left(V_{\rm{n},z} - V_{\rm{c},z} \right),
        \label{eq:2f_strat_mom_c}
\end{gather}
\begin{equation} \label{eq:2f_strat_pres_n}
        \frac{\partial P_{\rm{n},1}}{\partial t} = c_{\rm{S,n}}^{2} \frac{\partial \rho_{\rm{n},1}}{\partial t} + c_{\rm{S,n}}^{2} V_{\rm{n},z} \frac{d \rho_{\rm{n},0}}{d z} - V_{\rm{n},z} \frac{d P_{\rm{n},0}}{d z},
\end{equation}
\begin{equation} \label{eq:2f_strat_pres_c}
        \frac{\partial P_{\rm{c},1}}{\partial t} = c_{\rm{S,c}}^{2} \frac{\partial \rho_{\rm{c},1}}{\partial t} + c_{\rm{S,c}}^{2} V_{\rm{c},z} \frac{d \rho_{\rm{c},0}}{d z} - V_{\rm{c},z} \frac{d P_{\rm{c},0}}{d z},
\end{equation}   
\begin{equation} \label{eq:2f_strat_indu}
        \frac{\partial B_{1,x}}{\partial t} = -B_{0,x} \frac{\partial V_{\rm{c},z}}{\partial z} - V_{\rm{c},z} \frac{d B_{0,x}}{d z}.
\end{equation}
Notice that the background densities and pressures, $\rho_{\rm{n},0}$, $\rho_{\rm{c},0}$, $P_{\rm{n},0}$, and $P_{\rm{c},0}$, are all functions of height. Consequently, the Alfv\'en and sound speeds, $c_{\rm{A}}$, $c_{\rm{S},n}$, and $c_{\rm{S},c}$, and the collision coefficient $K_{\rm{col}}$ also depend on height.
    
The height variation of the background pressures of the neutral and the charged fluids are given by the conditions of (magneto-)hydrostatic equilibrium,
\begin{equation} \label{eq:2f_strat_pn0}
        \frac{d P_{\rm{n},0}}{d z} = - \frac{m_{\rm{H}} g}{k_{\rm{B}}T}P_{\rm{n},0},
\end{equation}
\begin{equation} \label{eq:2f_strat_pc0}
        \frac{d \left(P_{\rm{c},0} + P_{\rm{m},0} \right)}{d z} = - \frac{m_{\rm{H}} g}{2k_{\rm{B}} T}P_{\rm{c},0},
\end{equation}
where $P_{\rm{m},0} \equiv B_{0,x}^{2}/(2 \mu_{0})$ is the magnetic pressure. Ideal equation of state has been used. The solution of Eq. \ref{eq:2f_strat_pn0} is 
    \begin{equation} \label{eq:2f_strat_pn0sol}
        P_{\rm{n},0} = P_{\rm{n},0}(z_{0}) \exp \left(-\frac{m_{\rm{H}}g}{k_{\rm{B}}}\int_{z_{0}}^{z} \frac{1}{T(z')}dz'\right).
    \end{equation}
Equation \ref{eq:2f_strat_pc0} can be solved by assuming that the pressure of the charged fluid and the magnetic pressure have the same exponential dependence on height, 
\begin{equation} \label{eq:2f_strat_pc0sol}
        P_{\rm{c},0} = P_{\rm{c},0} \exp \left(-2 F(z)\right),
\end{equation}
\begin{equation} \label{eq:2f_strat_pm0sol}
        P_{\rm{m},0} = \frac{B_{0,x}(z)^{2}}{2\mu_{0}} = \frac{B_{0,x}(z_{0})^{2}}{2 \mu_{0}}\exp(-2F(z)) + C,
\end{equation}
with $C$ being an integration constant. By imposing $F(z)$ to satisfy $F(z) \geq 0$ and $F(z_{0}) = 0$, one obtains:
\begin{equation} \label{eq:2f_strat_fz}
        F(z)= \frac{m_{\rm{H}}g}{4k_{\rm{B}}}\frac{P_{\rm{c},0}(z_{0})}{P_{\rm{c},0}(z_{0})+B_{0,x}(z_{0})^{2}/(2\mu_{0})}\int_{z_{0}}^{z}\frac{1}{T(z')}dz'.
\end{equation}
    
Given the equilibrium conditions, linearized Eqs. \ref{eq:2f_strat_rho_n}--\ref{eq:2f_strat_indu} can be combined into two coupled differential equations for the vertical velocity component of each fluid:
\begin{equation} \label{eq:2f_strat_vnz}
        \frac{\partial^{2} V_{\rm{n},z}}{\partial t^{2}} = a_{\rm{n}}(z) \frac{\partial^{2} V_{\rm{n},z}}{\partial z^{2}} + b_{\rm{n}}(z) \frac{\partial V_{\rm{n},z}}{\partial z} + K_{\rm{col}} \rho_{\rm{c},0} \left(\frac{\partial V_{\rm{c},z}}{\partial t} - \frac{\partial V_{\rm{n},z}}{\partial t}\right),
\end{equation} 
\begin{equation} \label{eq:2f_strat_vcz}
        \frac{\partial^{2} V_{\rm{c},z}}{\partial t^{2}} = a_{\rm{c}}(z) \frac{\partial^{2} V_{\rm{c},z}}{\partial z^{2}} + b_{\rm{c}}(z) \frac{\partial V_{\rm{c},z}}{\partial z} + K_{\rm{col}} \rho_{\rm{n},0} \left(\frac{\partial V_{\rm{n},z}}{\partial t} - \frac{\partial V_{\rm{c},z}}{\partial t} \right),
\end{equation}
where
\begin{equation} \label{eq:2f_strat_ac_an}
        a_{\rm{c}}(z) = c_{\rm{S,c}}^{2} + c_{\rm{A}}^{2}, \quad a_{\rm{n}}(z) = c_{\rm{S,}}^{2},
\end{equation}
\begin{equation} \label{eq:2f_strat_bc_bn}
        b_{\rm{c}}(z) = \frac{1}{\rho_{\rm{c},0}} \frac{d \left(\rho_{\rm{c},0} a_{\rm{c}}\right)}{d z} \quad \text{and} \quad b_{\rm{n}}(z) = \frac{1}{\rho_{\rm{n},0}} \frac{d \left(\rho_{\rm{n},0}\right)}{d z}.
\end{equation}

\noindent Combining both velocity equations and assuming the perturbations in the form $\{V_{\rm{c},z}(z,t), V_{\rm{n},z}(z,t) \} = \{\tilde{V}_{\rm{c},z}(z), \tilde{V}_{\rm{n},z}(z) \} \exp \left(-i \omega t \right)$, the following expression is obtained:
\begin{gather}
        a_{\rm{c}} a_{\rm{n}}\frac{\partial^{4} \tilde{V}_{\rm{c},z}}{\partial z^{4}} + \left(a_{\rm{n}} b_{\rm{c}} + a_{\rm{c}} b_{\rm{n}} \right) \frac{\partial^{3} \tilde{V}_{\rm{c},z}}{\partial z^{3}} \nonumber \\
        + \left[b_{\rm{c}} b_{\rm{n}} + \omega^{2} \left(a_{\rm{c}} + a_{\rm{n}} \right) +i K_{\rm{col}} \omega \left(a_{\rm{c}} \rho_{\rm{c},0} + a_{\rm{n}} \rho_{\rm{n},0} \right) \right] \frac{\partial^{2} \tilde{V}_{\rm{c},z}}{\partial z^{2}} \nonumber \\
        + \omega \left[\omega \left(b_{\rm{c}} + b_{\rm{n}} \right) + i K_{\rm{col}} \left(b_{\rm{c}} \rho_{\rm{c},0} + b_{\rm{n}} \rho_{\rm{n},0} \right)\right] \frac{\partial \tilde{V}_{\rm{c},z}}{\partial z} \nonumber \\
        +\tilde{V}_{\rm{c},z} \omega^{3} \left[\omega + i K_{\rm{col}} \left(\rho_{\rm{c},0} + \rho_{\rm{n},0} \right) \right] = 0,
        \label{eq:2f_strat_d4vcz}
\end{gather}
which is a fourth-order differential equation with non-uniform coefficients. An approximate solution for this equation can be obtained by taking into account that the waves considered in this analysis are short-period waves. Therefore, a WKB approximation can be applied to this equation by assuming that the dependence of the velocity perturbations on height is
\begin{equation} \label{eq:2f_strat_wkb1}
        \{\tilde{V}_{\rm{n},z}, \tilde{V}_{\rm{c},z}\} = \{U_{\rm{n}} (z), U_{\rm{c}}(z)\} \cdot \exp \left[i \phi (z) \right],
\end{equation}
with
\begin{equation} \label{eq:2f_strat_wkb_phi}
        \phi(z) = - \int_{0}^{z} k(z') dz'.
\end{equation}
    
In addition, the velocity amplitude and the wavenumber gradients are assumed small and of the same order, so
\begin{equation} \label{eq:2f_strat_wkb2}
        \frac{\partial \phi}{\partial z} = - k = -k_{0} - \epsilon k_{1}(z) \quad \text{and} \quad U_{\rm{c,n}}(z) = U_{\rm{c,n},0} + \epsilon U_{\rm{c,n},1} (z),
\end{equation}
where $\epsilon$ is a small parameter. The second (and higher-order) derivatives of $k_{1}(z)$ and $U_{\rm{c,n},1}$ are set to zero.
    
After taking into account Eq.~\ref{eq:2f_strat_wkb2} to compute the derivatives of $\tilde{V}_{\rm{c},z}$, the following approximate dispersion relation is obtained from Eq. \ref{eq:2f_strat_d4vcz}: 
\begin{equation} \label{eq:2f_strat_wkbdr}
        \left(\omega^{2} - k^{2}a_{\rm{c}} - i k b_{\rm{c}} \right) \left(\omega^{2} - k^{2} a_{\rm{n}} - i k b_{\rm{n}}\right) - i \omega K_{\rm{col}} \rho_{0} \left(-\omega^{2} + k^{2} a + i k b\right) = 0,
\end{equation}
where $b = \left(\rho_{\rm{n},0} b_{\rm{n}} + \rho_{\rm{c},0} b_{\rm{c}}\right)/ \rho_{0}$, $a = \left(\rho_{\rm{n},0} a_{\rm{n}} + \rho_{\rm{c},0} a_{\rm{c}} \right) / \rho_{0}$, and $\rho_{0} = \rho_{\rm{c},0} + \rho_{\rm{n},0}$. Note that in the limit $K_{\rm{col}} \rho_{0} / \omega \gg 1$ (high collision limit) this dispersion relation reduces to the single-fluid expression, $-\omega^{2} + k^{2}a + i k b = 0$, while in the opposite limit, $K_{\rm{col}} \rho_{0} / \omega \gg 1$, the separate dispersion relations of charges and neutrals are recovered. Furthermore, if the effect of stratification is neglected, that is, if the coefficients $b_{\rm{c}}$, $b_{\rm{n}}$, and $b$ are set equal to zero, while $a_{\rm{c}}$, $a_{\rm{n}}$ and $a$ are now constant, Eq. \ref{eq:2f_strat_wkbdr} becomes the dispersion relation for fast magneto-acoustic waves propagating in a uniform atmosphere:
\begin{equation} \label{eq:2f_strat_dr_uni}
        \left(\omega^{2} - k^{2} a_{\rm{c}}\right) \left(\omega^{2} -k^{2} a_{\rm{n}} \right) + i \omega K_{\rm{col}} \rho_{0} \left(\omega^{2} - k^{2} a\right) = 0,
\end{equation}
which is equivalent to Eq.~\ref{eq:2f_mawaves_dr} if we set $k_{z} = 0$ in that equation. Note that in Section \ref{sec:2f_magnetoacoustic} the background magnetic field was oriented along the vertical direction while in the present analysis it is purely horizontal. 
    
    \begin{figure} [t!]
        \centering
        \includegraphics[width=0.7\hsize]{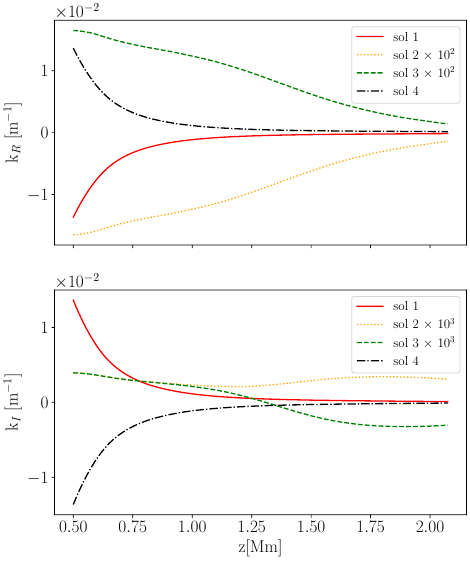}
        \caption{Solutions of the dispersion relation Eq.~\ref{eq:2f_strat_wkbdr} for waves with a period of 5 s propagating in the solar chromosphere. Upper and lower panels represent the real and the imaginary parts of the wavenumber as functions of height. Figure from \citet{PopescuBraileanu2019A&A...630A..79P}.}
        \label{fig:2f_strat_dr_sols}
    \end{figure}
    
The exact numerical solution of Eq.~\ref{eq:2f_strat_wkbdr} for the case of waves with periods of 5 s propagating in the solar atmosphere is presented in Fig. \ref{fig:2f_strat_dr_sols}. There are four different solutions, corresponding to a fourth order polynomial in $k$. Two of the solutions propagate upwards, with $k_{\rm{R}} > 0$, and two downwards, with $k_{\rm{R}} < 0$. One of the solutions propagating in each direction (black and red curves) has very large real and imaginary wavenumbers, meaning that it has a very small propagation speed and a very strong collisional damping. The combined effect of stratification and collisional interaction can be more clearly seen in the remaining modes (yellow and green curves). For instance, it can be seen that the upward propagating mode (green curve) has a positive $k_{\rm{I}}$ up to a height around $1.4$ Mm, which means that the amplitude of this wave increases as it moves from dense to rarefied regions. However, this wave is also affected by the collisional damping and at a certain height this effect is able to balance the amplitude growth due to stratification. Finally, at higher layers the damping due to collisions dominates and the imaginary part of the wavenumber becomes negative, so the amplitude of the perturbation decreases. For the wave propagating downwards ($k_{\rm{R}} < 0$, yellow curve), a positive $k_{\rm{I}}$ corresponds to a decrease in the amplitude. In this case, the gravitational stratification also produces a positive imaginary part of the wavenumber because the wave propagates from low density to high density layers.
    
As a next approximation, the derivatives of $\tilde{V}_{\rm{c},z}$ and $\tilde{V}_{\rm{n},z}$, computed before, can be inserted into the individual velocity equations, Eqs. \ref{eq:2f_strat_vnz}, and \ref{eq:2f_strat_vcz}, and after separating the 0th-order terms from the 1st-order terms, the following expressions for the velocity amplitudes can be obtained:
    \begin{equation} \label{eq:2f_strat_wkb_ampls}
        U_{\rm{c,n}}(z) = U_{\rm{c,n},0} \exp \left(\int_{0}^{z} \frac{d k}{d z} \frac{i a_{\rm{c,n}}}{b_{\rm{c,n}}-2 i k a_{\rm{c,n}}} dz'\right),
    \end{equation}
which provide the full profile of the velocity perturbations as functions of height. This way, by using Eq.~\ref{eq:2f_strat_wkb1} and the expression 
$\{V_{\rm{c},z}(z,t), V_{\rm{n},z}(z,t) \} = \{\tilde{V}_{\rm{c},z}(z), \tilde{V}_{\rm{n},z}(z) \} \exp \left(-i \omega t \right)$, the full temporal and spatial evolution of the waves can be reconstructed.
    
\begin{figure}[t!]
    \centering
    \includegraphics[width=\hsize]{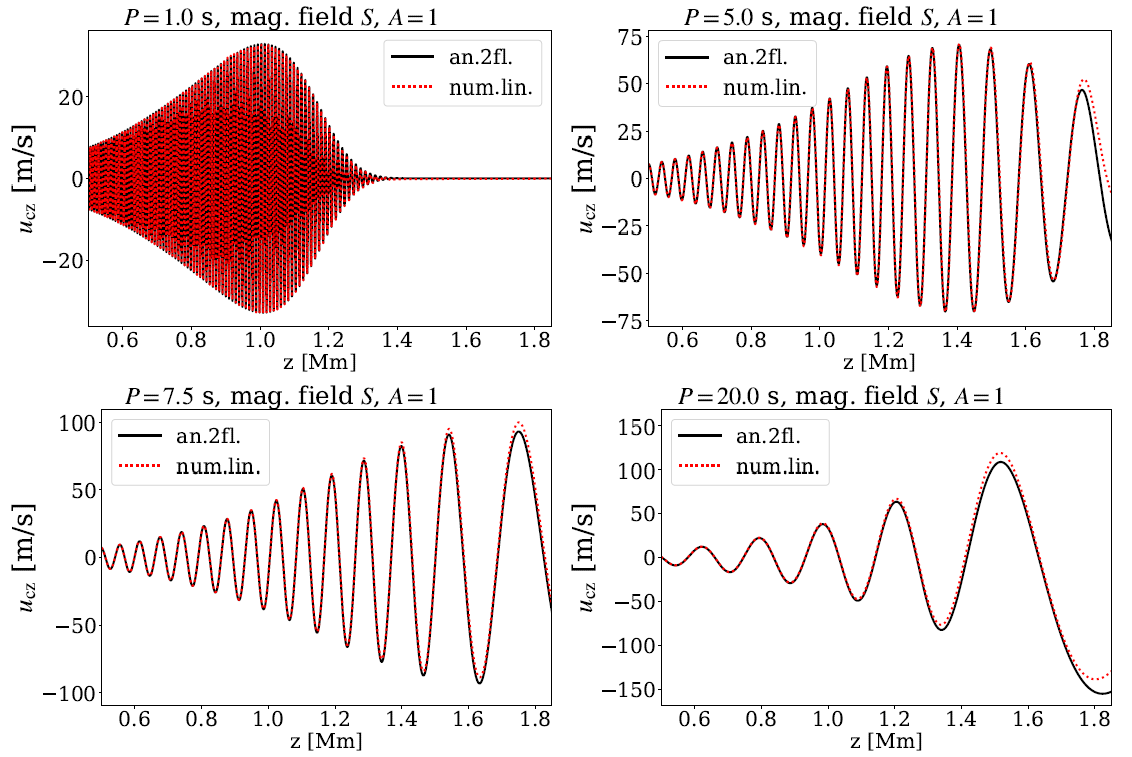}
     \caption{Comparison between results from the full numerical solution of the system of linear Eqs.~\ref{eq:2f_lin_rho_n}--\ref{eq:2f_lin_induc} (black lines) and approximate analytical solutions (red dotted lines) for fast magneto-acoustic wave propagating in the solar chromosphere. Panels show snapshots of the velocity of charges, $V_{\rm{c},z}$, as a function of height, $z$. Panel from left to right, from top to bottom corresponds to wave periods of 1, 5, 7.5, and 20 seconds, respectively. Figure from \citet{PopescuBraileanu2019A&A...630A..79P}.}
        \label{fig:2f_strat_comp}
\end{figure}
    
Figure~\ref{fig:2f_strat_comp} shows a comparison between the solutions computed from the dispersion relation following the procedure described above, and the results of numerically solving the full set of linear equations given by Eqs. \ref{eq:2f_lin_rho_n}--\ref{eq:2f_lin_induc}. It can be seen that the two solutions are in good agreement, except that the approximate solution is not sufficiently precise in the upper layers. Each panel of Fig. \ref{fig:2f_strat_comp} corresponds to a different period of the wave and it clearly shows how the amplitude of the velocity perturbations increases as the waves propagate upwards until it reaches the height where the collision damping dominates over the growth due to stratification and then the amplitude starts to decrease. Waves with shorter periods are more affected by the dissipation caused by the charged-neutral collisions.

\subsection{Plasma heating}

The linear theory of partially ionized plasmas shows that the charged-neutral collision interaction is a damping mechanism. As magnetohydrodynamic waves propagate, their amplitude reduces due to the charge-neutral collisions and, consequently, the kinetic and magnetic energies carried by those waves also decrease. However, the linear theory does not provide answers on what happens with the energy that is removed. Is it lost? Where does it go? It is transformed into another kind of energy? 
    
To provide an answer to these questions, it is necessary to resort to the full non-linear equations presented in Section \ref{sec:eqs_2fluid}. According to Eqs.~\ref{eq:2f_iene_n}--\ref{eq:2f_mcprime}, there is always an increase of the internal energy of each component of the plasma when colliding charged and neutral fluids have different velocities. Therefore,  the energy of magnetohydrodynamic waves is transformed into the internal energy, leading to plasma temperature increase, in a process known as frictional plasma heating.

The process of frictional heating is frequently invoked in the context of models of chromospheric heating by magnetohydrodynamic waves \citep{2021JGRA..12629097S}. 
For example, it has been considered in the models of dissipation and heating by Alfv\'en waves by, e.g.,  \citet{Goodman2011ApJ...735...45G,TuSong2013ApJ...777...53T,Soler2015ApJ...810..146S,MartinezGomez2018ApJ...856...16M}. This section follows the approach by \citet{Song2011JGRA..116.9104S} to study the process of plasma heating by Alfv\'en waves.
    
To study the energy of the whole system, evolution equations of the internal energy of neutrals and charges, Eqs.~\ref{eq:2f_iene_n} and \ref{eq:2f_iene_c}, are added up. Here we do not consider processes of thermal conduction, viscosity,  ionization and recombination. Thus, the following expression for the total internal energy of the two-fluid plasma is obtained:
\begin{equation} \label{eq:2f_iene2}
        \frac{\partial e}{\partial t} + \nabla \cdot \left(\bm{V}_{\rm{n}} e_{\rm{n}} + \bm{V}_{\rm{c}} e_{\rm{c}} \right) + P_{\rm{n}} \nabla \cdot \bm{V}_{\rm{n}} + P_{\rm{c}} \nabla \cdot \bm{V}_{\rm{c}} = Q,
\end{equation}
where $e = e_{\rm{n}} + e_{\rm{c}}$ and $Q$ is given by
\begin{equation} \label{eq:2f_nl_qheat}
        Q = \bm{J} \cdot \left(\bm{E} + \bm{V}_{\rm{c}} \times \bm{B} \right) + \nu_{\rm{cn}} \rho_{\rm{c}} |\bm{V}_{\rm{c}} - \bm{V}_{\rm{n}}|^{2}.
\end{equation}
The latter term represents the frictional heating of  plasma \citep[][]{Vasyliunas2005JGRA..110.2301V}. It contains the contribution from Ohmic heating, which is caused by collisions of ions and neutral particles with electrons, and from the frictional heating due to collisions between the charged and neutral species.

This form of Eq. \ref{eq:2f_nl_qheat} does not allow yet to link the plasma heating to the parameters of waves. In order to get better insights, \citet{Vasyliunas2005JGRA..110.2301V} performed several manipulations to the set of two-fluid non-linear equations. By neglecting gravity and pressure gradients, the momentum equations for neutrals and charges, Eq.~\ref{eq:momentum-2fluid-n} and \ref{eq:momentum-2fluid-c}, can be approximated as

\begin{equation} \label{eq:2f_nl_vn}
        \rho_{\rm{n}} \frac{\partial \bm{V}_{\rm{n}}}{\partial t} = \rho_{\rm{n}} \nu_{\rm{nc}} \left(\bm{V}_{\rm{c}} - \bm{V}_{\rm{n}} \right),
\end{equation}
\begin{equation} \label{eq:2f_nl_vc}
        \rho_{\rm{c}} \frac{\partial \bm{V}_{\rm{c}}}{\partial t} = \bm{J} \times \bm{B} - \rho_{\rm{c}} \nu_{\rm{cn}} \left(\bm{V}_{\rm{c}} - \bm{V}_{\rm{n}} \right).
\end{equation}
    
\noindent Assuming the velocity perturbations to be proportional to $\exp \left(-i \omega t\right)$, Eq. \ref{eq:2f_nl_vn} gives the difference of velocities between the two fluids,

\begin{equation} \label{eq:2f_nl_vdrift}
        \bm{V}_{\rm{c}} - \bm{V}_{\rm{n}} = - \frac{i \omega}{\nu_{\rm{nc}}}\frac{1}{1-i \omega /\nu_{\rm{nc}}} \bm{V}_{\rm{c}}.
\end{equation}

\noindent From Ohm's law, Eq. \ref{eq:2f_efield}, the Ohmic heating can be expressed as
\begin{equation} \label{eq:2f_ohmic}
        \bm{J}^{*} \cdot \left(\bm{E} + \bm{V}_{\rm{c}} \times \bm{B}\right) = \frac{m_{\rm{e}}\nu_{\rm{e}} |\bm{J}|^{2}}{n_{\rm{e}}e^{2}}, 
\end{equation}
where the symbol ``*'' denotes the complex conjugate and $\nu_{\rm{e}} = \nu_{\rm{ec}} + \nu_{\rm{en}}$.
    
\noindent Combining Eqs. \ref{eq:2f_nl_vc} and \ref{eq:2f_nl_vdrift}, the Lorentz force term becomes,

\begin{equation}
        \bm{J} \times \bm{B} = \rho_{\rm{n}} \nu_{\rm{nc}} \left(\bm{V}_{\rm{c}} - \bm{V}_{\rm{n}} \right) \left[1 + \frac{1}{\chi} - \frac{i \omega}{\nu_{\rm{nc}}} \frac{1}{\chi}\right].
\end{equation}
    
\noindent This equation is used to compute the product $\left(\bm{J} \times \bm{B} \right) \cdot \left(\bm{J} \times \bm{B} \right)^{*} = |\bm{J}|^{2} |\bm{B}|^{2}$ (where it has been taken into account that $\bm{J} \cdot \bm{B} = 0$), and obtain the expression:
\begin{equation} \label{eq:2f_nl_jj}
        |\bm{J}|^{2} = \bm{J} \cdot \bm{J}^{*} = \left(\frac{\rho_{\rm{n}} \nu_{\rm{nc}}}{B}\right)^{2} \left[\left(1 + \frac{1}{\chi}\right)^{2} +\left(\frac{\omega}{\chi \nu_{\rm{nc}}} \right)^{2} \right] |\bm{V}_{\rm{c}} - \bm{V}_{\rm{n}}|^{2}.
\end{equation}
    
\noindent By inserting Eq.~\ref{eq:2f_nl_jj} into Eq. \ref{eq:2f_ohmic}, the Ohmic heating becomes,
    \begin{equation} \label{eq:2f_nl_ohmic2}
        \bm{J}^{*} \cdot \left(\bm{E} + \bm{V}_{\rm{c}} \times \bm{B} \right) = \frac{\nu_{\rm{e}} \nu_{\rm{cn}}}{\Omega_{\rm{ce}} \Omega_{\rm{ci}}} \left[ \left(1 + \frac{1}{\chi}\right)^{2} + \left(\frac{\omega}{\chi\nu_{\rm{nc}}} \right)^{2} \right]\rho_{\rm{n}} \nu_{\rm{nc}} |\bm{V}_{\rm{c}} - \bm{V}_{\rm{n}}|^{2}.
    \end{equation}
    
\noindent Finally, summing the contributions of the Ohmic and the frictional heating and taking into account that $|\bm{V}_{\rm{c}} - \bm{V}_{\rm{n}}|^{2} = \left(\bm{V}_{\rm{c}} - \bm{V}_{\rm{n}} \right) \cdot \left(\bm{V}_{\rm{c}} - \bm{V}_{\rm{n}} \right)^{*}$, the time-averaged total heating rate is given by
\begin{equation} \label{eq:2f_nl_qheatav}
        \langle Q \rangle = \left\{1 + \frac{\nu_{\rm{e}} \nu_{\rm{cn}}}{\Omega_{\rm{ce}} \Omega_{\rm{ci}}} \left[\left(1+ \frac{1}{\chi}\right)^{2} + \left(\frac{\omega}{\chi \nu_{\rm{nc}}}\right)^{2}\right] \right\} \frac{\omega^{2}}{\nu_{\rm{nc}} \left(1 + \omega^{2}/\nu_{\rm{nc}}^{2}\right)}\rho_{\rm{n}} \langle \bm{V}_{\rm{c}}^{2} \rangle.
\end{equation}
In this expression, $\Omega_{\rm{ce}}$ and $\Omega_{\rm{ci}}$ are the electron and ion cyclotron frequencies. By defining $\kappa \equiv \nu_{\rm{e}} \nu/ \left(\Omega_{\rm{ce}} \Omega_{\rm{ci}} \right)$, with $\nu \equiv \nu_{\rm{cn}} + \nu_{\rm{nc}}$, the heating rate can also be expressed as,
%
%
\begin{equation} \label{eq:2f_nl_qheatav3}
        \langle Q \rangle = \frac{\chi \rho \omega^{2} \langle \bm{V}_{\rm{c}}^{2} \rangle}{\nu \left[1 + \left(1 + \chi \right)^{2}\omega^{2} / \nu^{2}\right]} \left[1 + \kappa \left(\frac{\chi + 1}{\chi}\right)\left(1 + \frac{\omega^{2}}{\nu^{2}}\right) \right].
\end{equation}
 
\begin{figure} [t!]
    \centering
    \includegraphics[width=0.7\hsize]{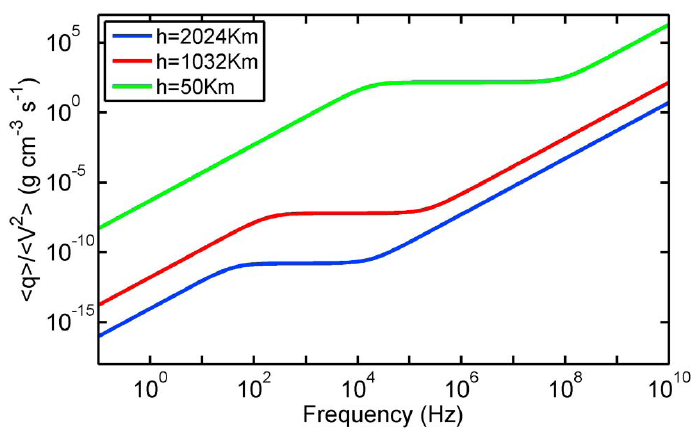}
    \caption{Heating rate divided by mean square velocity fluctuation, $\langle Q \rangle / \langle \bm{V}_c^{2} \rangle$ (in our notation), as a function of wave frequency, $\omega$, for different heights above the solar surface, and for different ionization degrees. Green, red, and blue lines correspond to the ionization degrees $\chi = 1.97 \times 10^{-4},$ $9.5 \times 10^{3}$ and $1.49$, and heights $h = 50 \ \rm{km}$, $1032 \ \rm{km}$, and $2024 \ \rm{km}$, respectively. Figure from \citet{Song2011JGRA..116.9104S}.}
    \label{fig:2f_heating}
\end{figure}
    
It is interesting to check the behavior of the total heating rate for certain ranges of the frequency of the perturbation. For instance, if the oscillation frequency is much lower than the collision frequency, $\omega \ll \nu/ \left(\chi + 1\right) = \nu_{\rm{nc}}$, Eq. \ref{eq:2f_nl_qheatav3} simplifies to
\begin{equation} \label{eq:2f_nl_qheat_app1}
        \langle Q \rangle = \frac{\rho \omega^{2} \langle \bm{V}_{\rm{c}}^{2} \rangle}{\nu} \left[\chi + \kappa \left(\chi + 1\right) \right],
\end{equation}
showing that in the limit of strong collision coupling the heating rate has a quadratic dependence on the frequency $\omega$. In the intermediate frequency range, $\nu_{\rm{nc}} \ll \omega \ll \nu$, we have
\begin{equation} \label{eq:2f_nl_qheat_app2}
        \langle Q \rangle = \rho \nu \langle \bm{V}_{\rm{c}}^{2} \rangle \frac{1 + \kappa}{\chi},
\end{equation}
and the heating rate is independent from $\omega$. Finally, in the high-frequency limit, $\omega \gg \nu$, the heating rate becomes,
\begin{equation} \label{eq:2f_nl_qheat_app3}
        \langle Q \rangle = \frac{\rho \omega^{2} \langle \bm{V}_{\rm{c}}^{2} \rangle}{\nu} \frac{\kappa}{\chi + 1},
\end{equation}
and the heating rate recovers its quadratic dependence on $\omega$. 
    
This three different regimes of the heating rate dependence on frequency are clearly visible in Fig. \ref{fig:2f_heating}, where the complete Eq.~\ref{eq:2f_nl_qheatav3} is used. This figure gives the results for three different heights in the solar atmosphere. According to these results, most of the collisional plasma heating occurs at the lower altitudes of the solar chromosphere, with higher-frequency waves producing the largest amounts of heating. \citet{Song2011JGRA..116.9104S} applied Eq. \ref{eq:2f_nl_qheatav3} to a 1-dimensional model of the solar chromosphere and computed the total frequency-integrated heating rate as $Q_{\rm{tot}}(z) = \int_{0}^{\infty} \langle Q(z,\omega) \rangle d\omega$, as a function of $z$. They found values of $Q_{\rm{tot}}$ of the order of $10^{-1} \ \rm{erg} \ \rm{cm^{-3}} \ \rm{s^{-1}}$ at the lower chromosphere and of the order of $10^{-2} \ \rm{erg} \ \rm{cm^{-3}} \ \rm{s^{-1}}$ at the middle and upper chromosphere. These values are consistent with the empirical estimates of the energy input required to balance the radiative losses \citep{Withbroe1977ARA&A..15..363W,Vernazza1981ApJS...45..635V}. \citet{Song2011JGRA..116.9104S} found that, for a given wave energy flux coming from the photosphere, the heating is stronger in regions with a weaker magnetic field. The reason is that in those regions the electron cyclotron motion becomes less important in comparison with the effect of collisions and, thus, the Joule heating is enhanced.

\begin{figure} [t!]
        \centering
        \includegraphics[width=\hsize]{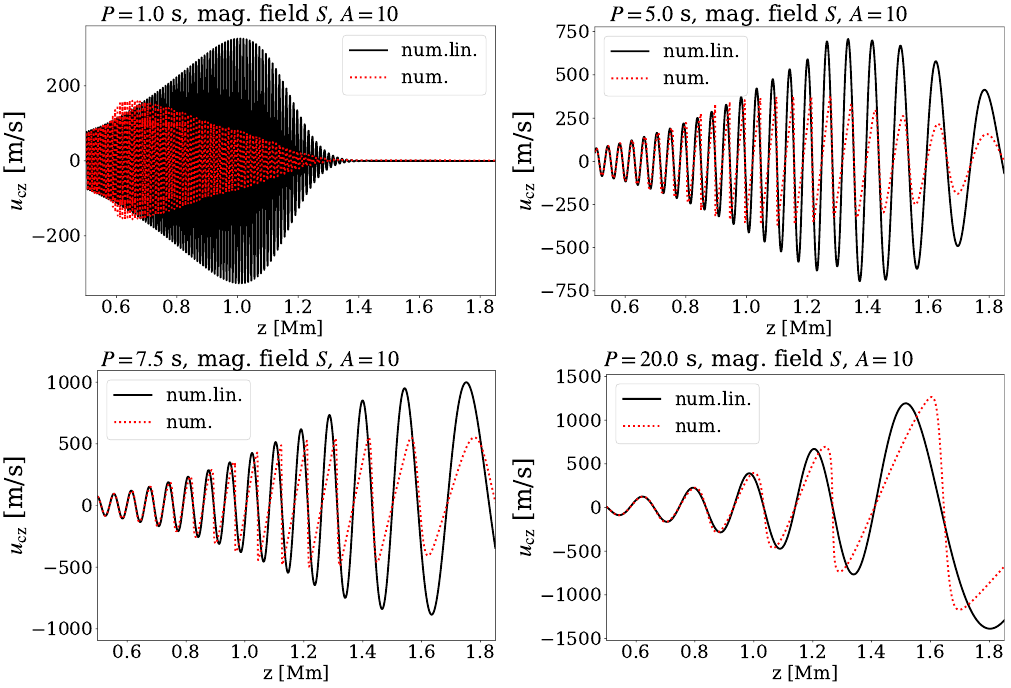}
        \caption{Results from numerical simulations of propagation of fast magneto-acoustic waves in the solar chromosphere in the linear regime (black lines) and non-linear regime (red dotted lines). Each panel represents the vertical $z$-component of the velocity of the charged fluid at a fixed time as a function of height above the solar surface for a different wave period ($P = 1, 5, 7.5, \text{and} \ 20$ s). Figure from \citet{PopescuBraileanu2019A&A...630A..79P}.}
        \label{fig:2f_waveshocks}
\end{figure}
       
\subsection{Multi-fluid shocks}

This chapter further explores the non-linear plasma dynamics under the two-fluid approximation and studies the formation and evolution of shocks using the mathematical framework described in Section \ref{sec:eqs_2fluid}.

\subsubsection{Propagation and dissipation of shocks}    

In order to illustrate the influence of non-linearities, we reconsider the model of fast magneto-acoustic waves propagation in a gravitationally stratified solar chromosphere permeated by a horizontal magnetic field, already discussed in Section \ref{sec:2f_stratified}, but now for waves with larger amplitudes. 

Figure \ref{fig:2f_waveshocks} shows the results from numerical simulations performed by    \citet{PopescuBraileanu2019A&A...630A..79P} for different periods of the driver that triggers the propagation of fast magneto-acoustic waves at the bottom of the chromosphere. It includes a comparison between the linear regime, where the set of Eqs.~\ref{eq:2f_lin_rho_n}--\ref{eq:2f_lin_induc} is solved (same data were already presented in Fig. \ref{fig:2f_strat_comp}), and the non-linear regime, where the full version of the two-fluid equations is considered, Eqs.~\ref{eq:continuity-2fluid-n}--\ref{eq:energy-2fluid-c}. In both linear and non-linear cases, one can observe a combination of two different effects: on the one hand, the growth of the amplitude due to the gravitational stratification and, on the other hand, the decrease of the amplitude at higher layers caused by the charged-neutral collisional damping. However, the non-linear regime (represented by the red dotted lines) has another feature that is not present in the linear one: the waves do not conserve their sinusoidal shapes but their fronts steepen as they propagate upwards. Due to the large amplitude of the perturbations, the phase speed of waves at the different parts of the wavefront is noticeably modified: crests propagate faster than valleys, and a saw-tooth profile is formed, characteristic for a shock wave. It can be seen that the shock formation process strongly varies with the period of the waves. The saw-tooth profiles appear at lower heights for shorter-period waves, while the longer-period ones remain in the linear regime up to higher layers of the atmosphere.
    
The formation of shocks clearly affects the attenuation of the waves. It enhances the effect of collisional damping, leading to a smaller amplitude of the waves compared to the linear regime, and modifying the height at which the waves reach their largest amplitudes. The reason for this enhancement is that the spatial scale of the shock fronts becomes closer to the mean charge-neutral collision free-path, which is the scale where the largest decoupling occurs and, thus, the damping due to collisions is more efficient, as discussed in Sections \ref{sec:2f_alfven} and \ref{sec:2f_magnetoacoustic}. As the waves propagate further upwards, the collisional damping becomes strong enough for the waves to (almost) recover their linear sinusoidal profiles due to the smallness of their amplitudes. This effect can be clearly appreciated at the top right and bottom panels of Fig. \ref{fig:2f_waveshocks}, where saw-tooth profiles appear at the intermediate layers of the atmosphere but are not present at the higher layers.
    
\begin{figure} [t!]
    \centering
    \includegraphics[width=\hsize]{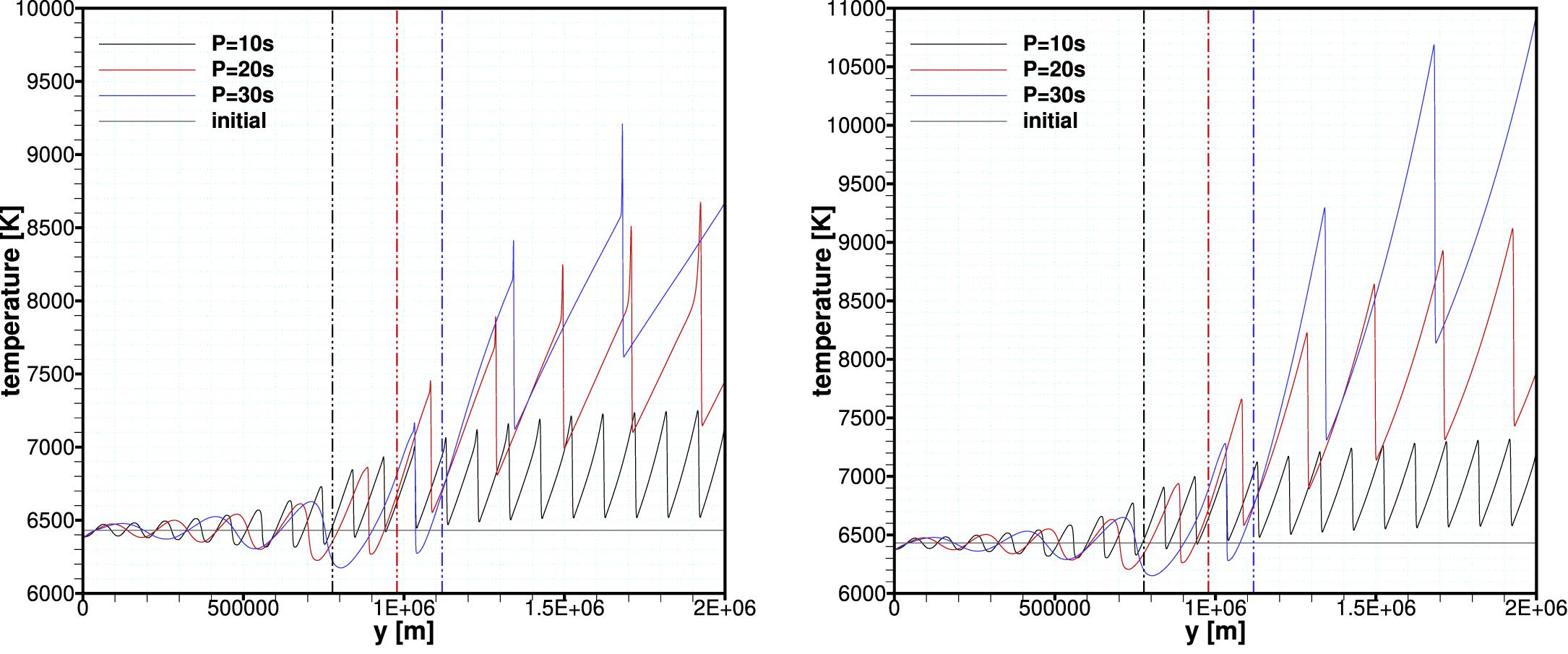}        \caption{Snapshots of the temperature fluctuations of charges from the two-fluid simulations of acoustic waves propagation in the solar chromosphere. Left panel: simulations including ionization/recombination and elastic collisions; right panel: only elastic collisions are included. Vertical dotted lines indicate approximate heights at which kinetic energy decay starts. Figure from \citet{Zhang2021ApJ...911..119Z}.}
        \label{fig:2f_shocks_heating}
\end{figure}

Similar results are obtained for the case of slow magneto-acoustic waves, propagating aligned to the magnetic field, for which the magnetic field only serves as a wave guide \citep{Zhang2021ApJ...911..119Z}. Figure \ref{fig:2f_shocks_heating}, obtained from two-fluid numerical simulations by\citet{Zhang2021ApJ...911..119Z}, shows that the development of shock fronts in acoustic waves is faster for shorter-period waves and that longer-period waves reach much larger amplitudes in the upper layers of the chromosphere. This figure also demonstrates that the dissipation of the wave energy causes an increment of the temperature of the plasma. The rise of temperature starts at lower heights for waves with shorter periods, as indicated by the vertical lines.
    
Up to this point, we have not discussed the influence of ionization and recombination on the propagation of waves, although the corresponding terms have been included into the general equations of the two-fluid model, Eqs.~\ref{eq:2f_sn}--\ref{eq:m-2fluid}. Only a few works have addressed this issue under a multi-fluid framework, using a rather simplified model \citep{2017ApJ...836..197M, Zhang2021ApJ...911..119Z}. The results of the simulations by \citet{Zhang2021ApJ...911..119Z}, including ionization and recombination are illustrated at the left panel of Fig. \ref{fig:2f_shocks_heating}, and can be compared to the same simulation but without taking these effects into account (right panel). It can be observed that a larger increase of the plasma temperature is obtained when ionization and recombination are not included in the model. According to  \citet{Zhang2021ApJ...911..119Z}, the inclusion of ionization and recombination enhances the decoupling between ions and neutrals and the collisional heating, as a consequence of the modifications of the equilibrium atmosphere conditions. This might seem contradictory in view of the smaller temperature increment achieved in the model with ionization/recombination. However, it has an easy explanation. The process of removing an electron from a neutral particle (ionization) requires a considerable amount of energy. Therefore, a significant part of the collisional heating is used into ionizing the neutral fluid, so there is less energy available to increase the temperature of the plasma. 

\subsubsection{Formation of shock sub-structure}

To conclude this section, we briefly discuss another interesting consequence that the charged-neutral interaction has on propagating shocks: the formation of internal sub-structure where each fluid has a very distinct behavior \citep{Snow2020A&A...637A..97S}. This effect cannot be appreciated in the results shown in Figs. \ref{fig:2f_waveshocks} and \ref{fig:2f_shocks_heating} because it occurs at very small scales and it requires a very good spatial resolution to be properly captured. To study how the charged-neutral interaction modifies the ``internal" structure of the shocks, \cite{Snow2019A&A...626A..46S} numerically solved the system of Eqs.~\ref{eq:continuity-2fluid-n}--\ref{eq:energy-2fluid-c}, without taking into account the processes of ionization or recombination, thermal conduction, or viscosity, and using the ideal version of the induction equation. They considered the following initial conditions (given in non-dimensional units) for their high-resolution numerical simulations

\begin{figure} [t!]
    \centering
    \includegraphics[width=\hsize]{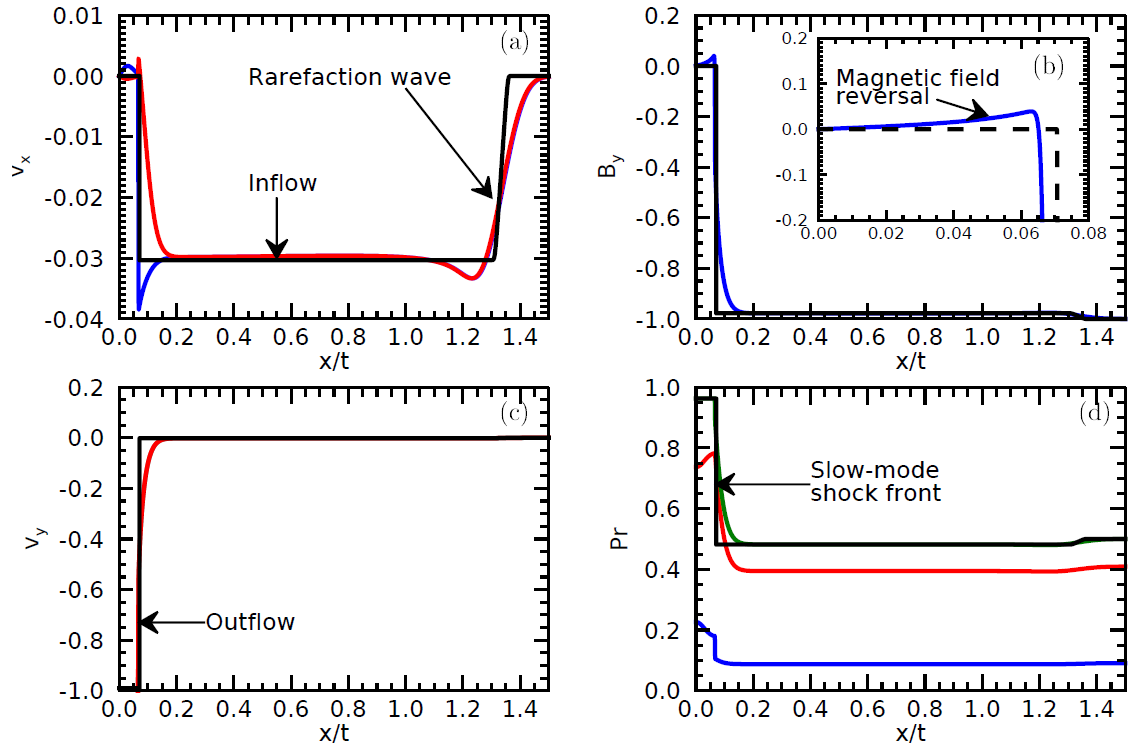}
    \caption{Snapshots of numerical simulations of stationary shock sub-structure in fully and partially ionized plasmas with $\beta = 1.0$, $B_{x} = 0.1$, and $\xi_{\rm{n}} = 0.9$. Black lines represent the single-fluid MHD case, red and blue lines represent the neutral and charged components, respectively, of the two-fluid case. Panel a): $x$- component of the velocity. Panel b): $y$-component of magnetic field. Panel c): "$y$-component of the velocity. Panel d): pressure; green line shows the total pressure, $P_{\rm{n}} + P_{\rm{c}}$. The two-fluid solution is plotted after 2500 collisional times. Adapted from \citet{Snow2019A&A...626A..46S}.}
    \label{fig:2f_shocks}
\end{figure}
    
    \begin{equation} \label{eq:2f_shocks_bx}
        B_{x} = 0.1,
    \end{equation}
    \begin{equation} \label{eq:2f_shocks_by}
        B_{y} = \left\{
        \begin{array}{c c}
        -1.0 & \quad \text{if x > 0} \\
        1.0 & \quad \text{if x < 0}
        \end{array} \right.,
    \end{equation}
    \begin{equation} \label{eq:2f_shocks_rhon}
        \rho_{\rm{n}} = \xi_{\rm{n}} \rho,
    \end{equation}
    \begin{equation} \label{eq:2f_shocks_rhoc}
        \rho_{\rm{c}} = \xi_{\rm{c}} \rho = \left(1 - \xi_{\rm{n}}\right) \rho,
    \end{equation}
    \begin{equation} \label{eq:2f_shocks_pn}
        P_{\rm{n}} = \frac{\xi_{\rm{n}}}{\xi_{\rm{n}} + 2 \xi_{\rm{c}}} P = \frac{\xi_{\rm{n}}}{\xi_{\rm{n}} + 2 \xi_{\rm{c}}} \beta \frac{B_{0}^{2}}{2},
    \end{equation}
     \begin{equation} \label{eq:2f_shocks_pc}
        P_{\rm{c}} = \frac{2\xi_{\rm{c}}}{\xi_{\rm{n}} + 2 \xi_{\rm{c}}} P = \frac{2\xi_{\rm{c}}}{\xi_{\rm{n}} + 2 \xi_{\rm{c}}} \beta \frac{B_{0}^{2}}{2},
    \end{equation}
Then, they compared the results for a partially ionized plasma with those for a fully ionized scenario. Figure \ref{fig:2f_shocks} shows the results for the case with $\beta = 1.0$, $B_{x} = 0.1$, and $\xi_{\rm{n}}=0.9$. It can be clearly seen that the shock transition for the partially ionized case (red and blue lines) has a larger complexity than the fully ionized one (black lines). The transitions between the pre- and post-shock values of the different variables in the fully ionized case are sharper while for the partially ionized case they occur along larger distances. For instance, the slow-mode shock front has a discontinuous jump in the MHD simulation but it presents a finite width with more complex sub-structure in the two-fluid simulation. In addition, the collision coupling leads to overshooting (sudden increase) in the neutral velocity and pressure, and  to a reversal of the magnetic field across the shock front. The magnetic field reversal is more prominent for larger values of the equilibrium neutral pressure. Although it is not shown in Fig. \ref{fig:2f_shocks}, an overshooting is also present in the neutral density, which is accompanied by a decrease in the total Alfv\'en speed (the Alfv\'en speed computed taking into account the total density of the plasma). 
    
These results are a clear example of how the neutral and the charged fluids have different dynamics at very small spatial scales and how single-fluid models are not able to properly describe the full behavior of the plasma. The use of multi-fluid framework is fundamental for analyzing the small-scale dynamics of partially ionized plasmas.

\section{Waves in hydrogen-helium plasmas}
    This section makes a further step in the multi-fluid modeling of solar waves and describes the effects of including Helium, additionally to Hydrogen, into the model.
According to Section \ref{sec:eqs_helium}, using the multi-fluid perspective, the inclusion of Helium can be done in various ways depending on the number of helium ionization states that are considered as separate fluids. Due to the complexity of the hydrogen-helium multi-fluid models, the present section focuses only on the study of Alfv\'en waves: subsection \ref{sec:3f_waves} focuses on the influence of neutral hydrogen and helium on the damping of Alfv\'en waves in the solar chromosphere, while subsection \ref{sec:5f_waves} analyses the effect of collisions on resonances and cut-off regions related to the presence of multiple ionized species in the plasma. 
    
\subsection{Three-fluid model} 
\label{sec:3f_waves}

Following \citet{Zaqarashvili2011A&A...534A..93Z}, we consider a plasma composed by three different fluids: charges, neutral hydrogen and neutral helium, and apply linearization to the set of Eqs.~\ref{eq:helium_continuity}--\ref{eq:helium_efield}. Taking into account that Alfv\'en waves are incompressible, the following condition must be fulfilled for all fluids,
\begin{equation} \label{eq:helium_divv0}
        \nabla \cdot \bm{V}_{\rm{s}} = 0 \quad \text{for} \quad s \in \{\rm{H}, \rm{He}, \rm{c} \}.
\end{equation}
This condition simplifies the set of equations as there is no need to include the continuity and pressure equations into consideration.
    
Let us assume that the equilibrium state is static and homogeneous, the background magnetic field is oriented along the $z$-direction, $\bm{B}_{0} = (0, 0, B_{0,z})^{t}$ and the waves propagate in $x-z$-plane, so that $\partial / \partial y = 0$, and they are polarized along the $y$-direction. The resulting set of linear equations for Alfv\'en waves in a hydrogen-helium plasma is given by
\begin{equation} \label{eq:helium_lin_dvcdt}
        \frac{\partial V_{\rm{c},y}}{\partial t} = \frac{B_{0,z}}{\mu_{0}\rho_{\rm{c},0}}\frac{\partial B_{1,y}}{\partial z} - \frac{\alpha_{\rm{H}} + \alpha_{\rm{He}}}{\rho_{\rm{c},0}} V_{\rm{c},y} + \frac{\alpha_{\rm{H}}}{\rho_{\rm{c},0}}V_{\rm{H},y} + \frac{\alpha_{\rm{He}}}{\rho_{\rm{c},0}}V_{\rm{He},y},
\end{equation}
\begin{equation} \label{eq:helium_lin_dvhdt}
        \frac{\partial V_{\rm{H},y}}{\partial t} = \frac{\alpha_{\rm{H}}}{\rho_{\rm{H},0}}V_{\rm{c},y} - \frac{\alpha_{\rm{H}} + \alpha_{\rm{HeH}}}{\rho_{\rm{H},0}}V_{\rm{H},y} + \frac{\alpha_{\rm{HeH}}}{\rho_{\rm{H},0}}V_{\rm{He},y},
\end{equation}
\begin{equation} \label{eq:helium_lin_dvhedt}
        \frac{\partial V_{\rm{He},y}}{\partial t} = \frac{\alpha_{\rm{He}}}{\rho_{\rm{He},0}}V_{\rm{c},y} - \frac{\alpha_{\rm{He}} + \alpha_{\rm{HeH}}}{\rho_{\rm{He},0}}V_{\rm{He},y} + \frac{\alpha_{\rm{HeH}}}{\rho_{\rm{He},0}}V_{\rm{H},y},
\end{equation}
\begin{equation} \label{eq:helium_lin_dbdt}
        \frac{\partial B_{1,y}}{\partial t} = B_{0,z} \frac{\partial V_{\rm{c},y}}{\partial t},
\end{equation}
where $V_{\rm{c},y}$, $V_{\rm{H},y}$, and $V_{\rm{He},y}$ are the velocity perturbations of the charged fluid, the neutral hydrogen and neutral helium, respectively, $B_{1,y}$ is the perturbation of magnetic field, and $\rho_{\rm{c},0}$, $\rho_{\rm{H},0}$, and $\rho_{\rm{He},0}$ are the equilibrium values of densities. In addition, the following definitions have been used:
\begin{equation} \label{eq:helium_ah_ahe}
        \alpha_{\rm{H}} = \alpha_{\rm{pH}} + \alpha_{\rm{He\sc{II}H}}, \ \ \ \alpha_{\rm{He}} = \alpha_{\rm{pHe}} + \alpha_{\rm{He\sc{II}}He}.
    \end{equation}

According to the original work by \citet{Zaqarashvili2011A&A...534A..93Z}, collisions between neutral hydrogen and neutral helium in solar prominence and spicule conditions do not have a strong effect on the propagation of Alfv\'en waves, since $\alpha_{\rm{HeH}} /\alpha_{\rm{pHe}} \sim n_{\rm{H}}/n_{\rm{p}} < 1$. Therefore, the terms proportional to $\alpha_{\rm{HeH}}$ can be dropped from Eqs. \ref{eq:helium_lin_dvcdt}--\ref{eq:helium_lin_dbdt}.
    \begin{figure} [t!]
        \centering
        \includegraphics[width=0.47\hsize]{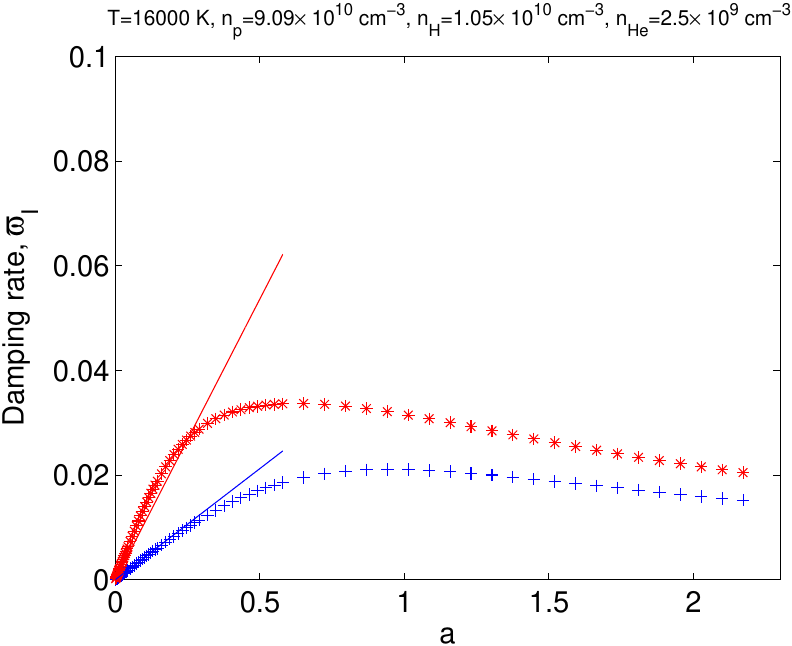}
        \includegraphics[width=0.47\hsize]{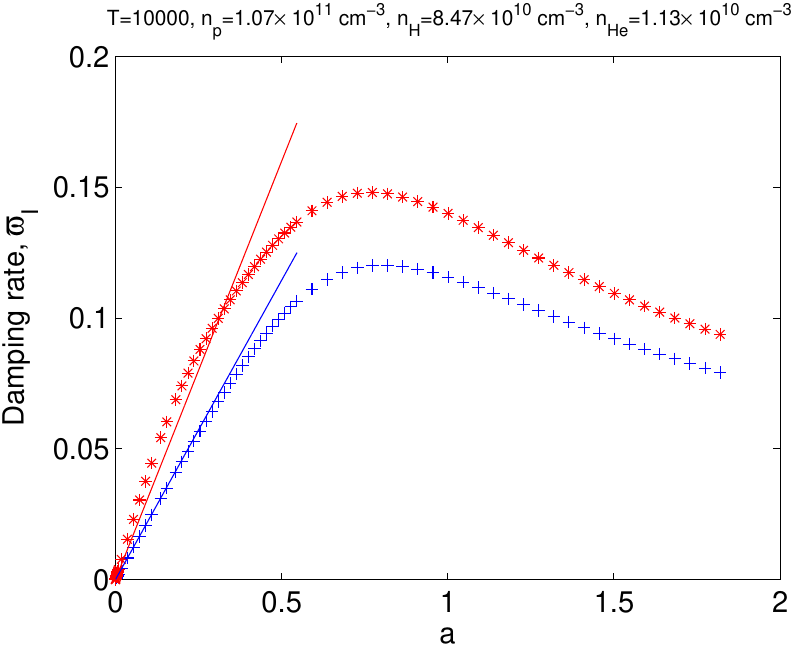} \\
        \includegraphics[width=0.47\hsize]{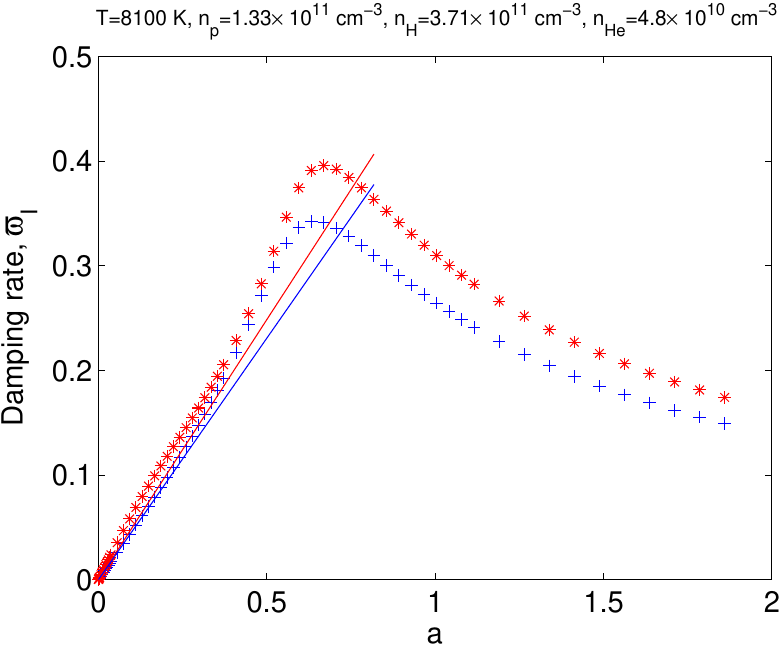}
        \includegraphics[width=0.47\hsize]{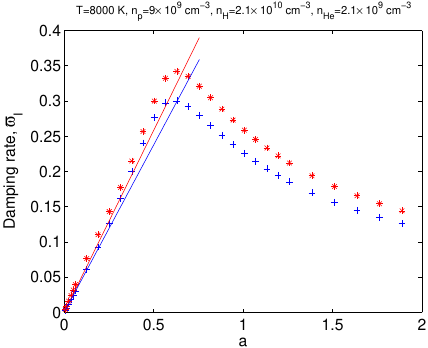}
        \caption{Normalized damping rate (imaginary part of the frequency) of Alfv\'en waves, $\varpi_{I} = \omega_{I} /\left(k_{z} c_{\rm{A}} \right)$ as a function of the Alfv\'en frequency, normalized by the hydrogen collision frequency, $a = k_{z}c_{\rm{A}}/\nu_{\rm{H}}$, for four different sets of plasma temperatures and number densities. Blue crosses correspond to the damping rates due to ion collision with neutral hydrogen only, while red asterisks include the effect of collisions with neutral helium particles. Red (blue) solid lines represent the damping rates derived in the single-fluid approach with (without) neutral helium. Figure from \citet{Zaqarashvili2011A&A...534A..93Z}.}
        \label{fig:helium_3f}
\end{figure}
    
A dispersion relation can be obtained by performing a Fourier and normal mode analysis, assuming the perturbations proportional to $\exp \left[i \left(k_{z}z - \omega t\right)\right]$. Then, the combination of Eqs. \ref{eq:helium_lin_dvcdt}--\ref{eq:helium_lin_dbdt} produces the following expression:
\begin{gather} 
        \omega^{4} + i \left(\frac{\alpha_{\rm{H}}}{\rho_{\rm{H}}} + \frac{\alpha_{\rm{He}}}{\rho_{\rm{He}}} + \frac{\alpha_{\rm{H}} + \alpha_{\rm{He}}}{\rho_{\rm{c}}} \right) \omega^{3} - \left(k_{z}^{2} c_{\rm{A}}^{2} + \frac{\alpha_{\rm{H}} \alpha_{\rm{He}} \rho}{\rho_{\rm{H}} \rho_{\rm{He}} \rho_{\rm{c}}}\right)\omega^{2} \nonumber \\
        -i k_{z}^{2}c_{\rm{A}}^{2} \left(\frac{\alpha_{\rm{H}}}{\rho_{\rm{H}}} + \frac{\alpha_{\rm{He}}}{\rho_{\rm{He}}} \right)\omega + \frac{\alpha_{\rm{H}} \alpha_{\rm{He}}}{\rho_{\rm{H}} \rho_{\rm{He}}}k_{z}^{2} c_{\rm{A}}^{2} = 0,
        \label{eq:helium_dr3f}
\end{gather}
where $\rho = \rho_{\rm{H}} + \rho_{\rm{He}} + \rho_{\rm{c}}$ is the total density of the plasma (note that, for simplicity, the subscript `0' has been dropped from the symbols for the equilibrium densities). This dispersion relation is a fourth order polynomial in the frequency $\omega$, which means that it has four different roots. Two of these roots correspond to Alfv\'en waves damped by the collisions between the three different species of the plasma. In addition there are two purely imaginary solutions, which correspond to evanescent modes associated with the neutral hydrogen and neutral helium.
    
To study the effect of helium of the propagation of Alfv\'en waves, it is illustrative to compare the predictions from Eq.~\ref{eq:helium_dr3f} with those obtained in the absence of helium. The dispersion relation describing the properties of Alfv\'en waves in a hydrogen-only two-fluid plasma can be obtained from Eq. \ref{eq:helium_dr3f} by setting $\alpha_{\rm{He}} = 0$,

\begin{equation} \label{eq:helium_dr_noHe}
        \omega^{3} + i \frac{\alpha_{\rm{H}}}{\rho_{\rm{H}}} \left(1 + \frac{\rho_{\rm{H}}}{\rho_{\rm{c}}}\right) \omega^{2} - k_{z}^{2} c_{\rm{A}}^{2} - i\frac{\alpha_{\rm{H}}}{\rho_{\rm{H}}}k_{z}^{2}c_{\rm{A}}^{2} = 0,
    \end{equation}
    
\noindent which is equivalent to Eqs. \ref{eq:2f_alfven_dr} or \ref{eq:2f_alfven_dr2}. The comparison of the damping rates of Alfv\'en waves computed from Eqs. \ref{eq:helium_dr3f} and \ref{eq:helium_dr_noHe} is presented in Fig. \ref{fig:helium_3f}. Each panel corresponds to a different set of plasma conditions (temperature and number densities), which represents a certain height in the solar chromosphere. The main conclusion that can be extracted from Fig. \ref{fig:helium_3f} is that, in general, the presence of neutral helium in the chromospheric plasma enhance the damping of Alfv\'en waves caused by elastic collisions. This effect is larger for frequencies closer to charge-neutral collision frequencies. 
    
\subsection{Five-fluid model} 
\label{sec:5f_waves}

Below, the five-fluid model detailed in Section \ref{sec:eqs_5f} is used to investigate the properties of Alfv\'en waves. In this model, all possible ionization states of hydrogen and helium are taken into account and treated as separate fluids. We follow the procedure described by \citet{MartinezGomez2017ApJ...837...80M}.
    
As a first step, the expression for the electric field, Eq.~\ref{eq:5f_electric2}, is introduced into the momentum equation, Eq.~\ref{eq:5f_momentum}, and in the Faraday's law to get the induction equation. Again, since we are only interested in Alfv\'en waves, we consider the condition of incompressibility, $\nabla \cdot \bm{V}_{\rm{s}} = 0$ for $s \in \{\rm{e}, \rm{p}, \rm{H}, \rm{He}, \rm{He\sc{II}}, \rm{He\sc{III}} \}$. Therefore, the terms related to the pressure gradients in the momentum equations are neglected, and there is no need to consider the continuity and pressure equations, Eqs. \ref{eq:5f_continuity} and \ref{eq:5f_pressures}.
    
After assuming a uniform and static background, the linear equation for the temporal evolution of the velocity perturbation of each ionization state $s \in \{\rm{p}, \rm{H}, \rm{He}, \rm{He\sc{II}}, \rm{He\sc{III}} \}$ becomes,

\begin{gather} 
    \frac{\partial \bm{V}_{\rm{s}}}{\partial t} = \frac{Z_{\rm{s}} e}{m_{\rm{s}}} \left[\left(\bm{V}_{\rm{s}} - \bm{V}_i \right) \times \bm{B}_{0} + \frac{\left(\nabla \times \bm{B_{1}}\right) \times \bm{B}_{0}}{e n_{\rm{e}} \mu_{0}} + \eta\mu_0 \bm{J}_{1} \right] \nonumber \\
        + \frac{Z_{\rm{s}}}{n_{\rm{e}}m_{\rm{s}}}\sum_{t \ne e} \alpha_{\rm{et}} \left(\bm{V}_{\rm{t}} - \bm{V}_i \right) + \sum_{t \ne s} \nu_{\rm{st}} \left(\bm{V}_{\rm{t}} - \bm{V}_{\rm{s}}\right),
        \label{eq:5f_vs}
\end{gather}

\noindent where the presence of gravity has been neglected, $\bm{g} = 0$. The weighted mean velocity of ions, $\bm{V}_i$, and the resistivity, $\eta$, are given by
\begin{equation} \label{eq:5f_vion}
        \bm{V}_i = \frac{Z_{\rm{p}}n_{\rm{p}} \bm{V}_{\rm{p}} + Z_{\rm{He\sc{II}}} n_{\rm{He\sc{II}}} \bm{V}_{\rm{He\sc{II}}} + Z_{\rm{He\sc{III}}} n_{\rm{He\sc{III}}} \bm{V}_{\rm{He\sc{III}}}}{n_{\rm{e}}},
\end{equation}

\begin{equation} \label{eq:5f_eta}
        \eta = \frac{\alpha_{\rm{ep}} + \alpha_{\rm{eH}} + \alpha_{\rm{eHe}} + \alpha_{\rm{eHe\textsc{II}}} + \alpha_{\rm{eHe\textsc{III}}}}{e^{2}n_{\rm{e}}^{2}\mu_0}.
\end{equation}
    
\noindent The linear induction equation is
\begin{gather}
        \frac{\partial \bm{B}_{1}}{\partial t} = \nabla \times \left[\bm{V}_i \times \bm{B}_{0} - \frac{\left(\nabla \times \bm{B}_{1} \right) \times \bm{B}_{0}}{e n_{\rm{e}} \mu_{0}} - \eta\mu_0 \bm{J_{1}} - \frac{1}{e n_{\rm{e}}} \sum_{t \ne e} \alpha_{\rm{et}} \left(\bm{V}_{\rm{t}} - \bm{V}_i \right)\right].
    \label{eq:5f_b1}
\end{gather}

  
\noindent The last term in Eq. \ref{eq:5f_vs} contains a dependence on the velocity of electrons, $\bm{V}_{\rm{e}}$ but this model does not include an evolution equation for $\bm{V}_{\rm{e}}$. To remove that explicit dependence on $\bm{V}_{\rm{e}}$, one can rewrite this collision term as

    \begin{equation} \label{eq:5f_nust1}
        \sum_{t \ne s} \nu_{\rm{st}} \left(\bm{V}_{\rm{t}} - \bm{V}_{\rm{s}} \right) = \sum_{t \ne s,e} \nu_{st} \left(\bm{V}_{\rm{s}} - \bm{V}_{\rm{s}} \right) + \nu_{\rm{se}} \left(\bm{V}_{\rm{e}} - \bm{V}_{\rm{s}} \right),
    \end{equation}
    
\noindent and then apply Eq. \ref{eq:5f_current} so that $\bm{V}_{\rm{e}} = \bm{V}_i - \bm{J}_{1} /(e n_{\rm{e}})$. Finally, applying Ampère's law one gets the following expression

\begin{equation} \label{eq:5f_nust2}
        \sum_{t \ne s} \nu_{\rm{st}} \left(\bm{V}_{\rm{s}} - \bm{V}_{\rm{t}} \right) = \sum_{t \ne s,e} \nu_{\rm{st}} \left(\bm{V}_{\rm{t}} - \bm{V}_{\rm{s}} \right) + \nu_{\rm{se}} \left(\bm{V}_i - \bm{V}_{\rm{s}} \right) - \frac{\nu_{\rm{se}} \nabla \times \bm{B}_{1}}{e n_{\rm{e}} \mu_{0}},
\end{equation}
which shows that, when resistivity is considered, even neutral species are affected by the magnetic field due to their coupling with electrons.
    
In order to derive the dispersion relation for Alfv\'en waves in this multi-fluid approach, one assumes that the background magnetic field is oriented along the $x$-direction, that is $\bm{B}_{0} = \{B_{x}, 0, 0\}$, and that the waves propagate along the same direction, so $\bm{k} = \{k_{x}, 0, 0\}$. After performing normal mode analysis and Fourier analysis in space, with perturbations proportional to $\exp \left(-i \omega t+i \bm{k} \cdot \bm{r}\right)$, one arrives to a set of twelve equations where the $y-$ and $z-$ components are coupled, unlike the case of two-fluid ideal Alfv\'en waves. The consideration of additional equations for ions causes Alfv\'en waves to be circularly polarized, instead of linearly polarized. 
    
To study the properties of circularly polarized waves it is useful to define the following variables:
\begin{equation} \label{eq:5f_polarized}
        V_{\rm{s},\pm} = V_{\rm{s},y} \pm i V_{\rm{s},z}, \quad B_{1,\pm} = B_{1,y} \pm i B_{1,z},
\end{equation}
where the ``+/-'' sign corresponds to the left/right-hand polarization ($L$/$R$ modes). Through this step, a system of twelve coupled equations is transformed into two independent systems of six equations, which are considerably easier to manage, and provide a clearer view on the properties of the waves. The resulting equations for the velocity and magnetic field perturbations are,
\begin{gather} 
    \omega V_{\rm{s},\pm} = \Omega_{\rm{s}} \left[\pm \left(V_{\rm{s},\pm} - V_{i,\pm} \right) - k_{x}\eta_H B_{1,\pm}\right] \mp i \eta\frac{Z_{\rm{s}}ek_{x}}{m_{\rm{s}}}B_{1,\pm} \nonumber
    \\+i \frac{Z_{\rm{s}}}{n_{\rm{e}}m_{\rm{s}}}\sum_{t \ne e} \alpha_{\rm{et}} \left(V_{\rm{t},\pm} - \bm{V}_{i,\pm}\right) + i \sum_{t \ne s,e} \nu_{\rm{st}} \left(V_{\rm{t},\pm} - V_{\rm{s},\pm} \right) \nonumber \\
    + i \nu_{\rm{se}} \left(V_{i,\pm} - V_{\rm{s},\pm}\right) \pm i\nu_{\rm{se}} k_{x}\eta_H B_{1,\pm},
        \label{eq:5f_vpm}
\end{gather}
\begin{gather}
        \omega B_{1,\pm} = - k_{x}B_{x}V_{\pm} -\left[i\eta k_{x}^{2} \pm k_{x}^{2}B_{x}\eta_H\right]B_{1,\pm} \pm i \frac{k_{x}}{e n_{\rm{e}}}\sum_{t \ne e} \alpha_{\rm{et}} \left(V_{\rm{t},\pm} - V_{\pm}\right),
        \label{eq:5f_bpm}
\end{gather}
\noindent where $\Omega_{\rm{s}} = Z_{\rm{s}} e B_{x}/m_{\rm{s}}$ is the cyclotron frequency of species $s$.


\begin{figure} [t!]
    \centering
    \includegraphics[width=0.9\hsize]{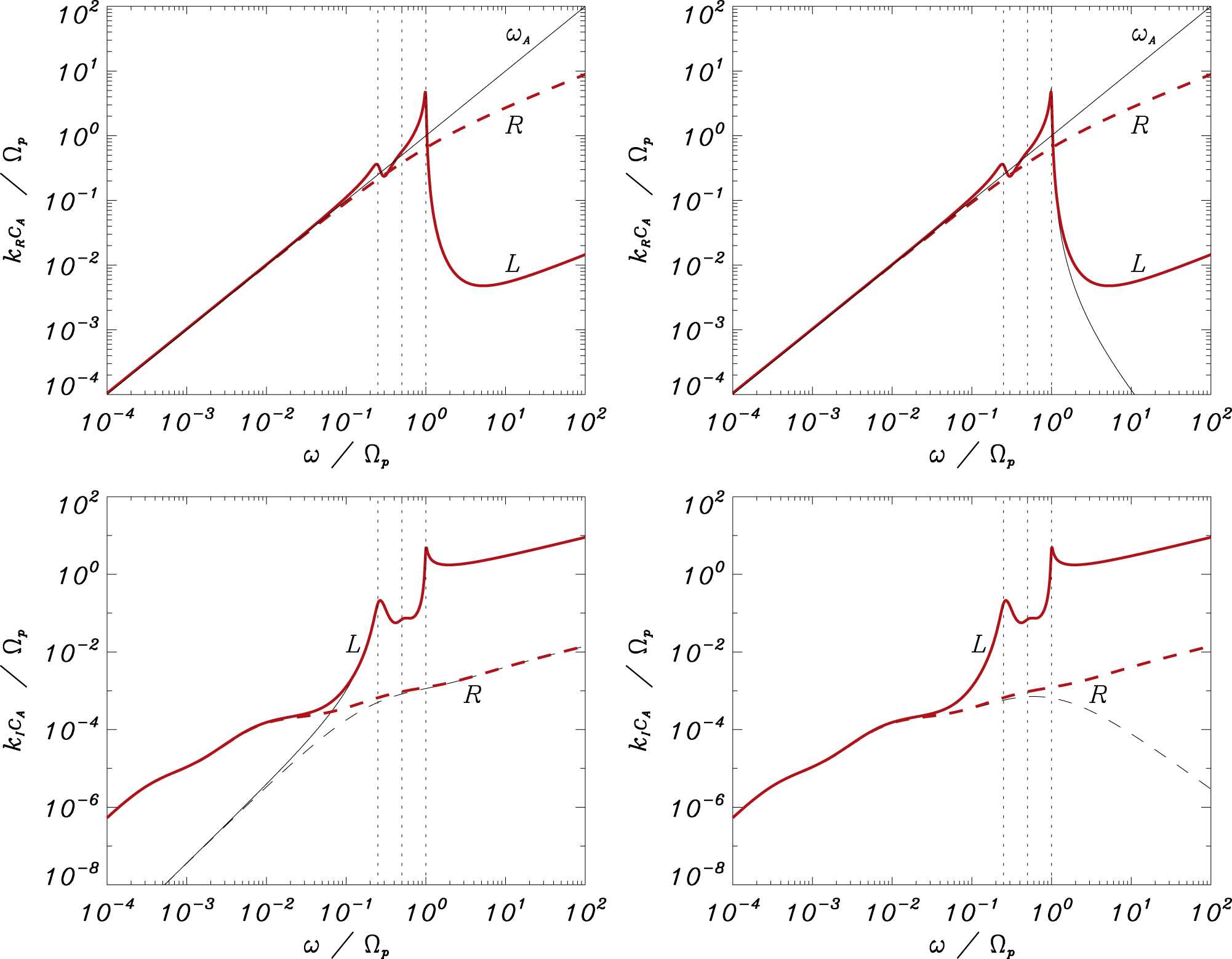}
    \caption{Real part (top) and imaginary part (bottom) of wavenumber for Alfv\'en waves excited by a periodic driver as functions of the driver frequency, $\omega$, normalized by the proton cyclotron frequency, $\Omega_{\rm{ci}}$ (in our notation). Left panels: solutions neglecting charged-neutral collisions (black lines), full solution (red lines). Right panels: solutions neglecting resistivity, i.e. collisions with electrons (black lines), full solution (red lines). The diagonal black lines is Alfv\'en frequency, $\omega_{\rm{A}}=k_x c_A$. The dotted vertical lines show the positions of the frequencies of cyclotron resonances, $\Omega_{\rm{p}} > \Omega_{\rm{He\textsc{III}}} > \Omega_{\rm{He\textsc{II}}} $. The labels $L$ and $R$ refer to left-handed and right-handed polarization. Figure from \citet{MartinezGomez2017ApJ...837...80M}.}
        \label{fig:5f_drsols}
\end{figure}

\begin{figure} [t!]
    \centering
    \includegraphics[width=0.9\hsize]{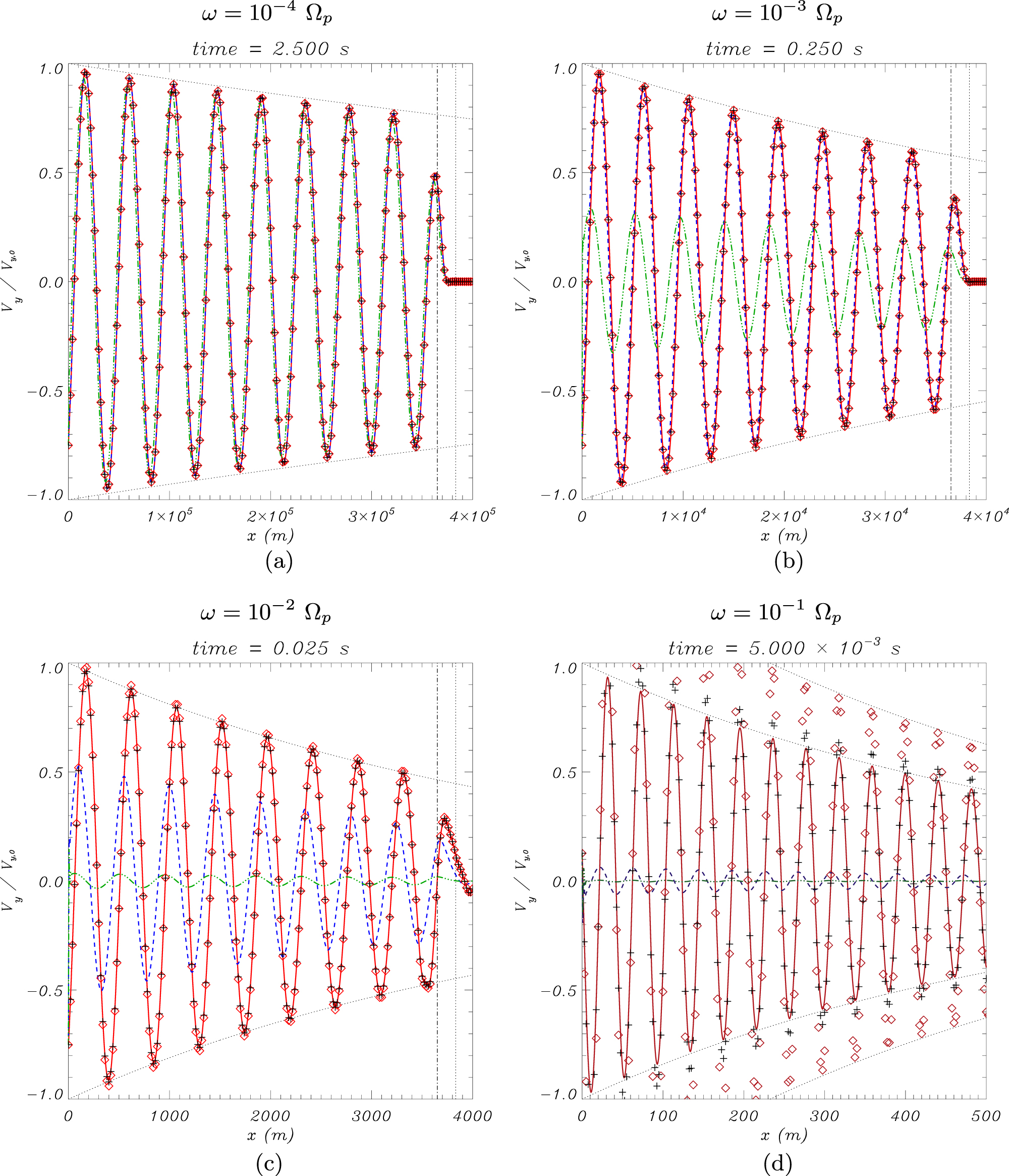}
    \caption{Snapshots from simulations of Alfv\'en waves generated by a periodic driver in a region with physical conditions of higher chromosphere. Panels a--d show the results for wave frequencies, $\omega=\{10^{-4}, 10^{-3}, 10^{-2}, 10^{-1}\} \Omega_{\rm{p}}$. Red solid, blue dashed, and green dashed-dotted lines represent the $y$-component of the velocity of protons, neutral hydrogen, and neutral helium. Red diamonds and the black crosses, represent the velocity for the singly and doubly ionized helium. The vertical dotted and dashed-dotted lines show the position of points moving at the Alfv\'en speed, and the modified Alfv\'en speed. The curved black dotted lines represent the collision damping predicted by the dispersion relation, Eq. \ref{eq:5f_dr}. Figure from \citet{MartinezGomez2017ApJ...837...80M}.}
        \label{fig:5f_sims}
\end{figure}

The dispersion relation for each polarization can be obtained by combining Eqs.~\ref{eq:5f_vpm} and \ref{eq:5f_bpm}. However, due to the large number of equations, it is easier to express the systems in a matrix form, and solve the matrix equations,

\begin{equation} \label{eq:5f_matrix}
        A_{\pm} \cdot \bm{u}_{\pm} = 0,
\end{equation}

\noindent where $\bm{u}_{\pm} = \{V_{\rm{p},\pm}, V_{\rm{He\sc{II}},\pm}, V_{\rm{He\sc{III}},\pm}, V_{\rm{H},\pm}, V_{\rm{He},\pm}, B_{1,\pm} \}$ are the vectors of unknowns and $A_{\pm}$ are the matrices of coefficients, which contain the dependence on the frequency, $\omega$, the wavenumber, $k_{x}$, and the parameters of the plasma, see the original work by \citet{MartinezGomez2017ApJ...837...80M}. 
The solution of Eq. \ref{eq:5f_matrix} is given by the characteristic equation of each matrix:
    \begin{equation} \label{eq:5f_dr}
        \mathcal{D}_{\pm}(\omega, k_{x}) \equiv \det A_{\pm} = 0.
    \end{equation}
    
The resulting dispersion relations are too complex to be shown here and exact analytical solutions are not practical. The exact solutions, discussed below, have been obtained numerically. Nevertheless, some information can yet be extracted by taking into account the general structure of the dispersion relations and the results from previous sections.
    
The dispersion relation for each polarization state is a convoluted polynomial of sixth order in  $\omega_{\pm}$, but only of the second order in $k_{x,\pm}$. Therefore, in contrast with the case of Alfv\'en waves in a single-fluid plasma, there is a clear asymmetry in the number of solutions depending on whether the wave driver is impulsive or periodic. When the dispersion relation is solved as a function of a real wavenumber, $k_{x}$ (impulsive driver), Eq. \ref{eq:5f_dr} has six different complex solutions for each polarization, $\omega_{\pm} = \omega_{\rm{R},\pm} + i \omega_{\rm{I},\pm}$. For low wavenumbers, two of the modes correspond to the forward and backward Alfv\'en waves. Other two modes are related to the presence of the two neutral species (H and He) and, although $\omega_{\rm{R},\pm}$ is not exactly zero, they have very low frequencies. The remaining solutions are high-frequency modes associated with the cyclotron motions of the ions \citep[see][]{MartinezGomez2016ApJ...832..101M,MartinezGomez2017ApJ...837...80M}. For high wavenumbers, one of the Alfv\'en waves turns into a whistler mode (with a frequency that keeps increasing with the wavenumber) while the other one becomes an ion-cyclotron wave, with a frequency that tends to the cyclotron frequency of singly-ionized helium, $\Omega_{\rm{He\sc{II)}}}$, \citep{2001paw..book.....C}. The nature of the neutral--related modes hardly changes, while the frequency of the two remaining solutions tends to the cyclotron frequency of the remaining ions, $\Omega_{\rm{He\sc{III}}}$ and $\Omega_{\rm{p}}$. Thus, in this high-wavenumber range, there is one ion-cyclotron mode for each ionized fluid. All the modes are damped due to collisions between the different species. The effect is more pronounced for the high-frequency modes, such as the whistler and ion-cylotron waves.
    
When a periodic driver is considered, and Eq.~\ref{eq:5f_dr} is solved as a function of a real frequency, $\omega$, each polarization has only two possible complex solutions, $k_{x,\pm} = k_{\rm{R},\pm} + i k_{\rm{I},\pm}$. Figure \ref{fig:5f_drsols} shows an example of this scenario, where the solutions have been numerically obtained for plasma parameters corresponding to an upper chromospheric region. In a multi-ion plasma with no collisions between its components, the left-hand polarized or ion-cyclotron modes have resonances when the frequency of the driver is equal to any of the cyclotron frequencies \citep{2001paw..book.....C}. It means that the wavenumber tends to infinity at those frequencies, leading to a zero phase speed, so the wave does not propagate. Thus, the effect of the driver is to increase the amplitude of the fluctuations of the ion fluid associated with the given cyclotron frequency at the location where the driver is applied. Furthermore, beyond each resonance, there is a cut-off region where the waves become evanescent and then a small region where the propagation is again allowed. 
As it can be seen at the top panels of Fig. \ref{fig:5f_drsols},  when collisions are taken into account, the wavenumber of the ion-cyclotron modes (denoted by the tag $L$) remains finite at the cyclotron frequencies, i.e., at $\omega = \Omega_{\rm{He\sc{II}}}$, $\omega = \Omega_{\rm{He\sc{III}}}$ and $\omega = \Omega_{\rm{p}}$. In addition, there is no evidence of the presence of cut-off regions where $k_{\rm{R}} = 0$. Therefore, it can be concluded that, one of the important effects of elastic collisions is to remove the cyclotron resonances and cut-off regions, i.e., waves are allowed to propagate for any driver frequency. However, the ion-cyclotron modes are still strongly damped by the collisional interaction, as shown in the bottom panels of Fig. \ref{fig:5f_drsols}.
    
As a final remark for this section, it is interesting to briefly abandon the analysis of the dispersion relations and come back to the full set of equations presented in Section \ref{sec:eqs_5f}. Figure \ref{fig:5f_sims} shows the results from several numerical simulations of Alfv\'en waves generated by a periodic driver. Each panel corresponds to a different frequency of the driver, $\omega$, and the different lines and symbols represent the perturbation of velocity of each of the five fluids of the plasma. It is clear how all the fluids are strongly coupled for a low driver frequency and how their behavior start to decouple as the frequency of the driver is increased. Figure \ref{fig:5f_sims} also shows that the amplitude of the waves decreases due to elastic collisions as they propagate, in good agreement with the predictions obtained from Eq.~\ref{eq:5f_dr}.

\section{Summary and outlook}
    This chapter focuses on describing the properties of waves in partially ionized plasmas with a different degree of collisional coupling between the plasma and neutral components. It describes the change of the physics of waves, compared to those in a classical ideal MHD case. It is shown in what way both linear and non-linear wave solutions are affected by plasma partial ionization, and how the gravitational stratification adds an additional flavor for waves propagating in a stratified solar atmosphere. The content of this chapter can be summarized in the following way:

\begin{itemize}
\item The complexity of the mathematical description of waves in partially ionized plasmas depends on the degree of collisional coupling. In general terms, the photosphere and low chromosphere, and in the dense cores of solar prominences, for waves with typical solar frequencies, a single-fluid approximation can be applied. In the middle-upper chromosphere and in the transition layer between prominences and corona, low collisional frequencies made multi-fluid approach to be necessary.

\item In the single-fluid approximation, the number of MHD modes in the system remains unchanged compared to ideal MHD and usual fast, slow and Alfv\'en modes can still be distinguished, but with modified properties. The main partial ionization effects that influence waves in this approximation are the ambipolar diffusion, modified Hall effect, and diamagnetic effect.

\item Fast and Alfv\'en waves can be significantly damped by ambipolar diffusion, while the slow mode is only weakly affected by this mechanism. Nevertheless, mechanisms such as radiative cooling and thermal conduction can be of the same importance for the damping of the fast and slow modes, depending on their wavenumber. The Alfv\'en mode is unaffected by the latter. 

\item In certain conditions, cutoff wavenumbers appear for the fast and Alfv\'en waves due to ambipolar diffusion, a similar effect to that of the Ohmic diffusion. However, once the Hall effect is taken into account, the hard cutoff disappears due to the distinct behavior of ions and electrons in this case. 

\item Modeling of fast and Alfv\'en waves in a gravitationally stratified atmosphere in a single fluid approximation revealed that the ambipolar diffusion produces significant damping of waves with periods of 1--10 s in the upper chromosphere. The damping is stronger for the shorter period waves.

\item Mode conversion in a stratified atmosphere is affected by the ambipolar and the Hall effects. The ambipolar diffusion reduces the amount of the energy going into the converted Alfv\'en wave since the fast wave energy flux in the conversion region is reduced. The Hall effect enables a new mechanism of mode transformation by coupling the fast and Alfv\'en waves through the so-called Hall window, typically located in the photosphere.

\item Resistive heating in shocks due to ambipolar diffusion can be an efficient chromospheric heating mechanism. Nevertheless, 1D computations show so far that compressional heating in shocks overcomes the efficiency of the resistive heating. Resistivity plays an important role by defining the width of the shock front over which the compression happens.

\item In the multi-fluid case, additional wave modes exist, compared to the usual MHD modes. These modes are related to the acoustic modes of the neutral fluid(s), and to the whistler modes and ion-cyclotron modes of the ionized fluid(s).

\item In the multi-fluid approximation, all the modes can be damped due to ion-neutral collisions, but to a different degree. In general, it holds that fast and Alfv\'en waves are damped more. Similarly, in the two-fluid approximation, both Alfv\'en and magneto-acoustic modes show the presence of cut-off regions, where the waves cannot propagate, affected by inter-particle collisions.

\item Fast and slow magneto-acoustic waves propagating in a gravitationally stratified atmosphere are affected by two competing effects: wave amplitude growth due to the stratification, and wave amplitude damping due to ion-neutral collisions in the two-fluid approximation. These competing effects extend the range of wave periods affected by ion-neutral effects up to 20--30 s. Since the wave with larger periods are able to propagate relatively undamped to the higher chromosphere, they suffer more significant ion-neutral decoupling, and consequently, produce more significant frictional heating. 

\item Frictional heating produces regimes where the heating rates scale either as a square of the wave frequency or linear with the wave frequency. Integrated frictional heating rates for frequencies of the order of $10^0-10^{10}$  Hz are large enough to balance chromospheric radiative losses.

\item Multi-fluid effects produce complex structures of the shock wave fronts. In general the shock transition is smoother in the two-fluid case compared to the purely MHD case. The multi-fluid shock transitions in the solar case have been only scarcely investigated.

\item The inclusion of helium in the multi-fluid description increases the amount of wave modes in the system. The collisions with helium atoms, the second most abundant element after hydrogen, increase the collisional damping of waves by 10--30\%, depending on the plasma conditions.

\end{itemize}
    



  \bibliography{1fluid_bib,2fluid_bib}

\begin{thebibliography}{92}
\expandafter\ifx\csname natexlab\endcsname\relax\def\natexlab#1{#1}\fi
\providecommand{\url}[1]{\texttt{#1}}
\providecommand{\href}[2]{#2}
\providecommand{\path}[1]{#1}
\providecommand{\DOIprefix}{doi:}
\providecommand{\ArXivprefix}{arXiv:}
\providecommand{\URLprefix}{URL: }
\providecommand{\Pubmedprefix}{pmid:}
\providecommand{\doi}[1]{\href{http://dx.doi.org/#1}{\path{#1}}}
\providecommand{\Pubmed}[1]{\href{pmid:#1}{\path{#1}}}
\providecommand{\bibinfo}[2]{#2}
\ifx\xfnm\relax \def\xfnm[#1]{\unskip,\space#1}\fi
\bibitem[{{Arber} et~al.(2016){Arber}, {Brady} and
  {Shelyag}}]{2016ApJ...817...94A}
\bibinfo{author}{{Arber}, T.D.}, \bibinfo{author}{{Brady}, C.S.},
  \bibinfo{author}{{Shelyag}, S.}, \bibinfo{year}{2016}.
\newblock \bibinfo{title}{{Alfv{\'e}n Wave Heating of the Solar Chromosphere:
  1.5D Models}}.
\newblock \bibinfo{journal}{ApJ} \bibinfo{volume}{817}, \bibinfo{pages}{94}.
\newblock \DOIprefix\doi{10.3847/0004-637X/817/2/94},
  \href{http://arxiv.org/abs/1512.05816}{{\tt arXiv:1512.05816}}.
\bibitem[{Balescu(1988)}]{Balescu}
\bibinfo{author}{Balescu, R.}, \bibinfo{year}{1988}.
\newblock \bibinfo{title}{{Transport processes in plasmas}}.
\newblock \bibinfo{publisher}{North Holland Publ., Amsterdam}.
\bibitem[{{Ballester} et~al.(2018){Ballester}, {Alexeev}, {Collados}, {Downes},
  {Pfaff}, {Gilbert}, {Khodachenko}, {Khomenko}, {Shaikhislamov}, {Soler},
  {V{\'a}zquez-Semadeni} and {Zaqarashvili}}]{2018SSRv..214...58B}
\bibinfo{author}{{Ballester}, J.L.}, \bibinfo{author}{{Alexeev}, I.},
  \bibinfo{author}{{Collados}, M.}, \bibinfo{author}{{Downes}, T.},
  \bibinfo{author}{{Pfaff}, R.F.}, \bibinfo{author}{{Gilbert}, H.},
  \bibinfo{author}{{Khodachenko}, M.}, \bibinfo{author}{{Khomenko}, E.},
  \bibinfo{author}{{Shaikhislamov}, I.F.}, \bibinfo{author}{{Soler}, R.},
  \bibinfo{author}{{V{\'a}zquez-Semadeni}, E.},
  \bibinfo{author}{{Zaqarashvili}, T.}, \bibinfo{year}{2018}.
\newblock \bibinfo{title}{{Partially Ionized Plasmas in Astrophysics}}.
\newblock \bibinfo{journal}{SSRv} \bibinfo{volume}{214}, \bibinfo{pages}{58}.
\newblock \DOIprefix\doi{10.1007/s11214-018-0485-6},
  \href{http://arxiv.org/abs/1707.07975}{{\tt arXiv:1707.07975}}.
\bibitem[{Bittencourt(1986)}]{Bittencourt}
\bibinfo{author}{Bittencourt, J.A.}, \bibinfo{year}{1986}.
\newblock \bibinfo{title}{Fundamentals of plasma physics}.
\newblock \bibinfo{publisher}{Pergamon Press}, \bibinfo{address}{Oxford}.
\bibitem[{{Bogdan} et~al.(2003){Bogdan}, {Carlsson}, {Hansteen}, {McMurry},
  {Rosenthal}, {Johnson}, {Petty-Powell}, {Zita}, {Stein}, {McIntosh} and
  {Nordlund}}]{2003ApJ...599..626B}
\bibinfo{author}{{Bogdan}, T.J.}, \bibinfo{author}{{Carlsson}, M.},
  \bibinfo{author}{{Hansteen}, V.H.}, \bibinfo{author}{{McMurry}, A.},
  \bibinfo{author}{{Rosenthal}, C.S.}, \bibinfo{author}{{Johnson}, M.},
  \bibinfo{author}{{Petty-Powell}, S.}, \bibinfo{author}{{Zita}, E.J.},
  \bibinfo{author}{{Stein}, R.F.}, \bibinfo{author}{{McIntosh}, S.W.},
  \bibinfo{author}{{Nordlund}, {\r{A}}.}, \bibinfo{year}{2003}.
\newblock \bibinfo{title}{{Waves in the Magnetized Solar Atmosphere. II. Waves
  from Localized Sources in Magnetic Flux Concentrations}}.
\newblock \bibinfo{journal}{ApJ} \bibinfo{volume}{599},
  \bibinfo{pages}{626--660}.
\newblock \DOIprefix\doi{10.1086/378512}.
\bibitem[{{Braginskii}(1965)}]{1965RvPP....1..205B}
\bibinfo{author}{{Braginskii}, S.I.}, \bibinfo{year}{1965}.
\newblock \bibinfo{title}{{Transport Processes in a Plasma}}.
\newblock \bibinfo{journal}{Reviews of Plasma Physics} \bibinfo{volume}{1},
  \bibinfo{pages}{205}.
\bibitem[{{Cally}(2001)}]{2001ApJ...548..473C}
\bibinfo{author}{{Cally}, P.S.}, \bibinfo{year}{2001}.
\newblock \bibinfo{title}{{Note on an Exact Solution for Magnetoatmospheric
  Waves}}.
\newblock \bibinfo{journal}{ApJ} \bibinfo{volume}{548},
  \bibinfo{pages}{473--481}.
\newblock \DOIprefix\doi{10.1086/318675}.
\bibitem[{{Cally}(2006)}]{2006RSPTA.364..333C}
\bibinfo{author}{{Cally}, P.S.}, \bibinfo{year}{2006}.
\newblock \bibinfo{title}{{Dispersion relations, rays and ray splitting in
  magnetohelioseismology}}.
\newblock \bibinfo{journal}{Philosophical Transactions of the Royal Society of
  London Series A} \bibinfo{volume}{364}, \bibinfo{pages}{333--349}.
\newblock \DOIprefix\doi{10.1098/rsta.2005.1702}.
\bibitem[{{Cally} and {Goossens}(2008)}]{2008SoPh..251..251C}
\bibinfo{author}{{Cally}, P.S.}, \bibinfo{author}{{Goossens}, M.},
  \bibinfo{year}{2008}.
\newblock \bibinfo{title}{{Three-Dimensional MHD Wave Propagation and
  Conversion to Alfv{\'e}n Waves near the Solar Surface. I. Direct Numerical
  Solution}}.
\newblock \bibinfo{journal}{Solar Phys.} \bibinfo{volume}{251},
  \bibinfo{pages}{251--265}.
\newblock \DOIprefix\doi{10.1007/s11207-007-9086-3},
  \href{http://arxiv.org/abs/0711.0498}{{\tt arXiv:0711.0498}}.
\bibitem[{{Cally} and {Khomenko}(2015)}]{2015ApJ...814..106C}
\bibinfo{author}{{Cally}, P.S.}, \bibinfo{author}{{Khomenko}, E.},
  \bibinfo{year}{2015}.
\newblock \bibinfo{title}{{Fast-to-Alfv{\'e}n Mode Conversion Mediated by the
  Hall Current. I. Cold Plasma Model}}.
\newblock \bibinfo{journal}{ApJ} \bibinfo{volume}{814}, \bibinfo{pages}{106}.
\newblock \DOIprefix\doi{10.1088/0004-637X/814/2/106},
  \href{http://arxiv.org/abs/1510.03927}{{\tt arXiv:1510.03927}}.
\bibitem[{{Cally} and {Khomenko}(2018)}]{2018ApJ...856...20C}
\bibinfo{author}{{Cally}, P.S.}, \bibinfo{author}{{Khomenko}, E.},
  \bibinfo{year}{2018}.
\newblock \bibinfo{title}{{Fast-to-Alfv{\'e}n Mode Conversion in the Presence
  of Ambipolar Diffusion}}.
\newblock \bibinfo{journal}{ApJ} \bibinfo{volume}{856}, \bibinfo{pages}{20}.
\newblock \DOIprefix\doi{10.3847/1538-4357/aaaf6a}.
\bibitem[{{Carbonell} et~al.(2004){Carbonell}, {Oliver} and
  {Ballester}}]{2004A&A...415..739C}
\bibinfo{author}{{Carbonell}, M.}, \bibinfo{author}{{Oliver}, R.},
  \bibinfo{author}{{Ballester}, J.L.}, \bibinfo{year}{2004}.
\newblock \bibinfo{title}{{Time damping of linear non-adiabatic
  magnetohydrodynamic waves in an unbounded plasma with solar coronal
  properties}}.
\newblock \bibinfo{journal}{A\&A} \bibinfo{volume}{415},
  \bibinfo{pages}{739--750}.
\newblock \DOIprefix\doi{10.1051/0004-6361:20034630}.
\bibitem[{{Christensen-Dalsgaard} et~al.(1996){Christensen-Dalsgaard},
  {Dappen}, {Ajukov}, {Anderson}, {Antia}, {Basu}, {Baturin}, {Berthomieu},
  {Chaboyer}, {Chitre}, {Cox}, {Demarque}, {Donatowicz}, {Dziembowski},
  {Gabriel}, {Gough}, {Guenther}, {Guzik}, {Harvey}, {Hill}, {Houdek},
  {Iglesias}, {Kosovichev}, {Leibacher}, {Morel}, {Proffitt}, {Provost},
  {Reiter}, {Rhodes}, {Rogers}, {Roxburgh}, {Thompson} and
  {Ulrich}}]{1996Sci...272.1286C}
\bibinfo{author}{{Christensen-Dalsgaard}, J.}, \bibinfo{author}{{Dappen}, W.},
  \bibinfo{author}{{Ajukov}, S.V.}, \bibinfo{author}{{Anderson}, E.R.},
  \bibinfo{author}{{Antia}, H.M.}, \bibinfo{author}{{Basu}, S.},
  \bibinfo{author}{{Baturin}, V.A.}, \bibinfo{author}{{Berthomieu}, G.},
  \bibinfo{author}{{Chaboyer}, B.}, \bibinfo{author}{{Chitre}, S.M.},
  \bibinfo{author}{{Cox}, A.N.}, \bibinfo{author}{{Demarque}, P.},
  \bibinfo{author}{{Donatowicz}, J.}, \bibinfo{author}{{Dziembowski}, W.A.},
  \bibinfo{author}{{Gabriel}, M.}, \bibinfo{author}{{Gough}, D.O.},
  \bibinfo{author}{{Guenther}, D.B.}, \bibinfo{author}{{Guzik}, J.A.},
  \bibinfo{author}{{Harvey}, J.W.}, \bibinfo{author}{{Hill}, F.},
  \bibinfo{author}{{Houdek}, G.}, \bibinfo{author}{{Iglesias}, C.A.},
  \bibinfo{author}{{Kosovichev}, A.G.}, \bibinfo{author}{{Leibacher}, J.W.},
  \bibinfo{author}{{Morel}, P.}, \bibinfo{author}{{Proffitt}, C.R.},
  \bibinfo{author}{{Provost}, J.}, \bibinfo{author}{{Reiter}, J.},
  \bibinfo{author}{{Rhodes}, E.~J., J.}, \bibinfo{author}{{Rogers}, F.J.},
  \bibinfo{author}{{Roxburgh}, I.W.}, \bibinfo{author}{{Thompson}, M.J.},
  \bibinfo{author}{{Ulrich}, R.K.}, \bibinfo{year}{1996}.
\newblock \bibinfo{title}{{The Current State of Solar Modeling}}.
\newblock \bibinfo{journal}{Science} \bibinfo{volume}{272},
  \bibinfo{pages}{1286--1292}.
\newblock \DOIprefix\doi{10.1126/science.272.5266.1286}.
\bibitem[{{Cowling}(1945)}]{1945RSPSA.183..453C}
\bibinfo{author}{{Cowling}, T.G.}, \bibinfo{year}{1945}.
\newblock \bibinfo{title}{{The Electrical Conductivity of an Ionized Gas in a
  Magnetic Field, with Applications to the Solar Atmosphere and the
  Ionosphere}}.
\newblock \bibinfo{journal}{Proceedings of the Royal Society of London Series
  A} \bibinfo{volume}{183}, \bibinfo{pages}{453--479}.
\newblock \DOIprefix\doi{10.1098/rspa.1945.0013}.
\bibitem[{{Cramer}(2001)}]{2001paw..book.....C}
\bibinfo{author}{{Cramer}, N.F.}, \bibinfo{year}{2001}.
\newblock \bibinfo{title}{{The Physics of Alfv{\'e}n Waves}}.
\bibitem[{{de Pontieu} and {Haerendel}(1998)}]{1998A&A...338..729D}
\bibinfo{author}{{de Pontieu}, B.}, \bibinfo{author}{{Haerendel}, G.},
  \bibinfo{year}{1998}.
\newblock \bibinfo{title}{{Weakly damped Alfven waves as drivers for
  spicules}}.
\newblock \bibinfo{journal}{A\&A} \bibinfo{volume}{338},
  \bibinfo{pages}{729--736}.
\bibitem[{{Draine}(1986)}]{Draine1986MNRAS.220..133D}
\bibinfo{author}{{Draine}, B.T.}, \bibinfo{year}{1986}.
\newblock \bibinfo{title}{{Multicomponent, reacting MHD flows}}.
\newblock \bibinfo{journal}{MNRAS} \bibinfo{volume}{220},
  \bibinfo{pages}{133--148}.
\newblock \DOIprefix\doi{10.1093/mnras/220.1.133}.
\bibitem[{{Ferraro} and {Plumpton}(1961)}]{1961itmf.book.....F}
\bibinfo{author}{{Ferraro}, V.C.A.}, \bibinfo{author}{{Plumpton}, C.},
  \bibinfo{year}{1961}.
\newblock \bibinfo{title}{{An introduction to magneto-fluid mechanics}}.
\bibitem[{{Fontenla} et~al.(1993){Fontenla}, {Avrett} and
  {Loeser}}]{1993ApJ...406..319F}
\bibinfo{author}{{Fontenla}, J.M.}, \bibinfo{author}{{Avrett}, E.H.},
  \bibinfo{author}{{Loeser}, R.}, \bibinfo{year}{1993}.
\newblock \bibinfo{title}{{Energy Balance in the Solar Transition Region. III.
  Helium Emission in Hydrostatic, Constant-Abundance Models with Diffusion}}.
\newblock \bibinfo{journal}{ApJ} \bibinfo{volume}{406}, \bibinfo{pages}{319}.
\newblock \DOIprefix\doi{10.1086/172443}.
\bibitem[{{Forteza} et~al.(2008){Forteza}, {Oliver} and
  {Ballester}}]{2008A&A...492..223F}
\bibinfo{author}{{Forteza}, P.}, \bibinfo{author}{{Oliver}, R.},
  \bibinfo{author}{{Ballester}, J.L.}, \bibinfo{year}{2008}.
\newblock \bibinfo{title}{{Time damping of non-adiabatic MHD waves in an
  unbounded partially ionised prominence plasma}}.
\newblock \bibinfo{journal}{A\&A} \bibinfo{volume}{492},
  \bibinfo{pages}{223--231}.
\newblock \DOIprefix\doi{10.1051/0004-6361:200810370}.
\bibitem[{{Forteza} et~al.(2007){Forteza}, {Oliver}, {Ballester} and
  {Khodachenko}}]{2007A&A...461..731F}
\bibinfo{author}{{Forteza}, P.}, \bibinfo{author}{{Oliver}, R.},
  \bibinfo{author}{{Ballester}, J.L.}, \bibinfo{author}{{Khodachenko}, M.L.},
  \bibinfo{year}{2007}.
\newblock \bibinfo{title}{{Damping of oscillations by ion-neutral collisions in
  a prominence plasma}}.
\newblock \bibinfo{journal}{A\&A} \bibinfo{volume}{461},
  \bibinfo{pages}{731--739}.
\newblock \DOIprefix\doi{10.1051/0004-6361:20065900}.
\bibitem[{{Goedbloed} and {Poedts}(2004)}]{Goedbloed2004prma.book.....G}
\bibinfo{author}{{Goedbloed}, J.P.H.}, \bibinfo{author}{{Poedts}, S.},
  \bibinfo{year}{2004}.
\newblock \bibinfo{title}{{Principles of Magnetohydrodynamics}}.
\bibitem[{{Gonz{\'a}lez-Morales} et~al.(2019){Gonz{\'a}lez-Morales}, {Khomenko}
  and {Cally}}]{2019ApJ...870...94G}
\bibinfo{author}{{Gonz{\'a}lez-Morales}, P.A.}, \bibinfo{author}{{Khomenko},
  E.}, \bibinfo{author}{{Cally}, P.S.}, \bibinfo{year}{2019}.
\newblock \bibinfo{title}{{Fast-to-Alfv{\'e}n Mode Conversion Mediated by Hall
  Current. II. Application to the Solar Atmosphere}}.
\newblock \bibinfo{journal}{ApJ} \bibinfo{volume}{870}, \bibinfo{pages}{94}.
\newblock \DOIprefix\doi{10.3847/1538-4357/aaf1a9},
  \href{http://arxiv.org/abs/1811.06565}{{\tt arXiv:1811.06565}}.
\bibitem[{{Gonz{\'a}lez-Morales} et~al.(2020){Gonz{\'a}lez-Morales},
  {Khomenko}, {Vitas} and {Collados}}]{2020A&A...642A.220G}
\bibinfo{author}{{Gonz{\'a}lez-Morales}, P.A.}, \bibinfo{author}{{Khomenko},
  E.}, \bibinfo{author}{{Vitas}, N.}, \bibinfo{author}{{Collados}, M.},
  \bibinfo{year}{2020}.
\newblock \bibinfo{title}{{Joint action of Hall and ambipolar effects in 3D
  magneto-convection simulations of the quiet Sun. I. Dissipation and
  generation of waves}}.
\newblock \bibinfo{journal}{A\&A} \bibinfo{volume}{642}, \bibinfo{pages}{A220}.
\newblock \DOIprefix\doi{10.1051/0004-6361/202037938},
  \href{http://arxiv.org/abs/2008.10429}{{\tt arXiv:2008.10429}}.
\bibitem[{{Goodman}(1996)}]{1996ApJ...463..784G}
\bibinfo{author}{{Goodman}, M.L.}, \bibinfo{year}{1996}.
\newblock \bibinfo{title}{{Heating of the Solar Middle Chromospheric Network
  and Internetwork by Large-Scale Electric Currents in Weakly Ionized Magnetic
  Elements}}.
\newblock \bibinfo{journal}{ApJ} \bibinfo{volume}{463}, \bibinfo{pages}{784}.
\newblock \DOIprefix\doi{10.1086/177290}.
\bibitem[{{Goodman}(2000)}]{2000ApJ...533..501G}
\bibinfo{author}{{Goodman}, M.L.}, \bibinfo{year}{2000}.
\newblock \bibinfo{title}{{On the Mechanism of Chromospheric Network Heating
  and the Condition for Its Onset in the Sun and Other Solar-Type Stars}}.
\newblock \bibinfo{journal}{ApJ} \bibinfo{volume}{533},
  \bibinfo{pages}{501--522}.
\newblock \DOIprefix\doi{10.1086/308635}.
\bibitem[{{Goodman}(2004)}]{2004A&A...416.1159G}
\bibinfo{author}{{Goodman}, M.L.}, \bibinfo{year}{2004}.
\newblock \bibinfo{title}{{On the efficiency of plasma heating by Pedersen
  current dissipation from the photosphere to the lower corona}}.
\newblock \bibinfo{journal}{A\&A} \bibinfo{volume}{416},
  \bibinfo{pages}{1159--1178}.
\newblock \DOIprefix\doi{10.1051/0004-6361:20031719}.
\bibitem[{{Goodman}(2011a)}]{2011ApJ...735...45G}
\bibinfo{author}{{Goodman}, M.L.}, \bibinfo{year}{2011}a.
\newblock \bibinfo{title}{{Conditions for Photospherically Driven Alfv{\'e}nic
  Oscillations to Heat the Solar Chromosphere by Pedersen Current
  Dissipation}}.
\newblock \bibinfo{journal}{ApJ} \bibinfo{volume}{735}, \bibinfo{pages}{45}.
\newblock \DOIprefix\doi{10.1088/0004-637X/735/1/45},
  \href{http://arxiv.org/abs/1410.8519}{{\tt arXiv:1410.8519}}.
\bibitem[{{Goodman}(2011b)}]{Goodman2011ApJ...735...45G}
\bibinfo{author}{{Goodman}, M.L.}, \bibinfo{year}{2011}b.
\newblock \bibinfo{title}{{Conditions for Photospherically Driven Alfv{\'e}nic
  Oscillations to Heat the Solar Chromosphere by Pedersen Current
  Dissipation}}.
\newblock \bibinfo{journal}{ApJ} \bibinfo{volume}{735}, \bibinfo{pages}{45}.
\newblock \DOIprefix\doi{10.1088/0004-637X/735/1/45},
  \href{http://arxiv.org/abs/1410.8519}{{\tt arXiv:1410.8519}}.
\bibitem[{{Goodman} and {Kazeminezhad}(2010)}]{2010ApJ...708..268G}
\bibinfo{author}{{Goodman}, M.L.}, \bibinfo{author}{{Kazeminezhad}, F.},
  \bibinfo{year}{2010}.
\newblock \bibinfo{title}{{Simulation of Magnetohydrodynamic Shock Wave
  Generation, Propagation, and Heating in the Photosphere and Chromosphere
  Using a Complete Electrical Conductivity Tensor}}.
\newblock \bibinfo{journal}{ApJ} \bibinfo{volume}{708},
  \bibinfo{pages}{268--287}.
\newblock \DOIprefix\doi{10.1088/0004-637X/708/1/268}.
\bibitem[{{Hildner}(1974)}]{1974SoPh...35..123H}
\bibinfo{author}{{Hildner}, E.}, \bibinfo{year}{1974}.
\newblock \bibinfo{title}{{The Formation of Solar Quiescent Prominences by
  Condensation}}.
\newblock \bibinfo{journal}{Solar Phys.} \bibinfo{volume}{35},
  \bibinfo{pages}{123--136}.
\newblock \DOIprefix\doi{10.1007/BF00156962}.
\bibitem[{Huba(2013)}]{Huba2013}
\bibinfo{author}{Huba, J.D.}, \bibinfo{year}{2013}.
\newblock \bibinfo{title}{{NRL PLASMA FORMULARY Supported by The Office of
  Naval Research}}.
\newblock \bibinfo{publisher}{Naval Research Laboratory},
  \bibinfo{address}{Washington, DC}.
\newblock \URLprefix \url{http://wwwppd.nrl.navy.mil/nrlformulary/}.
\bibitem[{{Hunana} et~al.(2022){Hunana}, {Passot}, {Khomenko},
  {Mart{\'\i}nez-G{\'o}mez}, {Collados}, {Tenerani}, {Zank}, {Maneva},
  {Goldstein} and {Webb}}]{2022ApJS..260...26H}
\bibinfo{author}{{Hunana}, P.}, \bibinfo{author}{{Passot}, T.},
  \bibinfo{author}{{Khomenko}, E.}, \bibinfo{author}{{Mart{\'\i}nez-G{\'o}mez},
  D.}, \bibinfo{author}{{Collados}, M.}, \bibinfo{author}{{Tenerani}, A.},
  \bibinfo{author}{{Zank}, G.P.}, \bibinfo{author}{{Maneva}, Y.},
  \bibinfo{author}{{Goldstein}, M.L.}, \bibinfo{author}{{Webb}, G.M.},
  \bibinfo{year}{2022}.
\newblock \bibinfo{title}{{Generalized Fluid Models of the Braginskii Type}}.
\newblock \bibinfo{journal}{ApJS} \bibinfo{volume}{260}, \bibinfo{pages}{26}.
\newblock \DOIprefix\doi{10.3847/1538-4365/ac5044},
  \href{http://arxiv.org/abs/2201.11561}{{\tt arXiv:2201.11561}}.
\bibitem[{{Judge}(2008)}]{2008ApJ...683L..87J}
\bibinfo{author}{{Judge}, P.}, \bibinfo{year}{2008}.
\newblock \bibinfo{title}{{An Explanation of the Solar Transition Region}}.
\newblock \bibinfo{journal}{ApJL} \bibinfo{volume}{683}, \bibinfo{pages}{L87}.
\newblock \DOIprefix\doi{10.1086/591470},
  \href{http://arxiv.org/abs/0807.1706}{{\tt arXiv:0807.1706}}.
\bibitem[{{Kazeminezhad} and {Goodman}(2006)}]{2006ApJS..166..613K}
\bibinfo{author}{{Kazeminezhad}, F.}, \bibinfo{author}{{Goodman}, M.L.},
  \bibinfo{year}{2006}.
\newblock \bibinfo{title}{{Magnetohydrodynamic Simulations of Solar
  Chromospheric Dynamics Using a Complete Electrical Conductivity Tensor}}.
\newblock \bibinfo{journal}{ApJS} \bibinfo{volume}{166},
  \bibinfo{pages}{613--633}.
\newblock \DOIprefix\doi{10.1086/506964}.
\bibitem[{{Khomenko} and {Collados}(2006)}]{2006ApJ...653..739K}
\bibinfo{author}{{Khomenko}, E.}, \bibinfo{author}{{Collados}, M.},
  \bibinfo{year}{2006}.
\newblock \bibinfo{title}{{Numerical Modeling of Magnetohydrodynamic Wave
  Propagation and Refraction in Sunspots}}.
\newblock \bibinfo{journal}{ApJ} \bibinfo{volume}{653},
  \bibinfo{pages}{739--755}.
\newblock \DOIprefix\doi{10.1086/507760}.
\bibitem[{{Khomenko} and {Collados}(2012)}]{2012ApJ...747...87K}
\bibinfo{author}{{Khomenko}, E.}, \bibinfo{author}{{Collados}, M.},
  \bibinfo{year}{2012}.
\newblock \bibinfo{title}{{Heating of the Magnetized Solar Chromosphere by
  Partial Ionization Effects}}.
\newblock \bibinfo{journal}{ApJ} \bibinfo{volume}{747}, \bibinfo{pages}{87}.
\newblock \DOIprefix\doi{10.1088/0004-637X/747/2/87},
  \href{http://arxiv.org/abs/1112.3374}{{\tt arXiv:1112.3374}}.
\bibitem[{{Khomenko} and {Collados}(2015)}]{2015LRSP...12....6K}
\bibinfo{author}{{Khomenko}, E.}, \bibinfo{author}{{Collados}, M.},
  \bibinfo{year}{2015}.
\newblock \bibinfo{title}{{Oscillations and Waves in Sunspots}}.
\newblock \bibinfo{journal}{Living Reviews in Solar Physics}
  \bibinfo{volume}{12}, \bibinfo{pages}{6}.
\newblock \DOIprefix\doi{10.1007/lrsp-2015-6}.
\bibitem[{{Khomenko} et~al.(2014){Khomenko}, {Collados}, {D{\'\i}az} and
  {Vitas}}]{Khomenko2014PhPl...21i2901K}
\bibinfo{author}{{Khomenko}, E.}, \bibinfo{author}{{Collados}, M.},
  \bibinfo{author}{{D{\'\i}az}, A.}, \bibinfo{author}{{Vitas}, N.},
  \bibinfo{year}{2014}.
\newblock \bibinfo{title}{{Fluid description of multi-component solar partially
  ionized plasma}}.
\newblock \bibinfo{journal}{Physics of Plasmas} \bibinfo{volume}{21},
  \bibinfo{pages}{092901}.
\newblock \DOIprefix\doi{10.1063/1.4894106},
  \href{http://arxiv.org/abs/1408.1871}{{\tt arXiv:1408.1871}}.
\bibitem[{{Khomenko} et~al.(2016){Khomenko}, {Collados} and
  {D{\'\i}az}}]{Khomenko2016ApJ...823..132K}
\bibinfo{author}{{Khomenko}, E.}, \bibinfo{author}{{Collados}, M.},
  \bibinfo{author}{{D{\'\i}az}, A.J.}, \bibinfo{year}{2016}.
\newblock \bibinfo{title}{{Observational Detection of Drift Velocity between
  Ionized and Neutral Species in Solar Prominences}}.
\newblock \bibinfo{journal}{ApJ} \bibinfo{volume}{823}, \bibinfo{pages}{132}.
\newblock \DOIprefix\doi{10.3847/0004-637X/823/2/132},
  \href{http://arxiv.org/abs/1604.01177}{{\tt arXiv:1604.01177}}.
\bibitem[{{Khomenko} et~al.(2021){Khomenko}, {Collados}, {Vitas} and
  {Gonz{\'a}lez-Morales}}]{2021RSPTA.37900176K}
\bibinfo{author}{{Khomenko}, E.}, \bibinfo{author}{{Collados}, M.},
  \bibinfo{author}{{Vitas}, N.}, \bibinfo{author}{{Gonz{\'a}lez-Morales},
  P.A.}, \bibinfo{year}{2021}.
\newblock \bibinfo{title}{{Influence of ambipolar and Hall effects on vorticity
  in three-dimensional simulations of magneto-convection}}.
\newblock \bibinfo{journal}{Philosophical Transactions of the Royal Society of
  London Series A} \bibinfo{volume}{379}, \bibinfo{pages}{20200176}.
\newblock \DOIprefix\doi{10.1098/rsta.2020.0176},
  \href{http://arxiv.org/abs/2009.09753}{{\tt arXiv:2009.09753}}.
\bibitem[{{Khomenko} et~al.(2018){Khomenko}, {Vitas}, {Collados} and {de
  Vicente}}]{2018A&A...618A..87K}
\bibinfo{author}{{Khomenko}, E.}, \bibinfo{author}{{Vitas}, N.},
  \bibinfo{author}{{Collados}, M.}, \bibinfo{author}{{de Vicente}, A.},
  \bibinfo{year}{2018}.
\newblock \bibinfo{title}{{Three-dimensional simulations of solar
  magneto-convection including effects of partial ionization}}.
\newblock \bibinfo{journal}{A\&A} \bibinfo{volume}{618}, \bibinfo{pages}{A87}.
\newblock \DOIprefix\doi{10.1051/0004-6361/201833048},
  \href{http://arxiv.org/abs/1807.01061}{{\tt arXiv:1807.01061}}.
\bibitem[{{Krall} and {Trivelpiece}(1973)}]{Krall+Trivelpiece1973}
\bibinfo{author}{{Krall}, N.A.}, \bibinfo{author}{{Trivelpiece}, A.W.},
  \bibinfo{year}{1973}.
\newblock \bibinfo{title}{{Principles of plasma physics}}.
\bibitem[{{Krasnoselskikh} et~al.(2010){Krasnoselskikh}, {Vekstein}, {Hudson},
  {Bale} and {Abbett}}]{2010ApJ...724.1542K}
\bibinfo{author}{{Krasnoselskikh}, V.}, \bibinfo{author}{{Vekstein}, G.},
  \bibinfo{author}{{Hudson}, H.S.}, \bibinfo{author}{{Bale}, S.D.},
  \bibinfo{author}{{Abbett}, W.P.}, \bibinfo{year}{2010}.
\newblock \bibinfo{title}{{Generation of Electric Currents in the Chromosphere
  via Neutral-Ion Drag}}.
\newblock \bibinfo{journal}{ApJ} \bibinfo{volume}{724},
  \bibinfo{pages}{1542--1550}.
\newblock \DOIprefix\doi{10.1088/0004-637X/724/2/1542},
  \href{http://arxiv.org/abs/1011.5834}{{\tt arXiv:1011.5834}}.
\bibitem[{{Kulsrud} and {Pearce}(1969)}]{Kulsrud1969ApJ...156..445K}
\bibinfo{author}{{Kulsrud}, R.}, \bibinfo{author}{{Pearce}, W.P.},
  \bibinfo{year}{1969}.
\newblock \bibinfo{title}{{The Effect of Wave-Particle Interactions on the
  Propagation of Cosmic Rays}}.
\newblock \bibinfo{journal}{ApJ} \bibinfo{volume}{156}, \bibinfo{pages}{445}.
\newblock \DOIprefix\doi{10.1086/149981}.
\bibitem[{{Leake} et~al.(2014){Leake}, {DeVore}, {Thayer}, {Burns}, {Crowley},
  {Gilbert}, {Huba}, {Krall}, {Linton}, {Lukin} and
  {Wang}}]{2014SSRv..184..107L}
\bibinfo{author}{{Leake}, J.E.}, \bibinfo{author}{{DeVore}, C.R.},
  \bibinfo{author}{{Thayer}, J.P.}, \bibinfo{author}{{Burns}, A.G.},
  \bibinfo{author}{{Crowley}, G.}, \bibinfo{author}{{Gilbert}, H.R.},
  \bibinfo{author}{{Huba}, J.D.}, \bibinfo{author}{{Krall}, J.},
  \bibinfo{author}{{Linton}, M.G.}, \bibinfo{author}{{Lukin}, V.S.},
  \bibinfo{author}{{Wang}, W.}, \bibinfo{year}{2014}.
\newblock \bibinfo{title}{{Ionized Plasma and Neutral Gas Coupling in the Sun's
  Chromosphere and Earth's Ionosphere/Thermosphere}}.
\newblock \bibinfo{journal}{SSRv} \bibinfo{volume}{184},
  \bibinfo{pages}{107--172}.
\newblock \DOIprefix\doi{10.1007/s11214-014-0103-1},
  \href{http://arxiv.org/abs/1310.0405}{{\tt arXiv:1310.0405}}.
\bibitem[{{Leake} et~al.(2012){Leake}, {Lukin}, {Linton} and
  {Meier}}]{2012ApJ...760..109L}
\bibinfo{author}{{Leake}, J.E.}, \bibinfo{author}{{Lukin}, V.S.},
  \bibinfo{author}{{Linton}, M.G.}, \bibinfo{author}{{Meier}, E.T.},
  \bibinfo{year}{2012}.
\newblock \bibinfo{title}{{Multi-fluid Simulations of Chromospheric Magnetic
  Reconnection in a Weakly Ionized Reacting Plasma}}.
\newblock \bibinfo{journal}{ApJ} \bibinfo{volume}{760}, \bibinfo{pages}{109}.
\newblock \DOIprefix\doi{10.1088/0004-637X/760/2/109},
  \href{http://arxiv.org/abs/1210.1807}{{\tt arXiv:1210.1807}}.
\bibitem[{{Leenaarts} et~al.(2007){Leenaarts}, {Carlsson}, {Hansteen} and
  {Rutten}}]{2007A&A...473..625L}
\bibinfo{author}{{Leenaarts}, J.}, \bibinfo{author}{{Carlsson}, M.},
  \bibinfo{author}{{Hansteen}, V.}, \bibinfo{author}{{Rutten}, R.J.},
  \bibinfo{year}{2007}.
\newblock \bibinfo{title}{{Non-equilibrium hydrogen ionization in 2D
  simulations of the solar atmosphere}}.
\newblock \bibinfo{journal}{A\&A} \bibinfo{volume}{473},
  \bibinfo{pages}{625--632}.
\newblock \DOIprefix\doi{10.1051/0004-6361:20078161},
  \href{http://arxiv.org/abs/0709.3751}{{\tt arXiv:0709.3751}}.
\bibitem[{{Maneva} et~al.(2017){Maneva}, {Alvarez Laguna}, {Lani} and
  {Poedts}}]{2017ApJ...836..197M}
\bibinfo{author}{{Maneva}, Y.G.}, \bibinfo{author}{{Alvarez Laguna}, A.},
  \bibinfo{author}{{Lani}, A.}, \bibinfo{author}{{Poedts}, S.},
  \bibinfo{year}{2017}.
\newblock \bibinfo{title}{{Multi-fluid Modeling of Magnetosonic Wave
  Propagation in the Solar Chromosphere: Effects of Impact Ionization and
  Radiative Recombination}}.
\newblock \bibinfo{journal}{ApJ} \bibinfo{volume}{836}, \bibinfo{pages}{197}.
\newblock \DOIprefix\doi{10.3847/1538-4357/aa5b83},
  \href{http://arxiv.org/abs/1611.08439}{{\tt arXiv:1611.08439}}.
\bibitem[{{Mart{\'\i}nez-G{\'o}mez} et~al.(2021){Mart{\'\i}nez-G{\'o}mez},
  {Popescu Braileanu}, {Khomenko} and {Hunana}}]{2021A&A...650A.123M}
\bibinfo{author}{{Mart{\'\i}nez-G{\'o}mez}, D.}, \bibinfo{author}{{Popescu
  Braileanu}, B.}, \bibinfo{author}{{Khomenko}, E.}, \bibinfo{author}{{Hunana},
  P.}, \bibinfo{year}{2021}.
\newblock \bibinfo{title}{{Simulations of the Biermann battery mechanism in
  two-fluid partially ionised plasmas}}.
\newblock \bibinfo{journal}{A\&A} \bibinfo{volume}{650}, \bibinfo{pages}{A123}.
\newblock \DOIprefix\doi{10.1051/0004-6361/202039113},
  \href{http://arxiv.org/abs/2104.06956}{{\tt arXiv:2104.06956}}.
\bibitem[{{Mart{\'\i}nez-G{\'o}mez} et~al.(2016){Mart{\'\i}nez-G{\'o}mez},
  {Soler} and {Terradas}}]{MartinezGomez2016ApJ...832..101M}
\bibinfo{author}{{Mart{\'\i}nez-G{\'o}mez}, D.}, \bibinfo{author}{{Soler}, R.},
  \bibinfo{author}{{Terradas}, J.}, \bibinfo{year}{2016}.
\newblock \bibinfo{title}{{Multi-fluid Approach to High-frequency Waves in
  Plasmas: I. Small-amplitude Regime in Fully Ionized Medium}}.
\newblock \bibinfo{journal}{ApJ} \bibinfo{volume}{832}, \bibinfo{pages}{101}.
\newblock \DOIprefix\doi{10.3847/0004-637X/832/2/101},
  \href{http://arxiv.org/abs/1609.06190}{{\tt arXiv:1609.06190}}.
\bibitem[{{Mart{\'\i}nez-G{\'o}mez} et~al.(2017){Mart{\'\i}nez-G{\'o}mez},
  {Soler} and {Terradas}}]{MartinezGomez2017ApJ...837...80M}
\bibinfo{author}{{Mart{\'\i}nez-G{\'o}mez}, D.}, \bibinfo{author}{{Soler}, R.},
  \bibinfo{author}{{Terradas}, J.}, \bibinfo{year}{2017}.
\newblock \bibinfo{title}{{Multi-fluid Approach to High-frequency Waves in
  Plasmas. II. Small-amplitude Regime in Partially Ionized Media}}.
\newblock \bibinfo{journal}{ApJ} \bibinfo{volume}{837}, \bibinfo{pages}{80}.
\newblock \DOIprefix\doi{10.3847/1538-4357/aa5eab},
  \href{http://arxiv.org/abs/1703.05093}{{\tt arXiv:1703.05093}}.
\bibitem[{{Mart{\'\i}nez-G{\'o}mez} et~al.(2018){Mart{\'\i}nez-G{\'o}mez},
  {Soler} and {Terradas}}]{MartinezGomez2018ApJ...856...16M}
\bibinfo{author}{{Mart{\'\i}nez-G{\'o}mez}, D.}, \bibinfo{author}{{Soler}, R.},
  \bibinfo{author}{{Terradas}, J.}, \bibinfo{year}{2018}.
\newblock \bibinfo{title}{{Multi-fluid Approach to High-frequency Waves in
  Plasmas. III. Nonlinear Regime and Plasma Heating}}.
\newblock \bibinfo{journal}{ApJ} \bibinfo{volume}{856}, \bibinfo{pages}{16}.
\newblock \DOIprefix\doi{10.3847/1538-4357/aab156},
  \href{http://arxiv.org/abs/1802.08134}{{\tt arXiv:1802.08134}}.
\bibitem[{{Mart{\'\i}nez-Sykora} et~al.(2012){Mart{\'\i}nez-Sykora}, {De
  Pontieu} and {Hansteen}}]{2012ApJ...753..161M}
\bibinfo{author}{{Mart{\'\i}nez-Sykora}, J.}, \bibinfo{author}{{De Pontieu},
  B.}, \bibinfo{author}{{Hansteen}, V.}, \bibinfo{year}{2012}.
\newblock \bibinfo{title}{{Two-dimensional Radiative Magnetohydrodynamic
  Simulations of the Importance of Partial Ionization in the Chromosphere}}.
\newblock \bibinfo{journal}{ApJ} \bibinfo{volume}{753}, \bibinfo{pages}{161}.
\newblock \DOIprefix\doi{10.1088/0004-637X/753/2/161},
  \href{http://arxiv.org/abs/1204.5991}{{\tt arXiv:1204.5991}}.
\bibitem[{{Meier}(2011)}]{2011PhDT.......208M}
\bibinfo{author}{{Meier}, E.T.}, \bibinfo{year}{2011}.
\newblock \bibinfo{title}{{Modeling Plasmas with Strong Anisotropy, Neutral
  Fluid Effects, and Open Boundaries}}.
\newblock Ph.D. thesis. University of Washington, Seattle.
\bibitem[{{Meier} and {Shumlak}(2012)}]{2012PhPl...19g2508M}
\bibinfo{author}{{Meier}, E.T.}, \bibinfo{author}{{Shumlak}, U.},
  \bibinfo{year}{2012}.
\newblock \bibinfo{title}{{A general nonlinear fluid model for reacting
  plasma-neutral mixtures}}.
\newblock \bibinfo{journal}{Physics of Plasmas} \bibinfo{volume}{19},
  \bibinfo{pages}{072508}.
\newblock \DOIprefix\doi{10.1063/1.4736975}.
\bibitem[{{Mihalas} and {Mihalas}(1984)}]{1984oup..book.....M}
\bibinfo{author}{{Mihalas}, D.}, \bibinfo{author}{{Mihalas}, B.W.},
  \bibinfo{year}{1984}.
\newblock \bibinfo{title}{{Foundations of radiation hydrodynamics}}.
\bibitem[{{Milne} et~al.(1979){Milne}, {Priest} and
  {Roberts}}]{1979ApJ...232..304M}
\bibinfo{author}{{Milne}, A.M.}, \bibinfo{author}{{Priest}, E.R.},
  \bibinfo{author}{{Roberts}, B.}, \bibinfo{year}{1979}.
\newblock \bibinfo{title}{{A model for quiescent solar prominences.}}
\newblock \bibinfo{journal}{ApJ} \bibinfo{volume}{232},
  \bibinfo{pages}{304--317}.
\newblock \DOIprefix\doi{10.1086/157290}.
\bibitem[{{Murtas} et~al.(2022){Murtas}, {Hillier} and
  {Snow}}]{2022arXiv220511091M}
\bibinfo{author}{{Murtas}, G.}, \bibinfo{author}{{Hillier}, A.},
  \bibinfo{author}{{Snow}, B.}, \bibinfo{year}{2022}.
\newblock \bibinfo{title}{{Collisional ionisation and recombination effects on
  coalescence instability in chromospheric partially ionised plasmas}}.
\newblock \bibinfo{journal}{arXiv e-prints} ,
  \bibinfo{pages}{arXiv:2205.11091}\href{http://arxiv.org/abs/2205.11091}{{\tt
  arXiv:2205.11091}}.
\bibitem[{{Pandey} and {Wardle}(2008)}]{2008MNRAS.385.2269P}
\bibinfo{author}{{Pandey}, B.P.}, \bibinfo{author}{{Wardle}, M.},
  \bibinfo{year}{2008}.
\newblock \bibinfo{title}{{Hall magnetohydrodynamics of partially ionized
  plasmas}}.
\newblock \bibinfo{journal}{MNRAS} \bibinfo{volume}{385},
  \bibinfo{pages}{2269--2278}.
\newblock \DOIprefix\doi{10.1111/j.1365-2966.2008.12998.x},
  \href{http://arxiv.org/abs/0707.2688}{{\tt arXiv:0707.2688}}.
\bibitem[{{Popescu Braileanu} and {Keppens}(2021)}]{2021A&A...653A.131P}
\bibinfo{author}{{Popescu Braileanu}, B.}, \bibinfo{author}{{Keppens}, R.},
  \bibinfo{year}{2021}.
\newblock \bibinfo{title}{{Effects of ambipolar diffusion on waves in the solar
  chromosphere}}.
\newblock \bibinfo{journal}{A\&A} \bibinfo{volume}{653}, \bibinfo{pages}{A131}.
\newblock \DOIprefix\doi{10.1051/0004-6361/202140872},
  \href{http://arxiv.org/abs/2105.10285}{{\tt arXiv:2105.10285}}.
\bibitem[{{Popescu Braileanu} et~al.(2019a){Popescu Braileanu}, {Lukin},
  {Khomenko} and {de Vicente}}]{PopescuBraileanu2019A&A...627A..25P}
\bibinfo{author}{{Popescu Braileanu}, B.}, \bibinfo{author}{{Lukin}, V.S.},
  \bibinfo{author}{{Khomenko}, E.}, \bibinfo{author}{{de Vicente}, {\'A}.},
  \bibinfo{year}{2019}a.
\newblock \bibinfo{title}{{Two-fluid simulations of waves in the solar
  chromosphere. I. Numerical code verification}}.
\newblock \bibinfo{journal}{A\&A} \bibinfo{volume}{627}, \bibinfo{pages}{A25}.
\newblock \DOIprefix\doi{10.1051/0004-6361/201834154},
  \href{http://arxiv.org/abs/1905.03559}{{\tt arXiv:1905.03559}}.
\bibitem[{{Popescu Braileanu} et~al.(2019b){Popescu Braileanu}, {Lukin},
  {Khomenko} and {de Vicente}}]{PopescuBraileanu2019A&A...630A..79P}
\bibinfo{author}{{Popescu Braileanu}, B.}, \bibinfo{author}{{Lukin}, V.S.},
  \bibinfo{author}{{Khomenko}, E.}, \bibinfo{author}{{de Vicente}, {\'A}.},
  \bibinfo{year}{2019}b.
\newblock \bibinfo{title}{{Two-fluid simulations of waves in the solar
  chromosphere. II. Propagation and damping of fast magneto-acoustic waves and
  shocks}}.
\newblock \bibinfo{journal}{A\&A} \bibinfo{volume}{630}, \bibinfo{pages}{A79}.
\newblock \DOIprefix\doi{10.1051/0004-6361/201935844},
  \href{http://arxiv.org/abs/1908.05262}{{\tt arXiv:1908.05262}}.
\bibitem[{{Priest}(2014)}]{Priest}
\bibinfo{author}{{Priest}, E.}, \bibinfo{year}{2014}.
\newblock \bibinfo{title}{{Magnetohydrodynamics of the Sun}}.
\newblock \DOIprefix\doi{10.1017/CBO9781139020732}.
\bibitem[{{Raboonik} and {Cally}(2019)}]{2019SoPh..294..147R}
\bibinfo{author}{{Raboonik}, A.}, \bibinfo{author}{{Cally}, P.S.},
  \bibinfo{year}{2019}.
\newblock \bibinfo{title}{{Hall-coupling of Slow and Alfv{\'e}n Waves at Low
  Frequencies in the Lower Solar Atmosphere}}.
\newblock \bibinfo{journal}{Solar Phys.} \bibinfo{volume}{294},
  \bibinfo{pages}{147}.
\newblock \DOIprefix\doi{10.1007/s11207-019-1544-1}.
\bibitem[{{Rosner} et~al.(1978){Rosner}, {Tucker} and
  {Vaiana}}]{1978ApJ...220..643R}
\bibinfo{author}{{Rosner}, R.}, \bibinfo{author}{{Tucker}, W.H.},
  \bibinfo{author}{{Vaiana}, G.S.}, \bibinfo{year}{1978}.
\newblock \bibinfo{title}{{Dynamics of the quiescent solar corona.}}
\newblock \bibinfo{journal}{ApJ} \bibinfo{volume}{220},
  \bibinfo{pages}{643--645}.
\newblock \DOIprefix\doi{10.1086/155949}.
\bibitem[{{Schunk}(1977)}]{Schunk1977RvGSP..15..429S}
\bibinfo{author}{{Schunk}, R.W.}, \bibinfo{year}{1977}.
\newblock \bibinfo{title}{{Mathematical Structure of Transport Equations for
  Multispecies Flows (Paper 7R0585)}}.
\newblock \bibinfo{journal}{Reviews of Geophysics and Space Physics}
  \bibinfo{volume}{15}, \bibinfo{pages}{429}.
\newblock \DOIprefix\doi{10.1029/RG015i004p00429}.
\bibitem[{{Shelyag} et~al.(2016){Shelyag}, {Khomenko}, {de Vicente} and
  {Przybylski}}]{2016ApJ...819L..11S}
\bibinfo{author}{{Shelyag}, S.}, \bibinfo{author}{{Khomenko}, E.},
  \bibinfo{author}{{de Vicente}, A.}, \bibinfo{author}{{Przybylski}, D.},
  \bibinfo{year}{2016}.
\newblock \bibinfo{title}{{Heating of the Partially Ionized Solar Chromosphere
  by Waves in Magnetic Structures}}.
\newblock \bibinfo{journal}{ApJL} \bibinfo{volume}{819}, \bibinfo{pages}{L11}.
\newblock \DOIprefix\doi{10.3847/2041-8205/819/1/L11},
  \href{http://arxiv.org/abs/1602.03373}{{\tt arXiv:1602.03373}}.
\bibitem[{{Snow} and {Hillier}(2019)}]{Snow2019A&A...626A..46S}
\bibinfo{author}{{Snow}, B.}, \bibinfo{author}{{Hillier}, A.},
  \bibinfo{year}{2019}.
\newblock \bibinfo{title}{{Intermediate shock sub-structures within a slow-mode
  shock occurring in partially ionised plasma}}.
\newblock \bibinfo{journal}{A\&A} \bibinfo{volume}{626}, \bibinfo{pages}{A46}.
\newblock \DOIprefix\doi{10.1051/0004-6361/201935326},
  \href{http://arxiv.org/abs/1904.12518}{{\tt arXiv:1904.12518}}.
\bibitem[{{Snow} and {Hillier}(2020)}]{Snow2020A&A...637A..97S}
\bibinfo{author}{{Snow}, B.}, \bibinfo{author}{{Hillier}, A.},
  \bibinfo{year}{2020}.
\newblock \bibinfo{title}{{Mode conversion of two-fluid shocks in a
  partially-ionised, isothermal, stratified atmosphere}}.
\newblock \bibinfo{journal}{A\&A} \bibinfo{volume}{637}, \bibinfo{pages}{A97}.
\newblock \DOIprefix\doi{10.1051/0004-6361/202037848},
  \href{http://arxiv.org/abs/2004.02550}{{\tt arXiv:2004.02550}}.
\bibitem[{{Soler} et~al.(2015a){Soler}, {Ballester} and
  {Zaqarashvili}}]{2015A&A...573A..79S}
\bibinfo{author}{{Soler}, R.}, \bibinfo{author}{{Ballester}, J.L.},
  \bibinfo{author}{{Zaqarashvili}, T.V.}, \bibinfo{year}{2015}a.
\newblock \bibinfo{title}{{Overdamped Alfv{\'e}n waves due to ion-neutral
  collisions in the solar chromosphere}}.
\newblock \bibinfo{journal}{A\&A} \bibinfo{volume}{573}, \bibinfo{pages}{A79}.
\newblock \DOIprefix\doi{10.1051/0004-6361/201423930},
  \href{http://arxiv.org/abs/1411.5887}{{\tt arXiv:1411.5887}}.
\bibitem[{{Soler} et~al.(2013a){Soler}, {Carbonell} and
  {Ballester}}]{Soler2013ApJS..209...16S}
\bibinfo{author}{{Soler}, R.}, \bibinfo{author}{{Carbonell}, M.},
  \bibinfo{author}{{Ballester}, J.L.}, \bibinfo{year}{2013}a.
\newblock \bibinfo{title}{{Magnetoacoustic Waves in a Partially Ionized
  Two-fluid Plasma}}.
\newblock \bibinfo{journal}{ApJS} \bibinfo{volume}{209}, \bibinfo{pages}{16}.
\newblock \DOIprefix\doi{10.1088/0067-0049/209/1/16},
  \href{http://arxiv.org/abs/1309.7204}{{\tt arXiv:1309.7204}}.
\bibitem[{{Soler} et~al.(2015b){Soler}, {Carbonell} and
  {Ballester}}]{Soler2015ApJ...810..146S}
\bibinfo{author}{{Soler}, R.}, \bibinfo{author}{{Carbonell}, M.},
  \bibinfo{author}{{Ballester}, J.L.}, \bibinfo{year}{2015}b.
\newblock \bibinfo{title}{{On the Spatial Scales of Wave Heating in the Solar
  Chromosphere}}.
\newblock \bibinfo{journal}{ApJ} \bibinfo{volume}{810}, \bibinfo{pages}{146}.
\newblock \DOIprefix\doi{10.1088/0004-637X/810/2/146},
  \href{http://arxiv.org/abs/1508.01497}{{\tt arXiv:1508.01497}}.
\bibitem[{{Soler} et~al.(2013b){Soler}, {Carbonell}, {Ballester} and
  {Terradas}}]{Soler2013ApJ...767..171S}
\bibinfo{author}{{Soler}, R.}, \bibinfo{author}{{Carbonell}, M.},
  \bibinfo{author}{{Ballester}, J.L.}, \bibinfo{author}{{Terradas}, J.},
  \bibinfo{year}{2013}b.
\newblock \bibinfo{title}{{Alfv{\'e}n Waves in a Partially Ionized Two-fluid
  Plasma}}.
\newblock \bibinfo{journal}{ApJ} \bibinfo{volume}{767}, \bibinfo{pages}{171}.
\newblock \DOIprefix\doi{10.1088/0004-637X/767/2/171},
  \href{http://arxiv.org/abs/1303.4297}{{\tt arXiv:1303.4297}}.
\bibitem[{{Soler} et~al.(2009){Soler}, {Oliver} and
  {Ballester}}]{2009ApJ...699.1553S}
\bibinfo{author}{{Soler}, R.}, \bibinfo{author}{{Oliver}, R.},
  \bibinfo{author}{{Ballester}, J.L.}, \bibinfo{year}{2009}.
\newblock \bibinfo{title}{{Magnetohydrodynamic Waves in a Partially Ionized
  Filament Thread}}.
\newblock \bibinfo{journal}{ApJ} \bibinfo{volume}{699},
  \bibinfo{pages}{1553--1562}.
\newblock \DOIprefix\doi{10.1088/0004-637X/699/2/1553},
  \href{http://arxiv.org/abs/0904.3013}{{\tt arXiv:0904.3013}}.
\bibitem[{{Song} and {Vasyli{\={u}}nas}(2011)}]{Song2011JGRA..116.9104S}
\bibinfo{author}{{Song}, P.}, \bibinfo{author}{{Vasyli{\={u}}nas}, V.M.},
  \bibinfo{year}{2011}.
\newblock \bibinfo{title}{{Heating of the solar atmosphere by strong damping of
  Alfv{\'e}n waves}}.
\newblock \bibinfo{journal}{Journal of Geophysical Research (Space Physics)}
  \bibinfo{volume}{116}, \bibinfo{pages}{A09104}.
\newblock \DOIprefix\doi{10.1029/2011JA016679}.
\bibitem[{{Spitzer}(1962)}]{Spitzer1962pfig.book.....S}
\bibinfo{author}{{Spitzer}, L.}, \bibinfo{year}{1962}.
\newblock \bibinfo{title}{{Physics of Fully Ionized Gases}}.
\bibitem[{{Srivastava} et~al.(2021){Srivastava}, {Ballester}, {Cally},
  {Carlsson}, {Goossens}, {Jess}, {Khomenko}, {Mathioudakis}, {Murawski} and
  {Zaqarashvili}}]{2021JGRA..12629097S}
\bibinfo{author}{{Srivastava}, A.K.}, \bibinfo{author}{{Ballester}, J.L.},
  \bibinfo{author}{{Cally}, P.S.}, \bibinfo{author}{{Carlsson}, M.},
  \bibinfo{author}{{Goossens}, M.}, \bibinfo{author}{{Jess}, D.B.},
  \bibinfo{author}{{Khomenko}, E.}, \bibinfo{author}{{Mathioudakis}, M.},
  \bibinfo{author}{{Murawski}, K.}, \bibinfo{author}{{Zaqarashvili}, T.V.},
  \bibinfo{year}{2021}.
\newblock \bibinfo{title}{{Chromospheric Heating by Magnetohydrodynamic Waves
  and Instabilities}}.
\newblock \bibinfo{journal}{Journal of Geophysical Research (Space Physics)}
  \bibinfo{volume}{126}, \bibinfo{pages}{e029097}.
\newblock \DOIprefix\doi{10.1029/2020JA029097},
  \href{http://arxiv.org/abs/2104.02010}{{\tt arXiv:2104.02010}}.
\bibitem[{{Tu} and {Song}(2013)}]{TuSong2013ApJ...777...53T}
\bibinfo{author}{{Tu}, J.}, \bibinfo{author}{{Song}, P.}, \bibinfo{year}{2013}.
\newblock \bibinfo{title}{{A Study of Alfv{\'e}n Wave Propagation and Heating
  the Chromosphere}}.
\newblock \bibinfo{journal}{ApJ} \bibinfo{volume}{777}, \bibinfo{pages}{53}.
\newblock \DOIprefix\doi{10.1088/0004-637X/777/1/53}.
\bibitem[{{Vasyli{\={u}}nas} and {Song}(2005)}]{Vasyliunas2005JGRA..110.2301V}
\bibinfo{author}{{Vasyli{\={u}}nas}, V.M.}, \bibinfo{author}{{Song}, P.},
  \bibinfo{year}{2005}.
\newblock \bibinfo{title}{{Meaning of ionospheric Joule heating}}.
\newblock \bibinfo{journal}{Journal of Geophysical Research (Space Physics)}
  \bibinfo{volume}{110}, \bibinfo{pages}{A02301}.
\newblock \DOIprefix\doi{10.1029/2004JA010615}.
\bibitem[{{Vernazza} et~al.(1981){Vernazza}, {Avrett} and
  {Loeser}}]{Vernazza1981ApJS...45..635V}
\bibinfo{author}{{Vernazza}, J.E.}, \bibinfo{author}{{Avrett}, E.H.},
  \bibinfo{author}{{Loeser}, R.}, \bibinfo{year}{1981}.
\newblock \bibinfo{title}{{Structure of the solar chromosphere. III. Models of
  the EUV brightness components of the quiet sun.}}
\newblock \bibinfo{journal}{ApJS} \bibinfo{volume}{45},
  \bibinfo{pages}{635--725}.
\newblock \DOIprefix\doi{10.1086/190731}.
\bibitem[{{Vranjes} and {Krstic}(2013)}]{Vranjes2013AA...554A..22V}
\bibinfo{author}{{Vranjes}, J.}, \bibinfo{author}{{Krstic}, P.S.},
  \bibinfo{year}{2013}.
\newblock \bibinfo{title}{{Collisions, magnetization, and transport
  coefficients in the lower solar atmosphere}}.
\newblock \bibinfo{journal}{A\&A} \bibinfo{volume}{554}, \bibinfo{pages}{A22}.
\newblock \DOIprefix\doi{10.1051/0004-6361/201220738},
  \href{http://arxiv.org/abs/1304.4010}{{\tt arXiv:1304.4010}}.
\bibitem[{{Wardle}(1999)}]{1999MNRAS.307..849W}
\bibinfo{author}{{Wardle}, M.}, \bibinfo{year}{1999}.
\newblock \bibinfo{title}{{The Balbus-Hawley instability in weakly ionized
  discs}}.
\newblock \bibinfo{journal}{MNRAS} \bibinfo{volume}{307},
  \bibinfo{pages}{849--856}.
\newblock \DOIprefix\doi{10.1046/j.1365-8711.1999.02670.x},
  \href{http://arxiv.org/abs/astro-ph/9809349}{{\tt arXiv:astro-ph/9809349}}.
\bibitem[{{Wiehr} et~al.(2021){Wiehr}, {Stellmacher}, {Balthasar} and
  {Bianda}}]{Wiehr2021ApJ...920...47W}
\bibinfo{author}{{Wiehr}, E.}, \bibinfo{author}{{Stellmacher}, G.},
  \bibinfo{author}{{Balthasar}, H.}, \bibinfo{author}{{Bianda}, M.},
  \bibinfo{year}{2021}.
\newblock \bibinfo{title}{{Velocity Difference of Ions and Neutrals in Solar
  Prominences}}.
\newblock \bibinfo{journal}{ApJ} \bibinfo{volume}{920}, \bibinfo{pages}{47}.
\newblock \DOIprefix\doi{10.3847/1538-4357/ac1791},
  \href{http://arxiv.org/abs/2108.13103}{{\tt arXiv:2108.13103}}.
\bibitem[{{Wiehr} et~al.(2019){Wiehr}, {Stellmacher} and
  {Bianda}}]{Wiehr2019ApJ...873..125W}
\bibinfo{author}{{Wiehr}, E.}, \bibinfo{author}{{Stellmacher}, G.},
  \bibinfo{author}{{Bianda}, M.}, \bibinfo{year}{2019}.
\newblock \bibinfo{title}{{Evidence for the Two-fluid Scenario in Solar
  Prominences}}.
\newblock \bibinfo{journal}{ApJ} \bibinfo{volume}{873}, \bibinfo{pages}{125}.
\newblock \DOIprefix\doi{10.3847/1538-4357/ab04a4},
  \href{http://arxiv.org/abs/1904.01536}{{\tt arXiv:1904.01536}}.
\bibitem[{{Withbroe} and {Noyes}(1977)}]{Withbroe1977ARA&A..15..363W}
\bibinfo{author}{{Withbroe}, G.L.}, \bibinfo{author}{{Noyes}, R.W.},
  \bibinfo{year}{1977}.
\newblock \bibinfo{title}{{Mass and energy flow in the solar chromosphere and
  corona.}}
\newblock \bibinfo{journal}{ARA\&A} \bibinfo{volume}{15},
  \bibinfo{pages}{363--387}.
\newblock \DOIprefix\doi{10.1146/annurev.aa.15.090177.002051}.
\bibitem[{{Zapi{\'o}r} et~al.(2022){Zapi{\'o}r}, {Heinzel} and
  {Khomenko}}]{Zapior2022ApJ...934...16Z}
\bibinfo{author}{{Zapi{\'o}r}, M.}, \bibinfo{author}{{Heinzel}, P.},
  \bibinfo{author}{{Khomenko}, E.}, \bibinfo{year}{2022}.
\newblock \bibinfo{title}{{Doppler-velocity Drifts Detected in a Solar
  Prominence}}.
\newblock \bibinfo{journal}{ApJ} \bibinfo{volume}{934}, \bibinfo{pages}{16}.
\newblock \DOIprefix\doi{10.3847/1538-4357/ac778a}.
\bibitem[{{Zaqarashvili} et~al.(2012){Zaqarashvili}, {Carbonell}, {Ballester}
  and {Khodachenko}}]{2012A&A...544A.143Z}
\bibinfo{author}{{Zaqarashvili}, T.V.}, \bibinfo{author}{{Carbonell}, M.},
  \bibinfo{author}{{Ballester}, J.L.}, \bibinfo{author}{{Khodachenko}, M.L.},
  \bibinfo{year}{2012}.
\newblock \bibinfo{title}{{Cut-off wavenumber of Alfv{\'e}n waves in partially
  ionized plasmas of the solar atmosphere}}.
\newblock \bibinfo{journal}{A\&A} \bibinfo{volume}{544}, \bibinfo{pages}{A143}.
\newblock \DOIprefix\doi{10.1051/0004-6361/201219763},
  \href{http://arxiv.org/abs/1207.5377}{{\tt arXiv:1207.5377}}.
\bibitem[{{Zaqarashvili} et~al.(2011){Zaqarashvili}, {Khodachenko} and
  {Rucker}}]{Zaqarashvili2011A&A...534A..93Z}
\bibinfo{author}{{Zaqarashvili}, T.V.}, \bibinfo{author}{{Khodachenko}, M.L.},
  \bibinfo{author}{{Rucker}, H.O.}, \bibinfo{year}{2011}.
\newblock \bibinfo{title}{{Damping of Alfv{\'e}n waves in solar partially
  ionized plasmas: effect of neutral helium in multi-fluid approach}}.
\newblock \bibinfo{journal}{A\&A} \bibinfo{volume}{534}, \bibinfo{pages}{A93}.
\newblock \DOIprefix\doi{10.1051/0004-6361/201117380},
  \href{http://arxiv.org/abs/1109.1154}{{\tt arXiv:1109.1154}}.
\bibitem[{{Zaqarashvili} et~al.(2013){Zaqarashvili}, {Khodachenko} and
  {Soler}}]{2013A&A...549A.113Z}
\bibinfo{author}{{Zaqarashvili}, T.V.}, \bibinfo{author}{{Khodachenko}, M.L.},
  \bibinfo{author}{{Soler}, R.}, \bibinfo{year}{2013}.
\newblock \bibinfo{title}{{Torsional Alfv{\'e}n waves in partially ionized
  solar plasma: effects of neutral helium and stratification}}.
\newblock \bibinfo{journal}{A\&A} \bibinfo{volume}{549}, \bibinfo{pages}{A113}.
\newblock \DOIprefix\doi{10.1051/0004-6361/201220272},
  \href{http://arxiv.org/abs/1211.1348}{{\tt arXiv:1211.1348}}.
\bibitem[{{Zhang} et~al.(2021){Zhang}, {Poedts}, {Lani}, {Ku{\'z}ma} and
  {Murawski}}]{Zhang2021ApJ...911..119Z}
\bibinfo{author}{{Zhang}, F.}, \bibinfo{author}{{Poedts}, S.},
  \bibinfo{author}{{Lani}, A.}, \bibinfo{author}{{Ku{\'z}ma}, B.},
  \bibinfo{author}{{Murawski}, K.}, \bibinfo{year}{2021}.
\newblock \bibinfo{title}{{Two-fluid Modeling of Acoustic Wave Propagation in
  Gravitationally Stratified Isothermal Media}}.
\newblock \bibinfo{journal}{ApJ} \bibinfo{volume}{911}, \bibinfo{pages}{119}.
\newblock \DOIprefix\doi{10.3847/1538-4357/abe7e8},
  \href{http://arxiv.org/abs/2011.13469}{{\tt arXiv:2011.13469}}.
\bibitem[{{Zhugzhda} and {Dzhalilov}(1984)}]{1984A&A...132...45Z}
\bibinfo{author}{{Zhugzhda}, I.D.}, \bibinfo{author}{{Dzhalilov}, N.S.},
  \bibinfo{year}{1984}.
\newblock \bibinfo{title}{{Magneto-acoustic-gravity waves on the Sun. I - Exact
  solution for an oblique magnetic field}}.
\newblock \bibinfo{journal}{A\&A} \bibinfo{volume}{132},
  \bibinfo{pages}{45--51}.

\end{thebibliography}

\Backmatter

 \bibliographystyle{elsarticle-harv}

\end{document}